\documentclass[12pt]{article}%
\usepackage[sort]{cite}
\usepackage{graphicx}
\usepackage{multicol}
\usepackage{amsfonts}
\usepackage{amssymb}
\usepackage{amsmath}
\usepackage{heck}
\usepackage{afterpage}
\usepackage{setspace}
\usepackage{verbatim}
\usepackage{color}
\usepackage{longtable}
\usepackage{standalone}
\usepackage{float}
\usepackage{subcaption}
\usepackage{epsfig}
\usepackage{epstopdf}
\usepackage{adjustbox}
\usepackage{tikz}
\usepackage[margin=1in]{geometry}
\usepackage{titletoc}
\usepackage{hyperref}%
\usepackage{braket}
\usepackage{mathtools}
\usepackage{bbm}
\providecommand{\U}[1]{\protect\rule{.1in}{.1in}}
\pdfoutput=1
\newsavebox{\mysavebox}

\hypersetup{colorlinks,citecolor=black,filecolor=black,linkcolor=black,urlcolor=black}
\numberwithin{equation}{section}

\newcommand{\ba}{\begin{eqnarray}}
\newcommand{\ea}{\end{eqnarray}}

\newcommand{\be}{\begin{equation}}
\newcommand{\ee}{\end{equation}}

\usepackage{scalerel}

\usepackage{tikz}
\usetikzlibrary{arrows}
\usetikzlibrary{arrows.meta}
\usetikzlibrary{shapes.geometric,calc,arrows, positioning,shapes.misc,decorations.markings}
\tikzset{
  big arrow/.style={
    decoration={markings,mark=at position 1 with {\arrow[scale=2,#1]{>}}},
    postaction={decorate},
    shorten >=0.4pt},
  big arrow/.default=black}

\pgfdeclarelayer{edgelayer}
\pgfdeclarelayer{nodelayer}
\pgfsetlayers{edgelayer,nodelayer,main}
\tikzstyle{none}=[inner sep=0pt]

\tikzstyle{NodeCross}=[draw, shape=circle, cross out, inner sep=0pt, minimum size=6pt,line width=0.25mm]
\tikzstyle{Circle}=[draw, shape=circle, black, fill=black, inner sep=0pt, minimum size=6pt]
\tikzstyle{Star}=[draw, shape=star, fill=black, star points=8, inner sep=0pt, minimum size=8pt]
\tikzstyle{CircleRed}=[draw, shape=circle, black, fill=red, inner sep=0pt, minimum size=4pt]
\tikzstyle{StarP}=[draw={rgb,255: red,128; green,0; blue,128}, shape=star, fill={rgb,256: red,128; green,0; blue,128}, star points=8, inner sep=0pt, minimum size=12pt]

\tikzstyle{DashedLine}=[-, densely dashed, line width=0.25mm]
\tikzstyle{DottedLine}=[-, dotted, line width=0.25mm]
\tikzstyle{ThickLine}=[-, line width=0.25mm]
\tikzstyle{ArrowLineRight}=[-, -{Stealth[scale=1.75]}, line width=0.1mm, scale=5]
\tikzstyle{ArrowLineRed}=[-, draw={rgb,255: red,191; green,0; blue,0}, -{Stealth[scale=1.75]}, line width=0.1mm, scale=5]
\tikzstyle{RedLine}=[-, draw={rgb,255: red,191; green,0; blue,0}, fill=none, line width=0.25mm]
\tikzstyle{DashedLineThin}=[-, densely dashed, line width=0.125mm, fill=none, draw=black]
\tikzstyle{DottedRed}=[-, dotted, draw={rgb,255: red,191; green,0; blue,0}, fill=none, line width=0.25mm]
\tikzstyle{DashedRed}=[-, densely dashed, draw={rgb,255: red,191; green,0; blue,0}, fill=none, line width=0.25mm]
\tikzstyle{BlueLine}=[-, draw={rgb,255: red,0; green,0; blue,191}, fill=none, line width=0.25mm]
\tikzstyle{ArrowLineBlue}=[-, draw={rgb,255: red,0; green,0; blue,191}, -{Stealth[scale=1.75]}, line width=0.1mm, scale=5]
\tikzstyle{GreenDoubleArrow}=[<->, draw={rgb,155: red,0; green,255; blue,0},  line width= 0.5mm, scale=5]
\tikzstyle{RedDoubleArrow}=[<->, draw={rgb,255: red,255; green,0; blue,0},  line width= 0.5mm, scale=5]

\begin{document}

\date{December 2022}

\title{Top Down Approach to \\[4mm] Topological Duality Defects}

\institution{PENN}{\centerline{$^{1}$Department of Physics and Astronomy, University of Pennsylvania, Philadelphia, PA 19104, USA}}
\institution{PENNmath}{\centerline{$^{2}$Department of Mathematics, University of Pennsylvania, Philadelphia, PA 19104, USA}}
\institution{NYU}{\centerline{$^{3}$Center for Cosmology and Particle Physics, New York University, New York, NY 10003, USA}}

\authors{
Jonathan J. Heckman\worksat{\PENN,\PENNmath}\footnote{e-mail: \texttt{jheckman@sas.upenn.edu}},
Max H\"ubner\worksat{\PENN}\footnote{e-mail: \texttt{hmax@sas.upenn.edu}},\\[4mm]
Ethan Torres\worksat{\PENN}\footnote{e-mail: \texttt{emtorres@sas.upenn.edu}},
Xingyang Yu\worksat{\NYU}\footnote{e-mail: \texttt{xy1038@nyu.edu}}, and
Hao Y. Zhang\worksat{\PENN}\footnote{e-mail: \texttt{zhangphy@sas.upenn.edu}}
}

\abstract{
Topological duality defects arise as codimension one
generalized symmetry operators in quantum field theories (QFTs) with a duality symmetry.
Recent investigations have shown that in the case of 4D $\mathcal{N} = 4$ Super Yang-Mills (SYM) theory,
an appropriate choice of (complexified) gauge coupling and global form of the gauge group can
lead to a rather rich fusion algebra for the associated defects, leading to examples of non-invertible symmetries. 
In this work we present a top down construction of these duality defects which generalizes to QFTs with lower supersymmetry, where other 0-form symmetries are often present. 
We realize the QFTs of interest via D3-branes probing $X$ a Calabi-Yau threefold cone with an isolated 
singularity at the tip of the cone. The IIB duality group descends to dualities of the 4D worldvolume theory. Non-trivial codimension one topological interfaces arise from configurations of 7-branes ``at infinity'' which implement a suitable $SL(2,\mathbb{Z})$ transformation when they are crossed. Reduction on the boundary topology $\partial X$ results in a 5D symmetry TFT. Different realizations of duality defects, such as the gauging of 1-form symmetries with certain mixed anomalies 
and half-space gauging constructions, simply amount to distinct choices of where to place the branch cuts in the 5D bulk.}

\maketitle
\tableofcontents
\enlargethispage{\baselineskip}

\setcounter{tocdepth}{2}

\newpage

\section{Introduction}

Dualities sit at the heart of some of the deepest insights into the non-perturbative dynamics of
quantum fields and strings. In the case of quantum field theories (QFTs) engineered via string theory,
these dualities can often be recast in terms of specific geometric transformations of the extra-dimensional
geometry. A particularly notable example of this sort is the famous $SL(2,\mathbb{Z})$ duality symmetry of
type IIB string theory which descends to a duality action on the QFTs realized on the worldvolume of
probe D3-branes.

Recently it has been appreciated that symmetries themselves can be generalized in a number of different
ways. In particular, in \cite{Gaiotto:2014kfa} it was argued that symmetries can be understood in
terms of corresponding topological operators (see also \cite{Gaiotto:2010be,Kapustin:2013qsa,Kapustin:2013uxa,Aharony:2013hda}).\footnote{For a partial list 
of recent work in this direction see e.g.,
\cite{Gaiotto:2014kfa,Gaiotto:2010be,Kapustin:2013qsa,Kapustin:2013uxa,Aharony:2013hda,
DelZotto:2015isa,Sharpe:2015mja, Heckman:2017uxe, Tachikawa:2017gyf,
Cordova:2018cvg,Benini:2018reh,Hsin:2018vcg,Wan:2018bns,
Thorngren:2019iar,GarciaEtxebarria:2019caf,Eckhard:2019jgg,Wan:2019soo,Bergman:2020ifi,Morrison:2020ool,
Albertini:2020mdx,Hsin:2020nts,Bah:2020uev,DelZotto:2020esg,Hason:2020yqf,Bhardwaj:2020phs,
Apruzzi:2020zot,Cordova:2020tij,Thorngren:2020aph,DelZotto:2020sop,BenettiGenolini:2020doj,
Yu:2020twi,Bhardwaj:2020ymp,DeWolfe:2020uzb,Gukov:2020btk,Iqbal:2020lrt,Hidaka:2020izy,
Brennan:2020ehu,Komargodski:2020mxz,Closset:2020afy,Thorngren:2020yht,Closset:2020scj,
Bhardwaj:2021pfz,Nguyen:2021naa,Heidenreich:2021xpr,Apruzzi:2021phx,Apruzzi:2021vcu,
Hosseini:2021ged,Cvetic:2021sxm,Buican:2021xhs,Bhardwaj:2021zrt,Iqbal:2021rkn,Braun:2021sex,
Cvetic:2021maf,Closset:2021lhd,Thorngren:2021yso,Sharpe:2021srf,Bhardwaj:2021wif,Hidaka:2021mml,
Lee:2021obi,Lee:2021crt,Hidaka:2021kkf,Koide:2021zxj,Apruzzi:2021mlh,Kaidi:2021xfk,Choi:2021kmx,
Bah:2021brs,Gukov:2021swm,Closset:2021lwy,Yu:2021zmu,Apruzzi:2021nmk,Beratto:2021xmn,Bhardwaj:2021mzl,
Debray:2021vob, Wang:2021vki,
Cvetic:2022uuu,DelZotto:2022fnw,Cvetic:2022imb,DelZotto:2022joo,DelZotto:2022ras,Bhardwaj:2022yxj,Hayashi:2022fkw,
Kaidi:2022uux,Roumpedakis:2022aik,Choi:2022jqy,
Choi:2022zal,Arias-Tamargo:2022nlf,Cordova:2022ieu, Bhardwaj:2022dyt,
Benedetti:2022zbb, Bhardwaj:2022scy,Antinucci:2022eat,Carta:2022spy,
Apruzzi:2022dlm, Heckman:2022suy, Baume:2022cot, Choi:2022rfe,
Bhardwaj:2022lsg, Lin:2022xod, Bartsch:2022mpm, Apruzzi:2022rei,
GarciaEtxebarria:2022vzq, Cherman:2022eml, Heckman:2022muc, Lu:2022ver, Niro:2022ctq, Kaidi:2022cpf,
Mekareeya:2022spm, vanBeest:2022fss, Antinucci:2022vyk, Giaccari:2022xgs, Bashmakov:2022uek,Cordova:2022fhg,
GarciaEtxebarria:2022jky, Choi:2022fgx, Robbins:2022wlr, Bhardwaj:2022kot, Bhardwaj:2022maz, Bartsch:2022ytj, Gaiotto:2020iye,Agrawal:2015dbf, Robbins:2021ibx, Robbins:2021xce,Huang:2021zvu,
Inamura:2021szw, Cherman:2021nox,Sharpe:2022ene,Bashmakov:2022jtl, Inamura:2022lun, Damia:2022bcd, Lin:2022dhv,Burbano:2021loy, Damia:2022rxw} and \cite{Cordova:2022ruw} for a recent review.}

The fact that generalized symmetry operators are topological also suggests that the ``worldvolume'' itself may support a 
non-trivial topological field theory. One striking consequence of this fact is that the product of 
two generalized symmetry operators may produce a sum of symmetry operators, i.e., there can be a non-trivial 
fusion category (i.e., multiple summands in the product), and this is closely tied to the appearance of ``non-invertible'' symmetry operators (see e.g., \cite{Thorngren:2019iar,Komargodski:2020mxz, Gaiotto:2020iye, Nguyen:2021naa, Heidenreich:2021xpr, Thorngren:2021yso, Agrawal:2015dbf, Robbins:2021ibx, Robbins:2021xce, Sharpe:2021srf, Koide:2021zxj, Huang:2021zvu,
Inamura:2021szw, Cherman:2021nox, Kaidi:2021xfk, Choi:2021kmx, Wang:2021vki, Bhardwaj:2022yxj,
Hayashi:2022fkw, Sharpe:2022ene, Choi:2022zal, Kaidi:2022uux, Choi:2022jqy, Cordova:2022ieu,
Bashmakov:2022jtl, Inamura:2022lun, Damia:2022bcd, Choi:2022rfe, Lin:2022dhv, Bartsch:2022mpm,
Lin:2022xod, Cherman:2022eml, Burbano:2021loy, Damia:2022rxw, Apruzzi:2022rei, GarciaEtxebarria:2022vzq, Heckman:2022muc, 
Niro:2022ctq, Kaidi:2022cpf, Mekareeya:2022spm, Antinucci:2022vyk, Giaccari:2022xgs, Bashmakov:2022uek,Cordova:2022fhg,
GarciaEtxebarria:2022jky, Choi:2022fgx, Bhardwaj:2022kot, Bhardwaj:2022maz, Bartsch:2022ytj}).

Now, one of the notable places where non-invertible symmetries have been observed is in the context of certain ``duality / triality defects''.\footnote{A word on terminology: in this work we use the stringy notion of a non-abelian $SL(2,\mathbb{Z})$ duality. In particular, we will be interested in operations which generate subgroups of $SL(2,\mathbb{Z})$ with order different than two. In what follows we shall sometimes 
refer to all of these as dualities even if the order is different than two. That being said, in our field theory examples we will explain when we are dealing with a specific duality / triality defect.}
At generic points of parameter space, a duality interchanges one description of a field theory with another. However, at special points in the parameter space (such as the critical point of the 2D Ising model), this duality operation simply sends one back to the same theory. In that case, one has a 0-form symmetry, and therefore one expects a codimension one generalized symmetry operator \cite{Kaidi:2021xfk, Choi:2021kmx, Choi:2022zal, Kaidi:2022uux, Kaidi:2022cpf}. 
An intriguing feature of these symmetry operators is that they can sometimes have a non-trivial fusion rule, indicating the presence of a non-invertible symmetry. In many cases, one can argue for the existence of such a non-invertible symmetry, even without knowing the full structure of the TFT localized on a duality defect.

Given the fact that many field theoretic dualities have elegant geometric characterizations, it is natural to ask whether these topological duality defects can be directly realized in terms of objects in string theory. In particular, one might hope that performing this analysis could provide additional insight into the associated worldvolume TFTs, and provide a systematic method for extracting the corresponding fusion rules for these generalized symmetry operators. A related point is that in many QFTs of interest, a weakly coupled Lagrangian description may be unavailable and so one must seek out alternative (often geometric) characterizations of these systems.

Along these lines, it was recently shown in \cite{Apruzzi:2022rei, GarciaEtxebarria:2022vzq,Heckman:2022muc} that for QFTs engineered via localized singularities / branes, generalized symmetry operators are obtained from ``branes at infinity''. The resulting defects are topological in the sense that they do not contribute to the stress energy tensor of the localized QFT. Starting from the topological terms of a brane ``at infinity'', one can then extract the resulting TFT on its worldvolume, and consequently, extract the resulting fusion rules for the associated generalized symmetry operators.

Our aim in this note will be to use this perspective to propose a general prescription for duality defects where the duality of the QFT is inherited from the $SL(2,\mathbb{Z})$ duality of type IIB strings. In particular, we focus on the case of 4D QFTs realized from D3-branes probing a localized singularity of a non-compact Calabi-Yau threefold $X$ which we assume has a conical topology, namely it can be written as a cone over $\partial X$: $\mathrm{Cone}(\partial X) = X$. Such QFTs have a marginal parameter $\tau$ descending from the axio-dilaton of type IIB string theory, and 2-form potentials of a bulk 5D TFT descending from the $SL(2,\mathbb{Z})$ doublet of 2-form potentials (RR and NS-NS) which governs the 1-form electric and magnetic symmetries of the 4D probe theory.

In this setting, the ``branes at infinity'' which implement a duality transformation are simply given by specific bound states of $(p,q)$ 7-branes. 
In a general IIB / F-theory background, a bound state of $(p,q)$ 7-branes acts on the axio-dilaton and $SL(2,\mathbb{Z})$ doublet of 2-form 
potentials $\mathcal{B}^j = (C_2 , B_2)$ as:
\begin{equation}
\tau \mapsto \frac{a \tau + b}{c \tau + d} \,\,\, \text{and} 
\left[
\begin{array}
[c]{c}%
C_{2}\\
B_{2}%
\end{array}
\right]  \mapsto\left[
\begin{array}
[c]{cc}%
a & b\\
c & d
\end{array}
\right]  \left[
\begin{array}
[c]{c}%
C_{2}\\
B_{2}%
\end{array}
\right] 
\end{equation}
in the obvious notation. At the level of topology this monodromy can be localized to a branch cut whose endpoints are physical, namely the locus of a bound state of 7-branes. 

Of particular significance are the specific monodromy transformations which leave fixed particular values of $\tau$. Geometrically, these are specified by constant axio-dilaton profiles for 7-branes, which are in turn given by specific Kodaira fibers which specify how the elliptic fiber of F-theory degenerates on the locus of the 7-brane. The full list is $II, III, IV, I_{0}^{\ast}, IV^{\ast}, III^{\ast}, II^{\ast}$, which respectively support the gauge algebras $\mathfrak{su}_1, \mathfrak{su}_{2}, \mathfrak{su_3}, \mathfrak{so}_8, \allowbreak \mathfrak{e}_6, \mathfrak{e}_7, \mathfrak{e}_8$. Putting all of this together, it is natural to expect that 
the duality defects of the QFT simply lift to appropriate 7-branes 
wrapped on all of $\partial X$.

Our main claim is that wrapping 7-branes on a ``cycle at infinity'' leads to topological duality / triality defects in the 
4D worldvolume theory of the probe D3-brane. One way to see this is to consider the dimensional reduction on the boundary five-manifold $\partial X$. This results in the 5D symmetry TFT of the 4D field theory (see \cite{Apruzzi:2021nmk} as well as \cite{Aharony:1998qu, Heckman:2017uxe}). In this 5D theory, 7-branes wrapped on $\partial X$ specify codimension two defects which fill out a three-manifold in the 4D spacetime. In this 5D TFT limit where all metric data has been decoupled, the reduction of the 7-brane ``at infinity'' can be pushed into the interior, and can equivalently be viewed as specifying a codimension two defect in the bulk. In particular, as codimension two objects, they come with a branch cut structure, and this in turn impacts the structure of anomalies both in the 5D bulk as well as the 3D TFT localized on the topological defect.

In particular, we find that the choice of where to terminate the other end of the branch cut emanating from the 7-branes has a non-trivial impact on the resulting structure of the TFT. For each choice of branch cut, we get a corresponding anomaly inflow to the 7-brane defect. Doing so, we show that one choice of a branch cut gives the constructions of \cite{Kaidi:2021xfk, Kaidi:2022cpf, Antinucci:2022vyk} for Kramers-Wannier-like duality defects, while another choice produces the half-space gauging construction of \cite{Choi:2021kmx, Choi:2022zal}.
One can also entertain ``hybrid'' configurations of branch cuts, and these also produce duality / triality defects. In an Appendix we also show how these considerations are compatible with dimensional reduction of topological terms present in the 8D worldvolume of the 7-branes. We emphasize that while these analyses also make use of the 5D symmetry TFT, our analysis singles out the role of codimension two objects (and their associated branch cuts) which descend from wrapped 7-branes. Indeed, this top down perspective allows us to unify different construction techniques.

In the field theory literature, the main examples of duality / triality defects have centered on $\mathcal{N} = 4$ SYM theory and closely related examples. In the present context where this QFT arises from D3-branes probing $\mathbb{C}^3$, we see that the main ingredients for duality / triality defects readily generalize to $\mathcal{N} = 1$ SCFTs as obtained from D3-branes probing $X$ a Calabi-Yau cone with a singularity. In that setting, the IIB duality group corresponds to a duality action which is present at a specific (tuned) subspace of the conformal manifold of the SCFT. In particular, dimensional reduction of 7-branes on $\partial X$ leads to precisely the same topological defects, and thus provides us with a generalization to QFTs with less supersymmetry. On the other hand, the full 5D symmetry TFT will in this case be more involved simply because the topology $\partial X$ can in general support more kinds of objects. For example, other 0-form symmetries are present in such systems, and crossing the associated local defects charged under these discrete 0-form symmetries  through a duality / triality wall leads to non-trivial transformation rules.

The rest of this paper is organized as follows. In section \ref{sec:SETUP} we present our general setup involving probe
D3-branes in a Calabi-Yau threefold. In particular, we show how boundary conditions ``at infinity'' specify the global form of the theory, and how duality / triality defects arise from 7-branes wrapped on the boundary geometry. In section \ref{sec:DefectTFT} we consider the 5D symmetry TFT obtained from dimensional reduction on the boundary $\partial X$. The 7-branes descend to codimension two objects with branch cuts, and the choice of how to terminate these branch cuts leads to different implementations of duality / triality defects. After this, in section \ref{sec:N4} we show that our top down considerations are compatible with the bottom up analyses in the field theory literature. Section \ref{sec:N=1} shows how these considerations generalize to systems with minimal supersymmetry. We present our conclusions and some directions for future work in section \ref{sec:CONC}. In Appendix \ref{app:other} we show how the various defects considered in the main body are implemented in other top down constructions. In Appendix \ref{app:minimalTFT7branes} we give a proposal for the relevant topological terms of a non-perturbative 7-brane which reduce to a suitable 3D TFT (after reduction on $\partial X$). Finally, in Appendix \ref{app:orbo} we give some further details on the special case of D3-branes probing $\mathbb{C}^{3} / \mathbb{Z}_3$.

\section{General Setup}\label{sec:SETUP}

We now present the general setup for implementing duality / triality interfaces and defects in the context of brane probes of singularities. The construction we present produces supersymmetric 4D quantum field theories $\mathfrak{T}^{(N)}_X$ realized as the world-volume theory of a stack of $N$ D3-branes probing a non-compact Calabi-Yau threefold $X$. 


The Calabi-Yau threefolds $X$ we are considering are of conical topology
\be 
X=\mathrm{Cone } \left( \partial X \right)
\ee 
with link $\partial X$, the asymptotic boundary of $X$. The topology of $\partial X$ therefore determines the topology of $X$ fully. The apex of the cone supports a real codimension six singularity. We introduce the radial coordinate $r\in \mathbb{R}_{\geq 0}$ so that the singularity sits at $r=0$ and the asymptotic boundary sits at $r=\infty$.

For example, $X=\mathbb{C}^3$ determines $\mathfrak{T}^{(N)}_X$ to be 4D $\mathcal{N}=4$ supersymmetric Yang-Mills theory. In cases of reduced holonomy, for example $X=\mathbb{C}^3/\Gamma$ with $\Gamma\subset SU(3)$, we preserve $\mathcal{N}=1$ supersymmetry. In all cases, $\mathfrak{T}^{(N)}_X$ is some quiver gauge theory, with quiver nodes specified by a basis of ``fractional branes'' which can be visualized at large volume (i.e., away from the orbifold point of moduli space) as a collection of D3-, D5- and D7-branes and their anti-brane counterparts wrapped on cycles in a resolution of $X$ \cite{Klebanov:1998hh, Uranga:1998vf,Aharony:1997ju}. Nodes are connected by oriented arrows which should be viewed as open strings stretching between the fractional branes. The gauge theory characterization is especially helpful at weak coupling, and serves to define the QFT in the first place. The quiver gauge theory comes with a collection of marginal couplings and we can consider tuning these parameters to ``strong coupling''. At such points in the conformal manifold, the gauge theory description is less useful, but we can still speak of the SCFT defined by the probe D3-branes.

In the quiver gauge theory, the IIB axio-dilaton descends to a particular choice of marginal couplings. Moreover, the celebrated $SL(2,\mathbb{Z})$ duality of IIB strings\footnote{The precise form of the duality group and its actions on fermions leads to some additional subtleties. For example, taking into account fermions, there is the metaplectic cover of $SL(2,\mathbb{Z})$ \cite{Pantev:2009de}, and taking into account reflections on the F-theory torus (associated with worldsheet orientation reversal and $(-1)^{F_L}$ parity, this enhances to the $\mathsf{Pin}^{+}$ cover of $GL(2,\mathbb{Z})$ \cite{Tachikawa:2018njr} (see also \cite{Debray:2021vob, Dierigl:2022reg}). These subtleties can appear if one carefully tracks the boson / fermion number of extended operators but in what follows we neglect this issue.} descends to a duality transformation at a specific point in the conformal manifold of the 4D SCFT \cite{Lawrence:1998ja, Kachru:1998ys} (for a recent discussion see, e.g., \cite{Garcia-Etxebarria:2016bpb}).

The other bulk supergravity fields of type IIB also play an important role in specifying the global structures of the field theory. Boundary conditions $P$ for such bulk fields at $\partial X$ determine an absolute theory $\mathfrak{T}^{(N)}_{X,P}$ from the relative theory $\mathfrak{T}^{(N)}_{X}$. In particular, such boundary conditions also determine the spectrum of extended objects ending at or contained within $\partial X$ which specify the defects and generalized symmetry operators of $\mathfrak{T}^{(N)}_{X,P}$ \cite{Gaiotto:2014kfa,GarciaEtxebarria:2019caf, Bhardwaj:2021mzl}.

For example, there is an $SL(2,\mathbb{Z})$ doublet $(C_2, B_2) = \mathcal{B}^{j}$ (RR and NS) of 2-form potentials which couple to D1- and F1-strings of the IIB theory respectively. Wrapping bound states of these objects compatible with $P$ along the radial direction in $X$ leads to heavy line defects of the 4D quiver gauge theory $\mathfrak{T}^{(N)}_{X,P}$. The spectrum of line defects then fixes the global form of the quiver gauge group.

\begin{figure}[t]
    \centering
    \scalebox{0.8}{
    \begin{tikzpicture}
	\begin{pgfonlayer}{nodelayer}
		\node [style=none] (0) at (-2, 2) {};
		\node [style=none] (1) at (-2, -2) {};
		\node [style=none] (2) at (-3, 2) {$r=\infty$};
		\node [style=none] (3) at (-3, -2) {$r=0$};
		\node [style=none] (10) at (0, 2.5) {$\partial X$};
		\node [style=none] (11) at (-3, -1.75) {};
		\node [style=none] (12) at (-3, 1.75) {};
		\node [style=none] (13) at (-3.75, 0) {$\mathbb{R}_{\geq0}$};
		\node [style=none] (15) at (2, 2) {};
		\node [style=none] (16) at (2, -2) {};
		\node [style=none] (17) at (-1.75, -2.5) {};
		\node [style=none] (18) at (1.75, -2.5) {};
		\node [style=none] (19) at (0, -3) {$\mathbb{R}_\perp$};
		\node [style=none] (20) at (3, -2) {$\ket{\mathfrak{T}^{(N)}_X}$};
		\node [style=none] (21) at (3, 2) {$\ket{P,D}$};
        \node [style=none] (22) at (0, 0) {\large $\frac{N}{2\pi} \int_{M_4\times \mathbb{R}_{\geq 0}}B_2\wedge dC_2$};
	\end{pgfonlayer}
	\begin{pgfonlayer}{edgelayer}
		\draw [style=ArrowLineRight] (11.center) to (12.center);
		\draw [style=ThickLine] (0.center) to (15.center);
		\draw [style=ThickLine] (1.center) to (16.center);
		\draw [style=ArrowLineRight] (17.center) to (18.center);
	\end{pgfonlayer}
\end{tikzpicture}}
    \caption{Sketch of the symmetry TFT \eqref{eq:5d TFT terms1}. We depict the half-plane $\mathbb{R}_{\geq 0}\times\mathbb{R}_\perp$ with coordinates $(r,x_\perp)$ where $\mathbb{R}_\perp$ is some direction parallel to the D3-brane worldvolume. The boundary conditions for the symmetry TFT are denoted $\ket{\mathfrak{T}^{(N)}_X},\ket{P,D}$ respectively. }
    \label{fig:SetUp0}
\end{figure}

The possible boundary conditions $P$ are determined by the symmetry TFT \cite{Freed:2012bs,Freed:2022qnc} which follows by reduction of the 10D Chern-Simons term in IIB supergravity,\footnote{Strictly speaking, one also needs to utilize the self-dual condition for $F_5=dC_4$ in order to get the correct coefficient in \eqref{eq:5d TFT terms1}. 
For further discussion on this point, see e.g., \cite{Belov:2006jd, Belov:2006xj}.} much as in references \cite{Aharony:1998qu, Apruzzi:2021nmk} (see also \cite{Heckman:2017uxe}):
\begin{equation}
    S_{\mathrm{(CS)}}=-\frac{1}{4\kappa^2}\int_{M_4\times X}C_4\wedge dB_2\wedge dC_2.
\end{equation}
The stack of D3-branes source $N$ units of 5-form flux threading $\partial X$ and therefore the symmetry TFTs of all quiver gauge theories under consideration contain the universal term
\begin{equation}\label{eq:5d TFT terms1}
    S_{(\mathrm{SymTFT}),0}= \frac{N}{4\pi}\int_{M_4\times \mathbb{R}_{\geq 0}} \epsilon_{ij} \mathcal{B}^{i} \cup d \mathcal{B}^{j} \,,
\end{equation}
where we have integrated over the link $\partial X$. Here we have introduce a manifestly $SL(2,\mathbb{Z})$ invariant presentation of the action using the 
two-index tensor $\epsilon_{ij}$ to raise and lower doublet indices. In our conventions, $\epsilon_{21}=-\epsilon_{12}=1$. In terms of the individual components of this $SL(2,\mathbb{Z})$ doublet, 
the equations of motion for the action \eqref{eq:5d TFT terms1} are
\begin{equation}\label{eq:ZN}
    N dB_2 = N dC_2 = 0,
\end{equation}
which constrains $B_2$ and $C_2$ to be $\mathbb{Z}_N$-valued 1-form symmetry background fields.\footnote{Note that these steps are identical to the derivation of a bulk topological term in $AdS_5$ \cite{Witten:1998wy}, while here the term lives along $M_4\times \mathbb{R}_{\geq 0}$.} In general, we will denote $\mathbb{Z}_N$-valued fields using the same notation as their $U(1)$ counterparts but are related by a rescaling. For example, in conventions where the NS-NS flux $\frac{1}{2\pi}H_3$ is integrally quantized we have\footnote{Another natural choice would be to take $\int_{Q_3}H_3\in \mathbb
{Z}$ for all 3-manfolds $Q_3$ in which case we would drop the factor of $2\pi$ on the RHS of \eqref{eq:rescaling}.}
\begin{equation}\label{eq:rescaling}
    B^{U(1)}_2=\frac{2\pi}{N}B^{\mathbb{Z}_N}_2
\end{equation}
where the holonomies $\int_{\Sigma_2}B^{\mathbb{Z}_N}_2=k \; \mathrm{mod}\; N$ for some Riemann surface $\Sigma$ and integer $k$. Notice that since \eqref{eq:rescaling} is only valid when the holonomies of the $U(1)$ field are $N^{\mathrm{th}}$ roots of unity, for a $\mathbb{Z}_N$-valued field one is free to take either the LHS or RHS as normalizations. We will drop the superscripts in the future making clear which convention we are using for discrete fields when it arises. For additional details on the structure of the defect group in this theory (via related top down constructions) see Appendix \ref{app:other}. 

The relative theory $\mathfrak{T}^{(N)}_X$ sets enriched Neumann boundary conditions at $r=0$ while at $r=\infty$ we have mixed Neumann-Dirichlet boundary conditions for the fields of the symmetry TFT.  These are respectively denoted as
\begin{equation}
      \ket{\mathfrak{T}^{(N)}_X}\,, \quad \ket{P,D}\,, \qquad \quad \braket{ P,D\,|\,\mathfrak{T}^{(N)}_X}=Z_{\mathfrak{T}^{(N)}_{X,P}}(D)
\end{equation}
and contract to give the partition function of the absolute theory $\mathfrak{T}^{(N)}_{X,P}$ with background fields determined by $P$ set to the values $D$. Here $D$ is a form profile and in particular does not carry $SL(2,\mathbb{Z})$ indices (see figure \ref{fig:SetUp0}).

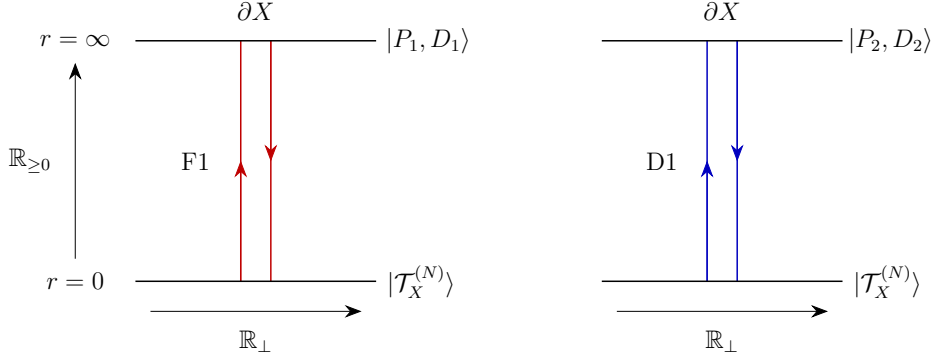
\begin{figure}[t]
    \centering
    \scalebox{0.8}{
\begin{tikzpicture}
	\begin{pgfonlayer}{nodelayer}
		\node [style=none] (0) at (-3, 2) {};
		\node [style=none] (1) at (8.75, 2) {};
		\node [style=none] (2) at (-3, -2) {};
		\node [style=none] (3) at (8.75, -2) {};
		\node [style=none] (4) at (-4, 2) {$r=\infty$};
		\node [style=none] (5) at (-4, -2) {$r=0$};
		\node [style=none] (6) at (-1.25, 0) {};
		\node [style=none] (7) at (-0.75, 0) {};
		\node [style=none] (8) at (-1.25, -2) {};
		\node [style=none] (9) at (-0.75, -2) {};
		\node [style=none] (10) at (-1.25, 2) {};
		\node [style=none] (11) at (-0.75, 2) {};
		\node [style=none] (12) at (6.5, 0) {};
		\node [style=none] (13) at (7, 0) {};
		\node [style=none] (14) at (7, -2) {};
		\node [style=none] (15) at (6.5, -2) {};
		\node [style=none] (16) at (6.5, 2) {};
		\node [style=none] (17) at (7, 2) {};
		\node [style=none] (18) at (-1, 2.5) {$\partial X$};
		\node [style=none] (20) at (-4, -1.625) {};
		\node [style=none] (21) at (-4, 1.625) {};
		\node [style=none] (22) at (-4.75, 0) {$\mathbb{R}_{\geq0}$};
		\node [style=none] (23) at (-2, 0) {F1};
		\node [style=none] (24) at (5.75, 0) {D1};
		\node [style=none] (25) at (1, 2) {};
		\node [style=none] (26) at (1, -2) {};
		\node [style=none] (27) at (4.75, 2) {};
		\node [style=none] (28) at (4.75, -2) {};
		\node [style=none] (29) at (6.75, 2.5) {$\partial X$};
		\node [style=none] (31) at (-2.75, -2.5) {};
		\node [style=none] (32) at (0.75, -2.5) {};
		\node [style=none] (33) at (-1, -3) {$\mathbb{R}_\perp$};
		\node [style=none] (34) at (5, -2.5) {};
		\node [style=none] (35) at (8.5, -2.5) {};
		\node [style=none] (36) at (6.75, -3) {$\mathbb{R}_\perp$};
		\node [style=none] (37) at (1.75, -2) {$\ket{\mathcal{T}^{(N)}_{X}}$};
		\node [style=none] (38) at (1.8675, 2) {$\ket{P_1,D_1}$};
		\node [style=none] (39) at (9.5, -2) {$\ket{\mathcal{T}^{(N)}_{X}}$};
		\node [style=none] (40) at (9.55, 2) {$\ket{P_2,D_2}$};
	\end{pgfonlayer}
	\begin{pgfonlayer}{edgelayer}
		\draw [style=ArrowLineRed] (8.center) to (6.center);
		\draw [style=ArrowLineRed] (11.center) to (7.center);
		\draw [style=RedLine] (10.center) to (8.center);
		\draw [style=RedLine] (11.center) to (9.center);
		\draw [style=ArrowLineBlue] (17.center) to (13.center);
		\draw [style=ArrowLineBlue] (15.center) to (12.center);
		\draw [style=BlueLine] (16.center) to (15.center);
		\draw [style=BlueLine] (17.center) to (14.center);
		\draw [style=ArrowLineRight] (20.center) to (21.center);
		\draw [style=ThickLine] (0.center) to (25.center);
		\draw [style=ThickLine] (2.center) to (26.center);
		\draw [style=ThickLine] (28.center) to (3.center);
		\draw [style=ThickLine] (27.center) to (1.center);
		\draw [style=ArrowLineRight] (31.center) to (32.center);
		\draw [style=ArrowLineRight] (34.center) to (35.center);
	\end{pgfonlayer}
\end{tikzpicture}
    }
    \caption{Boundary conditions and defects for $\mathfrak{T}_X^{(N)}$. We sketch the half-plane $\mathbb{R}_{\geq 0}\times \mathbb{R}_\perp$ parametrized by $(r,x_\perp)$. The polarization $P_1,P_2$ determine that Dirichlet boundary conditions are set for $B_2,C_2$ respectively $B_2|_{\partial X}=D_1$ and $C_2|_{\partial X}=D_2$. Line defects are realized by F1/D1-strings and correspond to Wilson and 't Hooft lines respectively. Our conventions are such that the left, radially outgoing strings are of charge $[0,-1]$ and $[-1,0]$ and the right, incoming strings are of charge $[0,1]$ and $[1,0]$ respectively.}
    \label{fig:F1D1}
\end{figure}

Consider for example $X=\mathbb{C}^3$ in which case \eqref{eq:5d TFT terms1} describes the full symmetry TFT. First note that \eqref{eq:ZN} makes it clear that we are discussing a theory with gauge algebra $\mathfrak{su}(N)$ rather than $\mathfrak{u}(N)$. This $U(1)$ factor is lifted via a Stueckelberg mechanism.\footnote{Intuitively, when we pick an origin for $\mathbb{C}^3$ we put the whole system in a ``box'' with a conformal boundary. This removes the center of mass degree of freedom for the system.} A standard set of boundary conditions include a purely electric or purely magnetic polarization via the boundary conditions:
\begin{align}\label{eq:electricbc}
 \textnormal{$B_2|_{\partial X}$ Dirichlet, $C_2|_{\partial X}$ Neumann} \quad  &\longleftrightarrow \quad \textnormal{Global electric 1-form symmetry}\\ \label{eq:magneticbc} 
 \textnormal{$B_2|_{\partial X}$ Neumann, $C_2|_{\partial X}$ Dirichlet} \quad  &\longleftrightarrow \quad \textnormal{Global magnetic 1-form symmetry}.
\end{align}
Concretely, we are considering $\mathcal{N} = 4$ SYM theory with gauge algebra $\mathfrak{su}(N)$. The electric polarization produces gauge group $SU(N)$ while the magnetic polarization produces gauge group $PSU(N) = SU(N) / \mathbb{Z}_N$. Given electric/magnetic boundary conditions we can stretch F1/D1 strings between the D3-branes and the asymptotic boundary to construct line defects in the 4D worldvolume theory (see figure \ref{fig:F1D1}).

One can also consider more general mixed boundary conditions. In general, $SL(2,\mathbb{Z}_N)$ duality transformations
\begin{equation}\label{eq:Mono}
~\begin{bmatrix} C_2 \\B_2 \end{bmatrix}\rightarrow \begin{bmatrix} a&b\\c&d \end{bmatrix}\begin{bmatrix} C_2 \\B_2 \end{bmatrix}, \qquad \mathbb{S}=\begin{bmatrix} 0&1\\-1&0 \end{bmatrix}\,, \quad \mathbb{T}=\begin{bmatrix} 1&1\\0&1 \end{bmatrix}
\end{equation}
where $ad-bc=1$, map between boundary conditions and group these into orbits. In the case $\mathcal{N}=4$ SYM with gauge algebra $\mathfrak{su}(N)$ and when $N$ has no square divisors,\footnote{See \cite{Bergman:2022otk} for details when dropping this assumption.} one can generate all possible mixed boundary conditions for $B_2$ and $C_2$, and thus all forms of the gauge group.

More generally, given a quiver gauge theory, the individual gauge group factors are all correlated due to bifundamental states which are charged under different centers of the gauge group. Indeed, note also that we can always move the D3-brane away from the singularity (i.e., to finite $r > 0$), and the center of this gauge group in the infrared will need to be compatible with the boundary conditions specified at $r = \infty$.

Let us also record our $SL(2,\mathbb{Z})$ conventions for strings and 5-branes here. The charge vector $Q$ of a $(p,q)$-string or $(p,q)$-5-brane has components, with $\epsilon_{12}=-\epsilon_{21}=-1$, following conventions laid out in \cite{Weigand:2018rez},
\begin{equation}
    Q_i=\epsilon_{ij}Q^j=[q,p]\,, \qquad Q^i=\begin{bmatrix}
        p \\ -q
    \end{bmatrix}\,,
\end{equation}
in particular S-duality maps Wilson lines $W$ and 't Hooft lines $H$ on the D3-brane worldvolume as $(W,H)\rightarrow (H,-W)$.

We further lay out our conventions for symmetry TFTs following \cite{Kaidi:2022uux,Kaidi:2022cpf}. The enriched Neumann boundary condition at $r=0$ is expanded as
\begin{equation}
    \ket{\mathfrak{T}_X^{(N)}}=\sum_{\mathbf{a}\in P} Z_{\mathfrak{T}_{X,P}^{(N)}}(a)\ket{\mathbf{a}}
\end{equation}
where $Z_{\mathfrak{T}_{P,X}^{(N)}}(a)$ is the partition function of the absolute theory derived from $\mathfrak{T}_X^{(N)}$ by choice of polarization $P$ and $a$ is a background field profile for the corresponding higher symmetry. In this paper we are mainly concerned with 1-form symmetries of gauge theories and here $P$ fixes the global form of the gauge group and $a$ is a 1-form symmetry background field. Further we denote a background field configuration by a vector $\mathbf{a}$ which is oriented in the corresponding defect group and has a form profile of $a$. Topological Dirichlet and Neumann boundary conditions at $r=\infty$ in 4D are respectively
\begin{equation}
\begin{aligned}    \ket{P,D}_{\mathrm{Dirichlet}}&=\sum_{\mathbf{a}\in P}\delta(D-a)\ket{\mathbf{a}} \\  \ket{P,E}_{\mathrm{Neumann}}&=\sum_{\mathbf{a}\in P}\exp\left(\frac{2\pi i}{N} \int E\cup a\right)\ket{\mathbf{a}}
    \end{aligned}
\end{equation}
where we have normalized fields to take values in $\mathbb{Z}_N$. We will mainly work with Dirichlet boundary conditions throughout and omit the index `Dirichlet' when it causes no confusion. Whenever two polarizations $P,P'$ are related by a discrete Fourier transform or equivalently by gauging we have the pairing
\begin{equation}
    \braket{\mathbf{a}|\mathbf{b}}=\exp\left(\frac{2\pi i}{N}\int a \cup b \right)\qquad \forall\, \mathbf{a}\in P,\mathbf{b}\in P'\,.
\end{equation}
More generally, boundary conditions in 4D can be stacked with counterterms, so we define 
\begin{equation}
\begin{aligned}    \ket{P_{G_r},D}_{\mathrm{Dirichlet}}&=\sum_{\mathbf{a}\in P_{G_k}}\delta(D-a)\exp\left(\frac{2\pi ir}{N}\int \frac{\mathcal{P}(D)}{2}\right)\ket{\mathbf{a}} 
    \end{aligned}
\end{equation}
where $\mathcal{P}$ is the Pontryagin square. Here we have labelled a polarization $P$ by the global form of the gauge group $G$ it realizes, and the subscript $r$ counts the number of stacked counterterms. For example $SU(2)_r$ denotes $SU(2)$ theory stacked with $r$ couterterms.

\subsection{Proposal for Topological Duality Interfaces/Operators}
\label{sec:Proposal}

{\renewcommand{\arraystretch}{1.35}
\begin{table}[t]
    \centering
    \begin{center}
\begin{tabular}{||c | c | c | c | c | c ||}
 \hline
 Fiber Type $\mathfrak{F}$ & Lines & Monodromy $ \rho $ & Refined Linking $p/2k$ & $(k,m)$ & $\tau$ \\ [0.5ex] 
 \hline\hline
 $II,\,\mathfrak{su}(1)$ & $-$ & {\footnotesize $\left(\begin{array}{cc}
     0 &  1 \\
     -1  & 1
 \end{array}\right)$ } & $-$ & $-$ & $e^{i\pi/3}$\\ 
 \hline
 $III,\,\mathfrak{su}(2)$ & $\mathbb{Z}_2$ & {\footnotesize $\left(\begin{array}{cc}
     0 &  1 \\
     -1  & 0
 \end{array}\right)$ } &  $\frac{3}{4}$ & $(2,3)$ & $e^{i\pi/2}$ \\
 \hline
 $IV,\,\mathfrak{su}(3)$ & $\mathbb{Z}_3$ & {\footnotesize $\left(\begin{array}{cc}
     0 &  1 \\
     -1  & -1
 \end{array}\right)$ } &    $\frac{4}{6}$ & $(3,4)$ &$e^{i\pi/3}$ \\
 \hline
 $I_{0}^{\ast},\,\mathfrak{so}(8)$ & $\mathbb{Z}_2\oplus \mathbb{Z}_2$ &  {\footnotesize $\left(\begin{array}{cc}
     -1 &  0 \\
     0  & -1
 \end{array}\right)$ } & {\footnotesize $\left(\begin{array}{cc}
     2/4 &  3/4 \\
     3/4  & 2/4
 \end{array}\right)$ } & $-$  & $\tau$ \\
 \hline
 $IV^{\ast},\,\mathfrak{e}_6$ & $\mathbb{Z}_3$ & {\footnotesize $\left(\begin{array}{cc}
     -1 &  -1 \\
     1  & 0
 \end{array}\right)$ } &  $\frac{2}{6}$ & $(3,2)$ &$e^{i\pi/3}$ \\
 \hline
 $III^{\ast},\,\mathfrak{e}_7$ & $\mathbb{Z}_2$ & {\footnotesize $\left(\begin{array}{cc}
     0 &  -1 \\
     1  & 0
 \end{array}\right)$ } &   $\frac{1}{4}$ & $(2,1)$ &$e^{i\pi/2}$ \\
 \hline
 $II^{\ast},\,\mathfrak{e}_8$ & $-$ &  {\footnotesize $\left(\begin{array}{cc}
     1 &  -1 \\
     1  & 0
 \end{array}\right)$ } & $-$ & $-$ & $e^{i\pi/3}$ \\
  \hline
\end{tabular}
\end{center}
    \caption{Elliptic data of 7-brane profiles with constant axio-dilaton $\tau$. Their group of lines is isomorphic to $\textnormal{coker}(\rho-1)$ which is isomorphic to $\mathbb{Z}_k$ except for fiber type $I_0^*,II,II^*$. The label $m$ of these lines is determined from the refined self-linking numbers $m/2k$ which gives the spin of non-trivial lines. The refined self-linking numbers compute via the Gordon-Litherland approach laid out in \cite{Apruzzi:2021nmk} employing the divisors (Kodaira thimbles) computed in \cite{Hubner:2022kxr} or alternatively via the quadratic refinement laid out in \cite{Gukov:2020btk} and the linking number computations in \cite{Cvetic:2021sxm}. Note in particular that in all cases $\textnormal{gcd}(k,m)=1$ and $mk\in 2\mathbb{Z}$.}
    \label{tab:Fibs}
\end{table}}

Consider the spacetime $M_4=M_3\times \mathbb{R}_\perp$ with $\mathbb{R}_\perp$ parametrized by the coordinate $x_\perp$. We now argue that 7-branes wrapped on $M_3\times\partial X$ at some point $\bar x_{\perp}\in \mathbb{R}_\perp$ realize topological duality/triality interfaces and operators. A subset of our constructions work only for 7-branes with constant axio-dilaton profile and we list these in table \ref{tab:Fibs} together with their topological data. 

\begin{figure}[t]
    \centering
    \scalebox{0.8}{\begin{tikzpicture}
	\begin{pgfonlayer}{nodelayer}
		\node [style=none] (0) at (-6, 1.5) {};
		\node [style=none] (1) at (-1, 1.5) {};
		\node [style=none] (2) at (-6, -1.5) {};
		\node [style=none] (3) at (-1, -1.5) {};
		\node [style=none] (4) at (0, 0) {$=$};
		\node [style=none] (5) at (1, 1.5) {};
		\node [style=none] (6) at (6, 1.5) {};
		\node [style=none] (7) at (1, -1.5) {};
		\node [style=none] (8) at (6, -1.5) {};
		\node [style=Star] (9) at (-3.5, 1.5) {};
		\node [style=Star] (10) at (3.5, 0) {};
		\node [style=none] (11) at (1, 0) {};
		\node [style=none] (12) at (2.25, -0.5) {Branch cut};
		\node [style=none] (13) at (3.5, 0.5) {7-branes};
		\node [style=none] (14) at (-3.5, 1) {7-branes};
		\node [style=none] (15) at (-5.25, 2) {$\ket{P_1,D_1}$};
		\node [style=none] (16) at (-1.75, 2) {$\ket{P_2,D_2}$};
		\node [style=none] (17) at (1.75, 2) {$\ket{P_2,D_2}$};
		\node [style=none] (18) at (5.25, 2) {$\ket{P_2,D_2}$};
		\node [style=none] (19) at (-7, -1.5) {$r=0$};
		\node [style=none] (20) at (-7, 1.5) {$r=\infty$};
		\node [style=none] (21) at (-7, 1) {};
		\node [style=none] (22) at (-7, -1) {};
		\node [style=none] (23) at (-5, -2) {};
		\node [style=none] (24) at (-2, -2) {};
		\node [style=none] (25) at (-3.5, -2.5) {$x_\perp$};
		\node [style=none] (26) at (2, -2) {};
		\node [style=none] (27) at (5, -2) {};
		\node [style=none] (28) at (3.5, -2.5) {$x_\perp$};
        \node [style=none] (29) at (1.5, 0.45) {$\mathbb{H}_{\leftarrow}$};
	\end{pgfonlayer}
	\begin{pgfonlayer}{edgelayer}
		\draw [style=ThickLine] (0.center) to (1.center);
		\draw [style=ThickLine] (3.center) to (2.center);
		\draw [style=ThickLine] (5.center) to (6.center);
		\draw [style=ThickLine] (8.center) to (7.center);
		\draw [style=DottedLine] (11.center) to (10);
		\draw [style=ArrowLineRight] (22.center) to (21.center);
		\draw [style=ArrowLineRight] (23.center) to (24.center);
		\draw [style=ArrowLineRight] (26.center) to (27.center);
	\end{pgfonlayer}
\end{tikzpicture}
}
    \caption{Case\,(1), 7-branes wrapped on $M_3\times \partial X$, we sketch the plane $\mathbb{R}_{\geq0}\times \mathbb{R}_\perp$. The topological boundary conditions $\ket{P_1,D_1}$ are the monodromy transform of the boundary conditions $\ket{P_2,D_2}$ and result from stacking the branch cut with the asymptotic boundary. The branch cut is supported on $\mathbb{H}_{\leftarrow}\times \partial X$ and runs parallel to the D3-branes. Conventions are such that the monodromy matrix $\rho$ acts crossing the branch cut top to bottom. }
    \label{fig:7BraneInfinity2}
\end{figure}
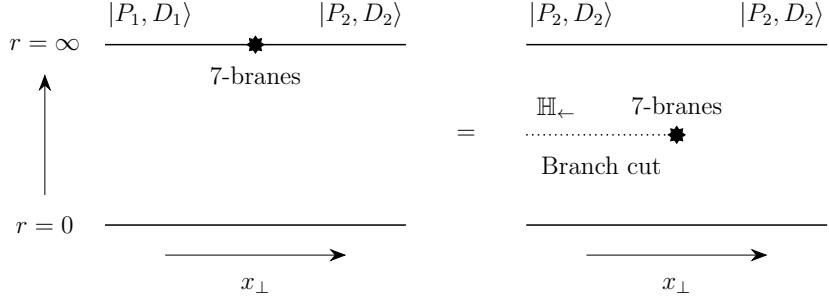

\begin{figure}[t]
    \centering
    \scalebox{0.8}{
    \begin{tikzpicture}
	\begin{pgfonlayer}{nodelayer}
		\node [style=none] (0) at (-6, 1.5) {};
		\node [style=none] (1) at (-1, 1.5) {};
		\node [style=none] (2) at (-6, -1.5) {};
		\node [style=none] (3) at (-1, -1.5) {};
		\node [style=none] (4) at (0, 0) {$=$};
		\node [style=none] (5) at (1, 1.5) {};
		\node [style=none] (6) at (6, 1.5) {};
		\node [style=none] (7) at (1, -1.5) {};
		\node [style=none] (8) at (6, -1.5) {};
		\node [style=Star] (9) at (-3.5, 1.5) {};
		\node [style=Star] (10) at (3.5, 0) {};
		\node [style=none] (11) at (3.5, -1.5) {};
		\node [style=none] (12) at (4.125, -0.75) {$\mathbb{H}_{\downarrow}$};
		\node [style=none] (13) at (3.5, 0.5) {7-brane};
		\node [style=none] (14) at (-3.5, 2) {7-brane};
		\node [style=none] (15) at (-5.5, 2) {$\ket{P_1,D_1}$};
		\node [style=none] (16) at (-1.5, 2) {$\ket{P_1,D_1}$};
		\node [style=none] (18) at (3.5, 2) {$\ket{P_1,D_1}$};
		\node [style=none] (19) at (-7, -1.5) {$r=0$};
		\node [style=none] (20) at (-7, 1.5) {$r=\infty$};
		\node [style=none] (21) at (-7, 1) {};
		\node [style=none] (22) at (-7, -1) {};
		\node [style=none] (23) at (-5, -2) {};
		\node [style=none] (24) at (-2, -2) {};
		\node [style=none] (25) at (-3.5, -2.5) {$x_\perp$};
		\node [style=none] (26) at (2, -2) {};
		\node [style=none] (27) at (5, -2) {};
		\node [style=none] (28) at (3.5, -2.5) {$x_\perp$};
		\node [style=Circle] (30) at (-3.5, -1.5) {};
		\node [style=Circle] (31) at (3.5, -1.5) {};
		\node [style=none] (32) at (-4.125, -1) {$M_3'$};
		\node [style=none] (33) at (2.875, -1) {$M_3'$};
        \node [style=none] (34) at (-2.875, 0) {$\mathbb{H}_{\downarrow}$};
	\end{pgfonlayer}
	\begin{pgfonlayer}{edgelayer}
		\draw [style=ThickLine] (0.center) to (1.center);
		\draw [style=ThickLine] (3.center) to (2.center);
		\draw [style=ThickLine] (5.center) to (6.center);
		\draw [style=ThickLine] (8.center) to (7.center);
		\draw [style=DottedLine] (11.center) to (10);
		\draw [style=ArrowLineRight] (22.center) to (21.center);
		\draw [style=ArrowLineRight] (23.center) to (24.center);
		\draw [style=ArrowLineRight] (26.center) to (27.center);
		\draw [style=DottedLine] (9) to (30);
	\end{pgfonlayer}
\end{tikzpicture}
    }
    \caption{Case\,(2), 7-branes wrapped on $M_3\times \partial X$, we sketch the plane $\mathbb{R}_{\geq0}\times \mathbb{R}_\perp$. There is a single set of boundary conditions $\ket{P_1,D_1}$. The branch cut is supported on $\mathbb{H}_{\downarrow}\times \partial X$ and runs perpendicular to the D3-branes. Conventions are such that the monodromy matrix $\rho$ acts crossing the branch cut left to right.}
    \label{fig:Case2}
\end{figure}
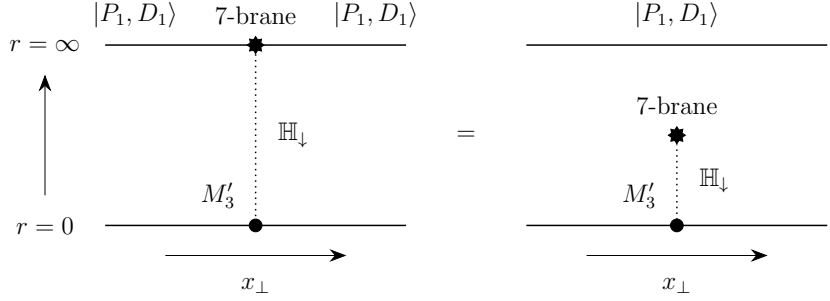

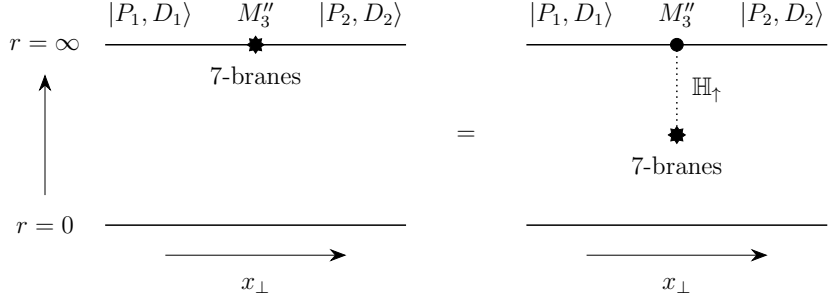
\begin{figure}
    \centering
    \scalebox{0.8}{
    \begin{tikzpicture}
	\begin{pgfonlayer}{nodelayer}
		\node [style=none] (0) at (-6, 1.5) {};
		\node [style=none] (1) at (-1, 1.5) {};
		\node [style=none] (2) at (-6, -1.5) {};
		\node [style=none] (3) at (-1, -1.5) {};
		\node [style=none] (4) at (0, 0) {$=$};
		\node [style=none] (5) at (1, 1.5) {};
		\node [style=none] (6) at (6, 1.5) {};
		\node [style=none] (7) at (1, -1.5) {};
		\node [style=none] (8) at (6, -1.5) {};
		\node [style=Star] (9) at (-3.5, 1.5) {};
		\node [style=Star] (10) at (3.5, 0) {};
		\node [style=none] (11) at (3.5, 1.5) {};
		\node [style=none] (12) at (4, 0.75) {$\mathbb{H}_{\uparrow}$};
		\node [style=none] (13) at (3.5, -0.5) {7-branes};
		\node [style=none] (14) at (-3.5, 2) {$M_3''$};
		\node [style=none] (15) at (-5.25, 2) {$\ket{P_1,D_1}$};
		\node [style=none] (16) at (-1.75, 2) {$\ket{P_2,D_2}$};
		\node [style=none] (18) at (-7, -1.5) {$r=0$};
		\node [style=none] (19) at (-7, 1.5) {$r=\infty$};
		\node [style=none] (20) at (-7, 1) {};
		\node [style=none] (21) at (-7, -1) {};
		\node [style=none] (22) at (-5, -2) {};
		\node [style=none] (23) at (-2, -2) {};
		\node [style=none] (24) at (-3.5, -2.5) {$x_\perp$};
		\node [style=none] (25) at (2, -2) {};
		\node [style=none] (26) at (5, -2) {};
		\node [style=none] (27) at (3.5, -2.5) {$x_\perp$};
		\node [style=Circle] (29) at (3.5, 1.5) {};
		\node [style=none] (30) at (-3.5, 1) {7-branes};
		\node [style=none] (31) at (3.5, 2) {$M_3''$};
		\node [style=none] (34) at (1.75, 2) {$\ket{P_1,D_1}$};
		\node [style=none] (35) at (5.25, 2) {$\ket{P_2,D_2}$};
	\end{pgfonlayer}
	\begin{pgfonlayer}{edgelayer}
		\draw [style=ThickLine] (0.center) to (1.center);
		\draw [style=ThickLine] (3.center) to (2.center);
		\draw [style=ThickLine] (5.center) to (6.center);
		\draw [style=ThickLine] (8.center) to (7.center);
		\draw [style=DottedLine] (11.center) to (10);
		\draw [style=ArrowLineRight] (21.center) to (20.center);
		\draw [style=ArrowLineRight] (22.center) to (23.center);
		\draw [style=ArrowLineRight] (25.center) to (26.center);
	\end{pgfonlayer}
\end{tikzpicture}
    }
    \caption{Case\,(3), 7-branes wrapped on $M_3\times \partial X$, we sketch the plane $\mathbb{R}_{\geq0}\times \mathbb{R}_\perp$. The 7-brane insertion gives rise to two boundary conditions $\ket{P_1,D_1},\ket{P_2,D_2}$. The branch cut is supported on $\mathbb{H}_{\uparrow} \times \partial X$ and runs perpendicular to the D3-branes. Conventions are such that the monodromy matrix $\rho$ acts crossing the branch cut right to left.}
    \label{fig:Case3}
\end{figure}

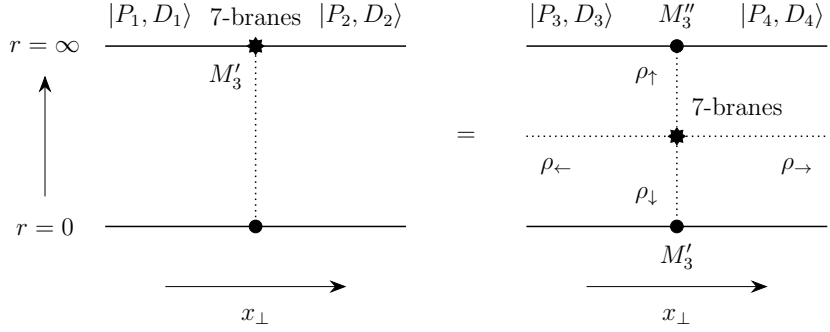
\begin{figure}
    \centering
    \scalebox{0.8}{\begin{tikzpicture}
	\begin{pgfonlayer}{nodelayer}
		\node [style=none] (0) at (-6, 1.5) {};
		\node [style=none] (1) at (-1, 1.5) {};
		\node [style=none] (2) at (-6, -1.5) {};
		\node [style=none] (3) at (-1, -1.5) {};
		\node [style=none] (4) at (0, 0) {$=$};
		\node [style=none] (5) at (1, 1.5) {};
		\node [style=none] (6) at (6, 1.5) {};
		\node [style=none] (7) at (1, -1.5) {};
		\node [style=none] (8) at (6, -1.5) {};
		\node [style=Star] (9) at (-3.5, 1.5) {};
		\node [style=Star] (10) at (3.5, 0) {};
		\node [style=none] (11) at (3.5, 1.5) {};
		\node [style=none] (13) at (4.5, 0.5) {7-branes};
		\node [style=none] (14) at (-3.5, 2) {7-branes};
		\node [style=none] (15) at (-5.25, 2) {$\ket{P_1,D_1}$};
		\node [style=none] (16) at (-1.75, 2) {$\ket{P_2,D_2}$};
		\node [style=none] (17) at (-7, -1.5) {$r=0$};
		\node [style=none] (18) at (-7, 1.5) {$r=\infty$};
		\node [style=none] (19) at (-7, 1) {};
		\node [style=none] (20) at (-7, -1) {};
		\node [style=none] (21) at (-5, -2.5) {};
		\node [style=none] (22) at (-2, -2.5) {};
		\node [style=none] (23) at (-3.5, -3) {$x_\perp$};
		\node [style=none] (24) at (2, -2.5) {};
		\node [style=none] (25) at (5, -2.5) {};
		\node [style=none] (26) at (3.5, -3) {$x_\perp$};
		\node [style=none] (28) at (-4, 1) {$M_3'$};
		\node [style=none] (29) at (3.5, 2) {$M_3''$};
		\node [style=none] (30) at (0, 2.5) {};
		\node [style=none] (31) at (1.75, 2) {$\ket{P_3,D_3}$};
		\node [style=none] (32) at (5.25, 2) {$\ket{P_4,D_4}$};
		\node [style=Circle] (33) at (3.5, 1.5) {};
		\node [style=Circle] (34) at (-3.5, -1.5) {};
		\node [style=none] (35) at (1, 0) {};
		\node [style=none] (36) at (3.5, -1.5) {};
		\node [style=none] (37) at (6, 0) {};
		\node [style=none] (38) at (3, -1) {$\rho_{\downarrow}$};
		\node [style=none] (39) at (5.5, -0.5) {$\rho_{\rightarrow}$};
		\node [style=none] (40) at (3, 1) {$\rho_{\uparrow}$};
		\node [style=none] (41) at (3.5, -2) {$M_3'$};
		\node [style=Circle] (42) at (3.5, -1.5) {};
		\node [style=none] (43) at (1.5, -0.5) {$\rho_{\leftarrow}$};
	\end{pgfonlayer}
	\begin{pgfonlayer}{edgelayer}
		\draw [style=ThickLine] (0.center) to (1.center);
		\draw [style=ThickLine] (3.center) to (2.center);
		\draw [style=ThickLine] (5.center) to (6.center);
		\draw [style=ThickLine] (8.center) to (7.center);
		\draw [style=DottedLine] (11.center) to (10);
		\draw [style=ArrowLineRight] (20.center) to (19.center);
		\draw [style=ArrowLineRight] (21.center) to (22.center);
		\draw [style=ArrowLineRight] (24.center) to (25.center);
		\draw [style=DottedLine] (9) to (34);
		\draw [style=DottedLine] (35.center) to (10);
		\draw [style=DottedLine] (10) to (37.center);
		\draw [style=DottedLine] (10) to (36.center);
	\end{pgfonlayer}
\end{tikzpicture}}
    \caption{Case\,(4) hybrid case of cases 1,2,3, the 7-branes wrapped on $M_3\times \partial X$, we sketch the plane $\mathbb{R}_{\geq0}\times \mathbb{R}_\perp$. The overall monodromy is $\rho=\rho_{\leftarrow}\rho_{\uparrow}\rho_{\rightarrow}\rho_{\downarrow}$. Each monodromy factor $\rho_{\bullet}$ has its own branch cut separately. The branch cuts labelled $\rho_{\downarrow},\rho_{\uparrow}$ intersect the D3-brane worldvolume and asymptotic boundary in $M_3',M_3''$, respectively.}
    \label{fig:Case4}
\end{figure}

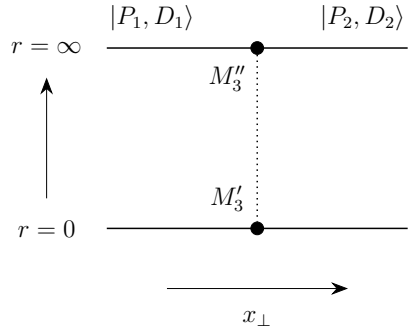
\begin{figure}
    \centering
    \scalebox{0.8}{\begin{tikzpicture}
	\begin{pgfonlayer}{nodelayer}
		\node [style=none] (0) at (-2.5, 1.5) {};
		\node [style=none] (1) at (2.5, 1.5) {};
		\node [style=none] (2) at (-2.5, -1.5) {};
		\node [style=none] (3) at (2.5, -1.5) {};
		\node [style=none] (6) at (-1.75, 2) {$\ket{P_1,D_1}$};
		\node [style=none] (7) at (1.75, 2) {$\ket{P_2,D_2}$};
		\node [style=none] (8) at (-3.5, -1.5) {$r=0$};
		\node [style=none] (9) at (-3.5, 1.5) {$r=\infty$};
		\node [style=none] (10) at (-3.5, 1) {};
		\node [style=none] (11) at (-3.5, -1) {};
		\node [style=none] (12) at (-1.5, -2.5) {};
		\node [style=none] (13) at (1.5, -2.5) {};
		\node [style=none] (14) at (0, -3) {$x_\perp$};
		\node [style=none] (15) at (-0.5, 1) {$M_3''$};
		\node [style=none] (17) at (-0.5, -1) {$M_3'$};
		\node [style=Circle] (18) at (0, 1.5) {};
		\node [style=Circle] (19) at (0, -1.5) {};
	\end{pgfonlayer}
	\begin{pgfonlayer}{edgelayer}
		\draw [style=ThickLine] (0.center) to (1.center);
		\draw [style=ThickLine] (3.center) to (2.center);
		\draw [style=ArrowLineRight] (11.center) to (10.center);
		\draw [style=ArrowLineRight] (12.center) to (13.center);
		\draw [style=DottedLine] (18) to (19);
	\end{pgfonlayer}
\end{tikzpicture}}
    \caption{S-duality is realized by cutting the symmetry TFT along the dotted line and gluing the pieces according to the desired S-duality transformation. No 7-branes are inserted. }
    \label{fig:Sduality}
\end{figure}

First consider a 7-brane wrapped at $r=\infty$ following the prescription in \cite{Heckman:2022muc}. This gives rise to a topological interface / symmetry operator in the 4D theory as the 7-brane is formally at infinite distance wrapped on a cycle of infinite volume. This decouples the non-topological interactions between the D3 and 7-brane and the non-topological degrees of freedom on the 7-brane worldvolume respectively \cite{Heckman:2022muc}. In this topological limit, for the 5D bulk theory, we can consider adding codimension two defects, i.e., the remnants of these 7-branes at infinity. Since everything is now treated as topological, we are free to insert these 7-branes anywhere in the interior, and as such we have different choices for where to extend the branch cut of this defect. We consider four distinct choices for the $SL(2,\mathbb{Z})$ monodromy branch cut:
\begin{enumerate}
    \item The branch cut is supported on $\mathbb{H}_{\leftarrow}\times \partial X$ with $\partial \mathbb{H}_{\leftarrow} =M_3$ and is oriented along $x_\perp$ parallel to the D3-branes with $x_\perp < \bar x_\perp$ along the asymptotic boundary. See the left subfigure in figure \ref{fig:7BraneInfinity2}. The branch cut ends at infinity. 
    \item The branch cut is supported on $\mathbb{H}_{\downarrow}\times \partial X$ with $\partial \mathbb{H}_{\downarrow}=M_3-M_3'$ (treated as a 3-chain) and is oriented radially inwards along $x_\perp=\bar x_\perp$ perpendicular to the D3-branes. See the left subfigure in figure \ref{fig:Case2}. The branch cut ends on $M_3'$, which is a subset of the D3-brane worldvolume, and supports an operator $\mathcal{D}(M_3')$ possibly coupled to background fields. While cuts can normally only begin / end on 7-branes, this can be made precise via a method of images procedure.\footnote{Without D3-branes, there is no physical meaning to “choosing a direction” for the branch cut. With D3-branes, however, there is an “end-of-the-world” QFT living at $r = 0$. The termination of the branch cut then makes sense via the method of images. We can then make a gauge choice to localize the effects of a branch cut in terms of an effective 3D TFT.} The 7-brane has constant axio-dilaton profile.
    \item The branch cut is supported on $\mathbb{H}_{\uparrow}\times \partial X$ with $\partial \mathbb{H}_{\uparrow}= M_3- M_3''$ (treated as a 3-chain) and is oriented radially outwards along $x_\perp=\bar x_\perp$ perpendicular to the D3-branes (see figure \ref{fig:Case3}). The branch cut ends on $M_3''$ which is contained in the asymptotic boundary. The 7-brane has constant axio-dilaton profile.
    \item Whenever the 7-brane monodromy matrix is not prime over $\mathbb{Z}_N$, i.e., it can be factored into more than one non-trivial factor in $SL(2,\mathbb{Z}_N)$, we can consider separate branch cuts for each factor. We can then
    consider hybrid configurations of cases 1,2,3 with each branch cut realizing one of the previous setups (see figure \ref{fig:Case4}). The 7-brane has constant axio-dilaton profile.
\end{enumerate}
The cases differ in the structure of the boundary conditions and cases 2,3,4 include additional operators supported on $M_3',M_3''$ absorbing the branch cut. In this picture, S-duality is realized by a vertically running branch cut without 7-brane insertions, see figure \ref{fig:Sduality}, starting and ending on $M_3',M_3''$. Of course we can also consider multiple 7-brane insertions.

We emphasize that the branchcut in all cases can be deformed arbitrarily without consequences. Branch cuts are not physical, but their endpoints are\footnote{The branch cut arises from a choice of gauge for the background connection $A_1$ of the $SL(2,\mathbb{Z})$ duality bundle as we explain momentarily. This localizes an anomaly flow which we can either absorbed by an end point (case 2,3) or flow along a half-line (case 1). Different branch cut choices give different choice of gauge, however, end points of branch cuts are gauge invariant.}, and the above cases differ precisely in how branch cut endpoints are realized.

We now discuss each case in turn. First, consider the setup for case 1. The branch cut is stacked with part of the asymptotic boundary, let us therefore move the topological 7-brane into the bulk. See the right subfigure in figure \ref{fig:7BraneInfinity2}. In the symmetry TFT formalism these two configurations are topologically equivalent. There is now a single 5D boundary condition $\ket{P_2,D_2}$ along the asymptotic boundary at $r=\infty$. Here $P_2$ denotes the choice of polarization, i.e., which combination of $B_2,C_2$ have Dirichlet boundary conditions imposed and $D_2$ denotes the boundary value for the condition, i.e., $(p_iB_2+q_iC_2)|_{r=\infty}=D_2$ for some collection of integer pairs $(p_i,q_i)$ specified by $P_2$.

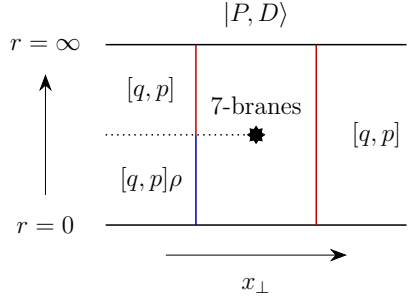
\begin{figure}[t]
    \centering
    \scalebox{0.8}{\begin{tikzpicture}
	\begin{pgfonlayer}{nodelayer}
		\node [style=none] (0) at (-2.5, 1.5) {};
		\node [style=none] (1) at (2.5, 1.5) {};
		\node [style=none] (2) at (-2.5, -1.5) {};
		\node [style=none] (3) at (2.5, -1.5) {};
		\node [style=Star] (4) at (0, 0) {};
		\node [style=none] (5) at (-2.5, 0) {};
		\node [style=none] (7) at (0, 0.5) {7-branes};
		\node [style=none] (8) at (0, 2) {$\ket{P,D}$};
		\node [style=none] (10) at (-1.5, -2) {};
		\node [style=none] (11) at (1.5, -2) {};
		\node [style=none] (12) at (0, -2.5) {$x_\perp$};
		\node [style=none] (13) at (1, 1.5) {};
		\node [style=none] (14) at (1, -1.5) {};
		\node [style=none] (15) at (-1, 1.5) {};
		\node [style=none] (16) at (-1, -1.5) {};
		\node [style=none] (17) at (-1, 0) {};
		\node [style=none] (18) at (2, 0) {$[q,p]$};
		\node [style=none] (19) at (-1.75, 0.75) {$[q,p]$};
		\node [style=none] (20) at (-1.75, -0.75) {$[q,p]\rho$};
		\node [style=none] (21) at (-3.5, -1.5) {$r=0$};
		\node [style=none] (22) at (-3.5, 1.5) {$r=\infty$};
		\node [style=none] (23) at (-3.5, 1) {};
		\node [style=none] (24) at (-3.5, -1) {};
	\end{pgfonlayer}
	\begin{pgfonlayer}{edgelayer}
		\draw [style=ThickLine] (0.center) to (1.center);
		\draw [style=ThickLine] (3.center) to (2.center);
		\draw [style=DottedLine] (5.center) to (4);
		\draw [style=ArrowLineRight] (10.center) to (11.center);
		\draw [style=RedLine] (13.center) to (14.center);
		\draw [style=RedLine] (15.center) to (17.center);
		\draw [style=BlueLine] (17.center) to (16.center);
		\draw [style=ArrowLineRight] (24.center) to (23.center);
	\end{pgfonlayer}
\end{tikzpicture}
}
    \caption{Half-space gauging via 7-brane insertion. Given a boundary condition $\ket{P,D}$ such that strings of type $[q,p]$ can terminate on $\partial X$ we find the admissible line defects on the D3-branes (located at $r=0$) to differ between the left and right hand side by a monodromy transformation. The D3-brane stack therefore experiences an effective change of polarization which amounts to a half-space gauging.}
    \label{fig:halfspacegauging2}
\end{figure}

Let us denote the monodromy matrix of the 7-brane by $\rho$. Crossing the branch cut, a string labelled by charges $[q,p]$ is transformed to a string with charges $[q,p]\rho$. Given a single topological boundary condition $\ket{P_2,D_2}$ which permits strings of charge $[q,p]$ to end on the boundary, it now follows that in the half-spaces $\bar x_\perp< x_\perp$ and $\bar x_\perp> x_\perp$ of the D3-brane worldvolume we have line defects whose charges are multiples of $[q,p]$ and $[q,p]\rho$, respectively (see figure \ref{fig:halfspacegauging2}).  Stacking the branch cut as shown in figure \ref{fig:7BraneInfinity2} (right to left) now clearly alters the boundary conditions by a monodromy transformation, more precisely its mod $N$ reduction $\rho\in SL(2,\mathbb{Z}_N)$. This change in boundary condition is a topological manipulation at the boundary, such as half-space gauging or the stacking of a counterterm
along $x_\perp<\bar x_\perp$ in $M_4$ and we propose:
\begin{equation}
    \textnormal{Topological Duality Interface $\mathcal{I}(M_3,\mathfrak{F})$}~~ \longleftrightarrow  
    \begin{array}{c}
    \textnormal{7-brane of Type $\mathfrak{F}$ on} \\ \textnormal{ $M_3\times \partial X$ with cut $\mathbb{H}_{\leftarrow}\times \partial X$} 
    \end{array}
\end{equation}
The interface $\mathcal{I}(M_3,\mathfrak{F})$ promotes to a topological defect operator $\mathcal{U}(M_3,\mathfrak{F})$ when the theories separated by the defect are dual. Due to the branch cut not intersecting the D3-brane stack we can employ arbitrary 7-branes in this construction.

Let us now consider the setup for case 2. As above, we can move the topological 7-brane into the bulk giving a topologically equivalent configuration, see figure \ref{fig:Case2} left to right. There is a single 5D boundary condition $\ket{P_1,D_1}$. Defects running between the D3 stack and the topological boundary do not cross the branch cut and the effective polarization on the brane is identical along $\bar x_\perp < x_\perp$ and $\bar x_\perp > x_\perp$. The branch cut intersecting the D3 worldvolume gives rise to a 0-form operator realizing an S-duality transformation on local\footnote{Standard S-duality acts on both local and non-local operators and is realized as in figure \ref{fig:Sduality} by a completely vertical branch cut. Such a branch cut cannot be deformed into the 5D bulk and connects two distinct boundary conditions $\ket{P_i,D_i}$ with $i=1,2$.} operators in 4D. Therefore, upon collapsing the 5D slab to 4D we obtain for all choices of 7-branes with constant axio-dilaton profile a topological defect operator in a given theory. We propose: 
\begin{equation}
    \textnormal{Topological Defect Operator $\mathcal{V}(M_3,\mathfrak{F})$}~~ \longleftrightarrow  
    \begin{array}{c}
    \textnormal{7-brane of Type $\mathfrak{F}$ on} \\ \textnormal{ $M_3\times \partial X$ with cut $\mathbb{H}_{\downarrow}\times \partial X$} 
    \end{array}
\end{equation}


Let us now consider the setup for case 3. Again we can move the topological 7-brane into the bulk giving a topologically equivalent configuration, see figure \ref{fig:Case3} left to right. There are now two boundary conditions $\ket{P_1,D_1},\ket{P_2,D_2}$ giving rise to distinct polarizations in $\bar{x}_\perp < x_\perp$ and $\bar{x}_\perp > x_\perp$. This change in polarization is realized by a codimension one operator in the asymptotic boundary. This operator acts on symmetry operators according to the monodromy matrix of the 7-brane. It does not act on local operators and we will not have much to say about this operator.

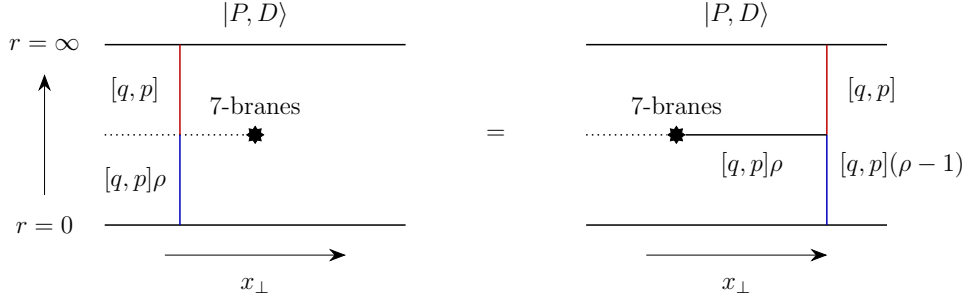
\begin{figure}[t]
    \centering
    \scalebox{0.8}{\begin{tikzpicture}
	\begin{pgfonlayer}{nodelayer}
		\node [style=none] (0) at (-6.5, 1.5) {};
		\node [style=none] (1) at (-1.5, 1.5) {};
		\node [style=none] (2) at (-6.5, -1.5) {};
		\node [style=none] (3) at (-1.5, -1.5) {};
		\node [style=Star] (4) at (-4, 0) {};
		\node [style=none] (5) at (-6.5, 0) {};
		\node [style=none] (6) at (-4, 0.5) {7-branes};
		\node [style=none] (7) at (-4, 2) {$\ket{P,D}$};
		\node [style=none] (9) at (-5.5, -2) {};
		\node [style=none] (10) at (-2.5, -2) {};
		\node [style=none] (11) at (-4, -2.5) {$x_\perp$};
		\node [style=none] (14) at (-5.25, 1.5) {};
		\node [style=none] (15) at (-5.25, -1.5) {};
		\node [style=none] (16) at (-5.25, 0) {};
		\node [style=none] (18) at (-6, 0.75) {$[q,p]$};
		\node [style=none] (19) at (-6, -0.75) {$[q,p]\rho$};
		\node [style=none] (20) at (-7.5, -1.5) {$r=0$};
		\node [style=none] (21) at (-7.5, 1.5) {$r=\infty$};
		\node [style=none] (22) at (-7.5, 1) {};
		\node [style=none] (23) at (-7.5, -1) {};
		\node [style=none] (24) at (1.5, 1.5) {};
		\node [style=none] (25) at (6.5, 1.5) {};
		\node [style=none] (26) at (1.5, -1.5) {};
		\node [style=none] (27) at (6.5, -1.5) {};
		\node [style=Star] (28) at (3, 0) {};
		\node [style=none] (29) at (1.5, 0) {};
		\node [style=none] (30) at (3, 0.5) {7-branes};
		\node [style=none] (31) at (4, 2) {$\ket{P,D}$};
		\node [style=none] (33) at (2.5, -2) {};
		\node [style=none] (34) at (5.5, -2) {};
		\node [style=none] (35) at (4, -2.5) {$x_\perp$};
		\node [style=none] (38) at (5.5, 1.5) {};
		\node [style=none] (39) at (5.5, -1.5) {};
		\node [style=none] (40) at (5.5, 0) {};
		\node [style=none] (44) at (6.25, 0.75) {$[q,p]$};
		\node [style=none] (45) at (4.25, -0.5) {$[q,p]\rho$};
		\node [style=none] (46) at (6.75, -0.5) {$[q,p](\rho-1)$};
		\node [style=none] (47) at (0, 0) {$=$};
	\end{pgfonlayer}
	\begin{pgfonlayer}{edgelayer}
		\draw [style=ThickLine] (0.center) to (1.center);
		\draw [style=ThickLine] (3.center) to (2.center);
		\draw [style=DottedLine] (5.center) to (4);
		\draw [style=ArrowLineRight] (9.center) to (10.center);
		\draw [style=RedLine] (14.center) to (16.center);
		\draw [style=BlueLine] (16.center) to (15.center);
		\draw [style=ArrowLineRight] (23.center) to (22.center);
		\draw [style=ThickLine] (24.center) to (25.center);
		\draw [style=ThickLine] (27.center) to (26.center);
		\draw [style=DottedLine] (29.center) to (28);
		\draw [style=ArrowLineRight] (33.center) to (34.center);
		\draw [style=RedLine] (38.center) to (40.center);
		\draw [style=BlueLine] (40.center) to (39.center);
		\draw [style=ThickLine] (40.center) to (28);
	\end{pgfonlayer}
\end{tikzpicture}
}
    \caption{Boundary condition $\ket{P,D}$ admitting $[q,p]$ strings to terminate at infinity. Passing a 4D line defect labelled by $[q,p]\rho$ through the symmetry defect $\mathcal{U}(M_3,\mathfrak{F})$ from left to right creates a topological surface operator of charge $[q,p](\rho-1)$ via Hanany-Witten brane creation. }
    \label{fig:HW}
\end{figure}

Next we discuss common properties of $\mathcal{I},\mathcal{V}$, independent of the choice of branch cuts, depending exclusively on the type $\mathfrak{F}$ of the 7-brane inserted. Hanany-Witten transition \cite{Hanany:1996ie} predicts the creation of a topological symmetry operator when line defects are dragged through $\mathcal{I}$, see figure \ref{fig:HW}.  Genuine defects constructed from asymptotic $[q,p]$ strings transform upon crossing $M_3$ to non-genuine defects with identical $[q,p]$ charge but with a topological operator of charge $[q,p](\rho-1)$ attached. The latter results from the corresponding string which attaches to the 7-branes stack and is otherwise embedded in the asymptotic boundary making it topological \cite{Heckman:2022muc}. Identical considerations hold for the operator $\mathcal{V}$. Such brane creation processes were already observed in D3/D5 systems in \cite{Apruzzi:2022rei}.

Let us now quantify the qualitative discussion above. The insertion of 7-branes turns on a non-trivial flat background for the $SL(2,\mathbb{Z})$ duality bundle. The topological nature of the symmetry TFT allows us to localize this background onto a choice of branch cut and in terms of the $SL(2,\mathbb{Z})$ doublet $(\mathcal{B}^i) = (C_2,B_2)$ then
\begin{equation}\label{eq:symtftcoupling}\begin{aligned}
    \mathcal{S}_{\mathrm{(SymTFT),0}}&=-\frac{N}{4\pi}\int_{M_5} \epsilon_{ij}\mathcal{B}^i \wedge  d \mathcal{B}^j \\[1em] 
     ~\xrightarrow[~\text{Insertion}~]{7-\text{brane}}~\qquad \mathcal{S}_{\mathrm{(SymTFT),1}}&=\mathcal{S}_{\mathrm{(SymTFT),0}}+\mathcal{S}_{(\mathrm{cut})} + \mathcal{S}_{\textnormal{(Defects)}}\,.
    \end{aligned}
\end{equation}
Here the term $\mathcal{S}_{\textnormal{(Defects)}}$ denotes the 3D TFT supported on $M_3\subset M_4$ resulting from wrapping 7-branes on $M_3\times \partial X$ in case 1 and in cases 2 and 3 it includes possibly an additional TFT supported on $M_3',M_3''$. We postpone the analysis of $\mathcal{S}_{\textnormal{(Defects)}}$ together with the discussion of boundary conditions for the bulk fields of the 5D symmetry TFT along such defects. The term $\mathcal{S}_{(\mathrm{cut})}$ denotes a counterterm localized to the branch cut $\mathbb{H}$ with $M_3 \subset \partial \mathbb{H}$.

\begin{figure}[t]
    \centering
    \scalebox{0.8}{
    \begin{tikzpicture}
	\begin{pgfonlayer}{nodelayer}
		\node [style=none] (0) at (-2, 0) {};
		\node [style=none] (1) at (2, 0) {};
		\node [style=none] (4) at (-3.25, 0) {Branch cut};
		\node [style=none] (5) at (0, 2) {};
		\node [style=none] (6) at (0, 0) {};
		\node [style=none] (7) at (-1, -2) {};
		\node [style=none] (8) at (1, -2) {};
		\node [style=none] (13) at (0.75, 2) {$S_{[q,p]}$};
		\node [style=none] (19) at (-1.75, -2) {$S_{[q,p]}$};
		\node [style=none] (20) at (2.25, -2) {$S_{[q,p](\rho-1)}$};
        \node [style=none] (21) at (0, -2.5) {};
	\end{pgfonlayer}
	\begin{pgfonlayer}{edgelayer}
		\draw [style=DottedLine] (0.center) to (1.center);
		\draw [style=RedLine] (5.center) to (6.center);
		\draw [style=RedLine] (6.center) to (7.center);
		\draw [style=ThickLine] (6.center) to (8.center);
	\end{pgfonlayer}
\end{tikzpicture}
    }
    \caption{5D surface symmetry operator stretching across the branch cut. $S_{[q,p]}$ transforms into $S_{[q,p]\rho}$ which can be decomposed into two surface symmetry operators as shown.}
    \label{fig:Surface}
\end{figure}
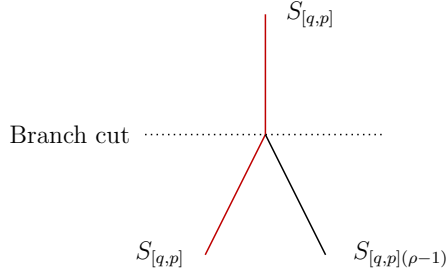

We now derive the 4D action $\mathcal{S}_{(\mathrm{cut})}$. Let us begin by determining the action on the branch cut attaching to a single D7-brane. The monodromy matrix is denoted $\mathbb{T}$ in \eqref{eq:Mono}. Clearly, a $B_2$ profile remains unchanged when crossing the branch cut and we now claim
\begin{equation}\label{eq:Cut3}
    \mathcal{S}_{(\mathrm{cut})}^{\mathrm{D7}}=\frac{2\pi i}{N}\int_\mathbb{H} \frac{\mathcal{P}(B_2)}{2} 
\end{equation}
where $\mathbb{H}$ denotes the branch cut. For a $(p,q)$ 7-brane we replace $B_2$ by $B_2^{[q,p]}=pB_2+qC_2$. Stacks of $k$ D7-branes then have action $k \mathcal{S}_{(\mathrm{cut})}^{\mathrm{D7}}$ and similarly for stacks of $(p,q)$ 7-branes.

We argue for \eqref{eq:Cut3} by considering its action on topological boundary conditions and via anomaly considerations. The former amounts to checking that the branch cut action indeed realizes the 0-form symmetry with monodromy matrix $\rho= \mathbb{T}$ on symmetry operators and defect operators in 5D. The latter shows that the branch cut carries off an anomaly sourced by the 7-brane insertion. In \cite{Antinucci:2022vyk} the action \eqref{eq:Cut3} is derived via condensation.

As a check, consider Dirichlet boundary conditions $\ket{P_{G_k},D}$ realizing $B_2|_{r=\infty}=D$ with global form $G$ and $k$ counterterms proportional to $\mathcal{P}(D)$ stacked. Then colliding the branch cut with the topological boundary amounts to acting as
\begin{equation}
    \ket{P_{G_k},D} \rightarrow \exp(\mathcal{S}_{(\mathrm{cut})})\ket{P_{G_{k}},D}=\ket{P_{G_{k+1}},D}
\end{equation}
which here stacks the boundary condition with the counterterm $(2\pi i/N)\mathcal{P}(D)/2$ without changing the polarization. This is consistent with the fact that defects for $P$ are Wilson lines constructed from fundamental strings which are not acted on upon crossing the branch cut. Therefore the polarization must remain unchanged and stacking the branch cut with the boundary can at most add counterterms. More generally this follows from T-duality \cite{Aharony:2013hda} and other cases are argued for identically, we give explicit computations in section \ref{sec:N4}.

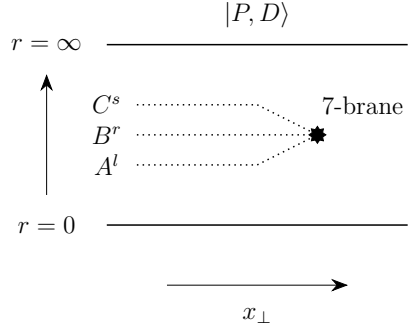
\begin{figure}
    \centering
    \scalebox{0.8}{
    \begin{tikzpicture}
	\begin{pgfonlayer}{nodelayer}
		\node [style=none] (0) at (-2.5, 1.5) {};
		\node [style=none] (1) at (2.5, 1.5) {};
		\node [style=none] (2) at (-2.5, -1.5) {};
		\node [style=none] (3) at (2.5, -1.5) {};
		\node [style=none] (4) at (0, 2) {$\ket{P,D}$};
		\node [style=none] (6) at (-3.5, -1.5) {$r=0$};
		\node [style=none] (7) at (-3.5, 1.5) {$r=\infty$};
		\node [style=none] (8) at (-3.5, 1) {};
		\node [style=none] (9) at (-3.5, -1) {};
		\node [style=none] (10) at (-1.5, -2.5) {};
		\node [style=none] (11) at (1.5, -2.5) {};
		\node [style=none] (12) at (0, -3) {$x_\perp$};
		\node [style=Star] (13) at (1, 0) {};
		\node [style=none] (14) at (0, 0.5) {};
		\node [style=none] (16) at (0, -0.5) {};
		\node [style=none] (17) at (-2, -0.5) {};
		\node [style=none] (18) at (-2, 0) {};
		\node [style=none] (19) at (-2, 0.5) {};
		\node [style=none] (20) at (-2.5, -0.5) {$A^l$};
		\node [style=none] (21) at (-2.5, 0) {$B^r$};
		\node [style=none] (22) at (-2.5, 0.5) {$C^s$};
		\node [style=none] (23) at (1.75, 0.5) {7-brane};
	\end{pgfonlayer}
	\begin{pgfonlayer}{edgelayer}
		\draw [style=ThickLine] (0.center) to (1.center);
		\draw [style=ThickLine] (3.center) to (2.center);
		\draw [style=ArrowLineRight] (9.center) to (8.center);
		\draw [style=ArrowLineRight] (10.center) to (11.center);
		\draw [style=DottedLine] (19.center) to (14.center);
		\draw [style=DottedLine] (17.center) to (16.center);
		\draw [style=DottedLine] (14.center) to (13);
		\draw [style=DottedLine] (16.center) to (13);
		\draw [style=DottedLine] (18.center) to (13);
	\end{pgfonlayer}
\end{tikzpicture}
    }
    \caption{The branch cut of attaching to any 7-brane insertion can be decomposed into branch cuts individually associated with $(p,q)$ 7-branes. }
    \label{fig:BCsplit}
\end{figure}

With this we can give the branch cut term for any 7-brane insertion (see figure \ref{fig:BCsplit}). Any 7-brane can be represented as a supersymmetric bound state of certain $(p,q)$ 7-branes denoted $A,B,C$, their $(p,q)$ charges are $(1,0),(3,1),(1,1)$ respectively. All 7-branes are then of the form $A^lB^rC^s$ and their branch cut therefore supports the operator
\begin{equation}\label{eq:BranchCutOperator}
\mathcal{O}_{(l,r,s)}=\mathcal{O}_A^l\mathcal{O}_B^r\mathcal{O}_C^s
\end{equation}
which is purely determined by the monodromy of the 7-brane and where
\begin{equation}\begin{aligned}
    \mathcal{O}_A&=\exp\left( \frac{2\pi i}{N}\int_{\mathbb{H}_A} \frac{\mathcal{P}(B_2)}{2}\right)\\
    \mathcal{O}_B&=\exp\left( \frac{2\pi i}{N}\int_{\mathbb{H}_B} \frac{\mathcal{P}(3B_2+C_2)}{2}\right)\\
    \mathcal{O}_C&=\exp\left( \frac{2\pi i}{N}\int_{\mathbb{H}_C} \frac{\mathcal{P}(B_2+C_2)}{2}\right)\,.
\end{aligned}\end{equation}
Here $B_2,C_2$ are normalized to $\mathbb{Z}_N$-valued forms and $\mathbb{H}_A,\mathbb{H}_B,\mathbb{H}_C$ denote the branch cuts localizing the monodromy of the respective brane stack. Note that it is not possible in general to present $\mathcal{O}_{(l,r,s)}$ as a single exponential, the fields $B_2,C_2$ are conjugate and do not commute. However, it is possible to present it as a TFT \cite{Antinucci:2022vyk}.

Let us discuss possible anomalies. Note first that we have the boundary condition $B_2|_{\mathrm{D7}}=0$ at the D7 locus. This follows from noting that $B_2|_{\mathrm{D7}}\neq 0$ implies that along infinitesimal loops linking the D7-brane we have $C_2\rightarrow C_2'\approx C_2+B_2|_{\mathrm{D7}}$ leading to a discontinuous, ill-defined profile for $C_2$ contradicting the D7-brane solution. In the M-theory dual picture this is equivalent to the torus cycle corresponding to $B_2$ shrinking at the D7-brane locus. Backgrounds with only $C_2$ turned on are invariant under monodromy and therefore we have no constraint on the $C_2$ profile along the D7-brane.

When considering a stack of $k$ D7-branes the boundary condition is $kB_2|_{\mathrm{D7}}=0$ and together with $B_2$ taking values in $\mathbb{Z}_N$ we have $\textnormal{gcd}(k,N)B_2|_{\mathrm{D7}}=0$. Now $B_2|_{\mathrm{D7}}$ no longer vanishes generically but takes values which are multiples of $N/\textnormal{gcd}(k,N)$. The $C_2$ profile remains unconstrained. 

Next, consider the background transformation $\mathcal{B}\rightarrow \mathcal{B}+d\lambda$ where $\lambda$ is twisted by the monodromy of the 7-brane, i.e. $d\lambda$ is subject to the same boundary conditions along the D7-brane as $\mathcal{B}_2$. The Pontryagin square term $\mathcal{P}(B_2)$ gives rise to a boundary term iff $\textnormal{gcd}(k,N)\neq 1$. This anomaly must be absorbed by the 3D TFT associated with the wrapped D7-branes. 

The 3D TFT has non-trivial lines precisely when $\textnormal{gcd}(k,N)\neq 1$. For this consider the dual M-theory geometry of a D7-brane whose normal geometry is given by an elliptic fibration $Z$ over $\mathbb{C}$ with one Kodaira $I_k$ singularity. The boundary $\partial Z$ of this normal geometry is elliptically fibered over a circle and has first homology groups $H_1(\partial Z)=\mathbb{Z}\oplus\mathbb{Z}\oplus \mathbb{Z}_k$. One free factor corresponds to the base circle and can be neglected in our discussion. This leaves us with $\mathbb{Z}\oplus \mathbb{Z}_k$. Of these, only the torsional generator of $\mathbb{Z}_k$ collapses at the $I_k$ singularity, sweeping out a non-compact two-cycle in the process. Wrapping an M2 brane on this cycle constructs a Wilson line. In 5D we work modulo $N$ and overall this gives a defect group of lines isomorphic to $\mathbb{Z}_{\textnormal{gcd}(k,N)}$ for the 3D TFT. In section \ref{sec:DefectTFT} we show that these lines organize into a 3D TFT with an anomaly sourcing the one carried by the branch cut. 

More generally the Pontryagin square terms on individual elementary branch cuts are anomalous under background transformations and we have a local anomaly flowing along each cut. Summing over branch cuts, the net anomaly is then necessarily absorbed by the 7-brane which can be viewed as an edge mode to the theories localized on the branch cuts. Alternatively, the overall anomaly is non-vanishing whenever the 7-brane sources an anomaly.

Let us therefore discuss when 7-brane insertions source anomalies in greater generality by studying the branch cut terms. For this, let us first consider in the 5D symmetry TFT the surface symmetry operators acting on the surface defects constructed from $(p,q)$-strings. They are:\footnote{Our definition differs from the definition $S_{(p,q)}'(\Sigma_2)\equiv S_{(p,0)}(\Sigma_2)S_{(0,q)}(\Sigma_2)$ as given in \cite{Kaidi:2022cpf} by a phase with argument proportional to the self-linking number of the surface $\Sigma_2$. This relative phase follows from the Baker-Campbell-Hausdorff formula and noting that $B_2,C_2$ are conjugate variables in the 5D TFT.}
\begin{equation}
    S_{[q,p]}(\Sigma_2)=\exp\left(2\pi i\oint_{\Sigma_2} (pB_2+qC_2)\right)\,, 
\end{equation}
with integers $p,q$ modulo $N$. Similarly to strings they are transformed when stretching across the $SL(2,\mathbb{Z})$ branch cut. Consider a surface $\Sigma_2$ separated by the branch cut $\mathbb{H}$ into two components $\Sigma_2 =\Sigma_2^+\cup \Sigma_2^-$, then
\begin{equation}
    S_{[q,p]}(\Sigma_2^{+})S_{[q,p]\rho}(\Sigma_2^{-})=S_{[q,p]}(\Sigma_2)S_{[q,p](\rho-1)}(\Sigma_2^{-})
\end{equation}
whenever the self-linking number of $\Sigma_2$ in the 5D bulk vanishes, see figure \ref{fig:Surface}. This is interpreted as the intersection of $S_{[q,p]}(\Sigma_2)$ with the branch cut sourcing $S_{[q,p](\rho-1)}(\Sigma_2^{-})$. Correspondingly, in terms of the $\mathbb{Z}_N$-valued 1-form symmetry background fields this relation can be expressed as
\begin{equation}\label{eq:dualityznaction}
    \delta \left( (\rho-1) \begin{bmatrix}
        C_2 \\ B_2
    \end{bmatrix}\right) = A_1\cup \begin{bmatrix}
        C_2 \\ B_2
    \end{bmatrix}
\end{equation}
where $A_1$ is the background field of the $SL(2,\mathbb{Z})$ bundle proportional to the Poincar\'e dual of the branch cut. A priori, $A_1$ takes values in $\mathbb{Z}_n$ where $n$ is the order of $\rho\in SL(2,\mathbb{Z})$, or $U(1)$ when $n$ is infinite, but in \eqref{eq:dualityznaction} this $A_1$ takes values in in $\mathbb{Z}_{\mathrm{gcd}(n,N)}$. In other words, $A_1$ takes values in $\mathbb{Z}_{\mathrm{gcd}(4,N)}$ or $\mathbb{Z}_{\mathrm{gcd}(3,N)}$ for duality\footnote{One might ask whether one should call this order four operation a ``quadrality'' operator. On local operators it is indeed order two, which accounts for the terminology. This subtlety between $2$ versus $4$ will show up later in our analysis of mixed anomalies.} and triality defects respectively\footnote{There are also hexality defects which are furnished by 7-branes of Type $II$ or $II^*$ whose background field in the 4D relative theory is $\mathbb{Z}_{\mathrm{gcd}(6,N)}$-valued. We come back to these theories at the end of the section to show that their mixed anomalies with the 1-form symmetry is always trivial.}, and similar to \eqref{eq:rescaling} we have a relation between normalizations of discrete-valued fields as (assuming $n$ is finite)
\begin{equation}\label{eq:rescaling2}
    A^{\mathbb{Z}_n}_1=\frac{1}{\mathrm{gcd}(n,N)}A^{\mathbb{Z}_{\mathrm{gcd}(k,N)}}_1
\end{equation}
where $A^{\mathbb{Z}_\ell}_1$ has holonomies that take values $p \; \mathrm{mod}\; \ell$.

Two related points to take into account are that one must first choose a polarization to have an invertible anomaly TFT, and in general $\rho-1$ does not have an inverse in $\mathbb{Z}_N$ coefficients.\footnote{In the cases when $(\rho-1)$ does have an inverse we have that $\mathrm{gcd}(2,N)=0$ or $\mathrm{gcd}(3,N)=0$ making the anomaly trivial in the first place.} We are motivated then to consider polarization choices that are duality/triality invariant, which is consistent with the fact that otherwise, the topological defect implementing the duality or triality defect is an interface between separate theories rather than a symmetry operator which may have a mixed 't Hooft anomaly. Recall that the background 1-form field for a given polarization spans a 1-dimensional subspace of $\mathrm{Span}(C_2,B_2)^\mathrm{T}$ and we say that a polarization is duality or triality invariant if there are non-trivial solutions to the following equation, posed over $\mathbb{Z}_N$,
\begin{equation}\label{eq:Eigenvalue}
    (\rho-1)\mathcal{B}_2=0\,
\end{equation}
which only has non-trivial solutions in $\mathbb{Z}_K$ subgroups of $\mathbb{Z}_N$ where $K=\mathrm{gcd}(k,N)$ and $k$ is the coefficient appearing in Table \ref{tab:Fibs}.\footnote{There are no solutions if $K$ and $N$ are coprime.} We denote such eigenvector solutions as $\mathcal{B}_2^\rho$ which is the product of an $SL(2,\mathbb{Z})$ vector and the form profile $B_2^\rho$. 

Consider for example the case of dualities that $N$ is even. We denote the lift to the full $\mathbb{Z}^{(1)}_N$-background field by $B^\rho_2$ as well where the two are related as
\begin{equation}\label{eq:rescale3}
    (B^\rho_2)^{\mathbb{Z}_{N}}=\frac{1}{2}(B^\rho_2)^{\mathbb{Z}_{2}}
\end{equation}
when the LHS only takes values in the $\mathbb{Z}_2$ subgroup of $\mathbb{Z}_N$ generated by $N/2 \; \mathrm{mod} \; N$. It will be clear that the anomaly will be invariant under this choice of lift.

We can decompose the vector space as $\mathrm{ker}(\rho-1)\oplus \mathrm{ker}(\rho-1)^\perp$ where the latter is generated by vectors $v$ such that $(\rho-1)v=v$. This allows us to define $(\mathcal{B}^\rho_2)^\perp$ which due to the $SL(2,\mathbb{Z})$ invariant pairing $\epsilon_{ij}$ in \eqref{eq:symtftcoupling}, allows us to rewrite that coupling schematically as $\int \mathcal{B}_2^\rho \cup \delta (\mathcal{B}^\rho_2)^\perp$. Substituting \eqref{eq:dualityznaction} then allows us to derive the mixed 't Hooft anomalies between 0-form duality/triality symmetries and $\mathbb{Z}^{(1)}_N$ 1-form symmetries:\footnote{In $\mathcal{S}_{\mathrm{duality}}$, observe that we take a gcd with respect to $2$ rather then $N$ because the operation is order two on local operators.}
\begin{equation}\label{eq:mixed anomaly term in the bulk}
    \begin{aligned}
       \mathcal{S}_{\mathrm{duality}}&= \frac{2\pi}{\mathrm{gcd}(2,N)}\int A_1\cup \frac{\mathcal{P}(B^\rho_2)}{2}  \\
    \mathcal{S}_{\mathrm{triality}} &= \frac{2\pi}{\mathrm{gcd}(3,N)}\int A_1\cup \frac{\mathcal{P}(B^\rho_2)}{2}
    \end{aligned}
\end{equation}
where we implemented the refinement $B_2^\rho \cup B_2^\rho \rightarrow \mathcal{P}(B_2^\rho)/2$ with $\mathcal{P}$ the Pontryagin square operation following \cite{Gaiotto:2014kfa,Kapustin:2014gua,Benini:2018reh}. The normalization of discrete-valued fields is motivated to match with \cite{Kaidi:2021xfk} which, in our presentation, means that we are using the LHS of \eqref{eq:rescaling2} for the normalization of $A_1$ and the RHS of \eqref{eq:rescaling} for the normalization of the $\mathbb{Z}_2$ or $\mathbb{Z}_3$-valued field $B^\rho_2$.\footnote{For the duality case, only the image of $A_1$ in the quotient $\mathbb{Z}_4/\mathbb{Z}_2\simeq \mathbb{Z}_2$ couples to $B^\rho_2$ which explains the factor $\mathrm{gcd}(2,N)$ rather than $\mathrm{gcd}(4,N)$.} 

We have that $A_1$ is Poincar\'e dual to the branch cut $\mathbb{H}$ and therefore
\begin{equation}\label{eq:CT}
    \begin{aligned}
    \textnormal{Duality Interface:}&\quad   S_{\mathrm{cut}}= \frac{2\pi}{\mathrm{gcd}(2,N)}\int_\mathbb{H} \frac{\mathcal{P}(B^\rho_2)}{2}  \\
    \textnormal{Triality Interface:}&\quad S_{\mathrm{cut}}= \frac{2\pi}{\mathrm{gcd}(3,N)}\int_\mathbb{H} \frac{\mathcal{P}(B^\rho_2)}{2}\,.
    \end{aligned}
\end{equation}
We conclude that upon inserting a 7-brane giving rise to the anomaly \eqref{eq:mixed anomaly term in the bulk}, we can localize this anomaly to a single branch cut with action \eqref{eq:CT}. Conversely, in cases with no net anomaly the anomaly localized to individual branch cuts cancels and therefore the branch cut term does not admit a presentation as a simple exponential.

\begin{figure}
    \centering
    \scalebox{0.8}{\begin{tikzpicture}
	\begin{pgfonlayer}{nodelayer}
		\node [style=none] (0) at (-2.5, 1.5) {};
		\node [style=none] (1) at (2.5, 1.5) {};
		\node [style=none] (2) at (-2.5, -1.5) {};
		\node [style=none] (3) at (2.5, -1.5) {};
		\node [style=none] (7) at (0, 2) {$\ket{P,D}$};
		\node [style=none] (13) at (-3.5, -1.5) {$r=0$};
		\node [style=none] (14) at (-3.5, 1.5) {$r=\infty$};
		\node [style=none] (15) at (-3.5, 1) {};
		\node [style=none] (16) at (-3.5, -1) {};
		\node [style=none] (17) at (-2.5, 0) {};
		\node [style=none] (18) at (2.5, 0) {};
		\node [style=none] (19) at (3.25, 0) {$\mathcal{O}^{[q,p]}$};
		\node [style=none] (20) at (0, -0.5) {$M_4'$};
		\node [style=none] (21) at (0, 0.25) {};
		\node [style=none] (22) at (0, 1.25) {};
		\node [style=none] (23) at (0, -2) {$\ket{\mathfrak{T}^{(N)}_X}$};
	\end{pgfonlayer}
	\begin{pgfonlayer}{edgelayer}
		\draw [style=ThickLine] (0.center) to (1.center);
		\draw [style=ThickLine] (3.center) to (2.center);
		\draw [style=ArrowLineRight] (16.center) to (15.center);
		\draw [style=DottedLine] (17.center) to (18.center);
		\draw [style=ArrowLineRight] (21.center) to (22.center);
	\end{pgfonlayer}
\end{tikzpicture}
}
    \caption{The operator $\mathcal{O}^{[q,p]}$ is defined on $M_4'$ in the 5D bulk which is homotopic to either boundary. The operator acts on the topological boundary condition $\ket{P,D}$ by colliding $M_4'$ with the corresponding boundary component. }
    \label{fig:BulkOps}
\end{figure}
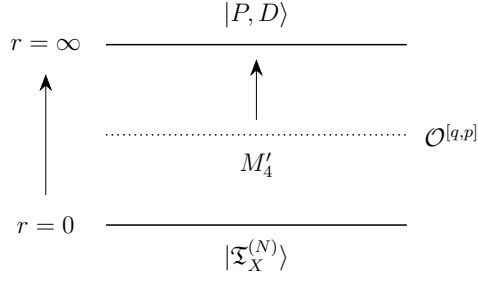

Let us now discuss another class of codimension one bulk operators of the 5D symmetry TFT. We use this to generalize the discussion of counterterm actions considered above. Along these lines, fix $B_2^{[q,p]}$ as the combination of doublet fields which pulls back to a $(p,q)$ 7-brane (i.e., is compatible with this choice of monodromy). 
Then, introduce the operator:
\begin{equation}
 \mathcal{O}^{[q,p]}=\exp\left( \frac{2\pi i}{N}\int_{M_4'}\frac{\mathcal{P}(B_2^{[q,p]})}{2}\right)\,.
\end{equation}
where $M_4'$ a copy of $M_4$ deformed into the bulk of the 5D TFT. It runs horizontally, parallel to the D3-brane worldvolume. More generally we could consider four-manifolds with boundary in conjunction with various edge modes.  Here $B_2^{[q,p]}=pB_2+qC_2$ is the form supported on the branch cut of a $(p,q)$ 7-brane. When contracting the 5D slab to 4D the branch cut is layered with the topological boundary condition and so the unitary operator $\mathcal{O}^{[q,p]}$ realizes an isomorphism on the vector space of boundary conditions generated by $\{\ket{P_i,D_i}\}$ (see figure \ref{fig:BulkOps}). Let us write
\begin{equation}\label{eq:Opq}
    \bra{P_i,D_i}\mathcal{O}^{[q,p]}= \bra{P_i^{[q,p]},D_i}\,.
\end{equation}
Note that whenever the $SL(2,\mathbb{Z})$ vector $[q,p]$ is contained in the polarization $P$ then the corresponding topological boundary condition is an eigenvector and $\mathcal{O}^{[q,p]}$ acts by stacking the boundary condition with a counterterm whose profile is determined by the Dirichlet boundary condition. However, more generally $\mathcal{O}^{[q,p]}$ is not diagonal, in particular there is no operator $\mathcal{O}^{[q,p]}$ which acts via counterterm stacking on all topological boundary conditions. 

We close this section by emphasizing that nothing about this discussion requires a large $N$ limit, the existence of a holographic dual, or having $\mathcal{N} = 4$ supersymmetry.

\section{Defect TFT}
\label{sec:DefectTFT}

We now discuss the 3D TFT supported on the defect constructed by wrapping some 7-brane on $ \partial X$ and its coupling to the bulk 5D symmetry TFT. The main idea will be to treat the wrapped 7-branes as codimension two defects in the 5D theory. See Appendix \ref{app:minimalTFT7branes} for a discussion of how dimensional reduction of the 7-brane on $\partial X$ can result in such topological terms.

\subsection{Minimal Abelian TFT from Defects}\label{ssec:linkingminimaltheory}

Consider the setup of the previous section consisting of a stack of 7-branes of type $\mathfrak{F}$ with monodromy matrix $\rho$ compactified on $\partial X$ to 3D in the presence of $N$ D3-branes located at the apex of $X$, the cone over $\partial X$. This produces a 3D TFT $\mathcal{T}$ supported on $M_3$ coupled to the worldvolume of the D3-brane stack. 

Let us begin by considering case 1 with a horizontal branch cut (along $\mathbb{H}_{\leftarrow}$) for which there is a single topological boundary condition along the asymptotic boundary. The 7-brane is located at $\bar x_\perp$ and separates the D3-worldvolume into two half-spaces $\bar x_\perp>  x_\perp$ and $\bar x_\perp<  x_\perp$.

\begin{figure}[t]
    \centering
    \scalebox{0.8}{
    \begin{tikzpicture}
	\begin{pgfonlayer}{nodelayer}
		\node [style=none] (0) at (-6.5, 1.5) {};
		\node [style=none] (1) at (-1.5, 1.5) {};
		\node [style=none] (2) at (-6.5, -1.5) {};
		\node [style=none] (3) at (-1.5, -1.5) {};
		\node [style=Star] (4) at (-4, 0) {};
		\node [style=none] (5) at (-6.5, 0) {};
		\node [style=none] (6) at (-4, 0.5) {7-branes};
		\node [style=none] (7) at (-4, 2) {$\ket{P,D}$};
		\node [style=none] (8) at (-5.5, -2) {};
		\node [style=none] (9) at (-2.5, -2) {};
		\node [style=none] (10) at (-4, -2.5) {$x_\perp$};
		\node [style=none] (11) at (-5.5, 1.5) {};
		\node [style=none] (12) at (-5.5, -1.5) {};
		\node [style=none] (13) at (-5.5, 0) {};
		\node [style=none] (14) at (-6.25, 0.75) {$[q,p]$};
		\node [style=none] (15) at (-6.25, -0.75) {$[q,p]\rho$};
		\node [style=none] (16) at (-8, -1.5) {$r=0$};
		\node [style=none] (17) at (-8, 1.5) {$r=\infty$};
		\node [style=none] (18) at (-8, 1) {};
		\node [style=none] (19) at (-8, -1) {};
		\node [style=none] (21) at (-2.5, 1.5) {};
		\node [style=none] (22) at (-2.5, -1.5) {};
		\node [style=none] (23) at (-1.75, 0.75) {$[q,p]$};
		\node [style=none] (24) at (-2.5, 0) {};
		\node [style=none] (25) at (1.5, 1.5) {};
		\node [style=none] (26) at (6.5, 1.5) {};
		\node [style=none] (27) at (1.5, -1.5) {};
		\node [style=none] (28) at (6.5, -1.5) {};
		\node [style=Star] (29) at (4, 0) {};
		\node [style=none] (30) at (1.5, 0) {};
		\node [style=none] (32) at (4, 2) {$\ket{P,D}$};
		\node [style=none] (33) at (2.5, -2) {};
		\node [style=none] (34) at (5.5, -2) {};
		\node [style=none] (35) at (4, -2.5) {$x_\perp$};
		\node [style=none] (36) at (4.5, 1.5) {};
		\node [style=none] (37) at (3.5, -1.5) {};
		\node [style=none] (38) at (3.5, 0) {};
		\node [style=none] (39) at (2.75, 0.75) {$[q,p]$};
		\node [style=none] (40) at (2.75, -0.75) {$[q,p]\rho$};
		\node [style=none] (46) at (4.5, -1.5) {};
		\node [style=none] (47) at (5.5, 0) {$[q,p]$};
		\node [style=none] (48) at (4.5, 0) {};
		\node [style=none] (49) at (0, 0) {$=$};
		\node [style=none] (50) at (4, 1) {};
	\end{pgfonlayer}
	\begin{pgfonlayer}{edgelayer}
		\draw [style=ThickLine] (0.center) to (1.center);
		\draw [style=ThickLine] (3.center) to (2.center);
		\draw [style=DottedLine] (5.center) to (4);
		\draw [style=ArrowLineRight] (8.center) to (9.center);
		\draw [style=RedLine] (11.center) to (13.center);
		\draw [style=BlueLine] (13.center) to (12.center);
		\draw [style=ArrowLineRight] (19.center) to (18.center);
		\draw [style=RedLine] (21.center) to (22.center);
		\draw [style=ArrowLineBlue] (12.center) to (13.center);
		\draw [style=ArrowLineRed] (21.center) to (24.center);
		\draw [style=ThickLine] (25.center) to (26.center);
		\draw [style=ThickLine] (28.center) to (27.center);
		\draw [style=DottedLine] (30.center) to (29);
		\draw [style=ArrowLineRight] (33.center) to (34.center);
		\draw [style=BlueLine] (38.center) to (37.center);
		\draw [style=ArrowLineBlue] (37.center) to (38.center);
		\draw [style=RedLine] (46.center) to (48.center);
		\draw [style=RedLine, in=0, out=90] (48.center) to (50.center);
		\draw [style=RedLine, in=90, out=180] (50.center) to (38.center);
		\draw [style=ArrowLineRed, in=90, out=0] (50.center) to (48.center);
	\end{pgfonlayer}
\end{tikzpicture}
    }
    \caption{Collision of line defect of charge $[q,p]$ with line defect $-[q,p]\rho$. Conversely, line defects of charge $[q,p](1-\rho)$ are lines not inherent to the 3D TFT $\mathcal{T}$.}
    \label{fig:Screening}
\end{figure}
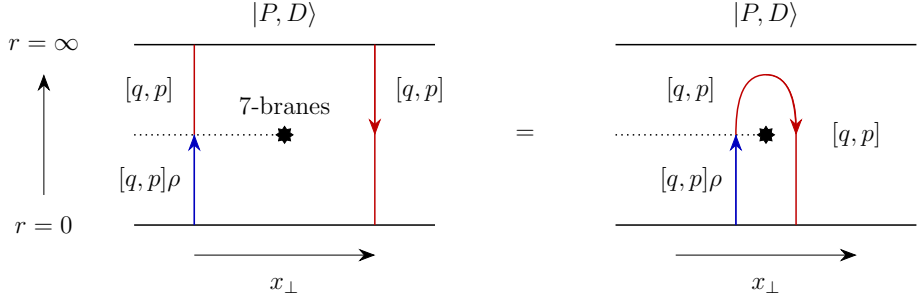

Previously we studied, via the Hanany-Witten transition, the consequence of dragging lines in $P_i$ through $M_3$ for $i=1,2$. Now consider two lines of equal but opposite asymptotic charge in each of the half-spaces and collide these. This produces a line operator and is interpreted as an open string running between the 7-branes and the D3-branes. If the lines have charges $\pm[q,p]$ then the latter is a string of charge $[q,p](1-\rho)$ (see figure \ref{fig:Screening}). These are precisely the lines not inherent to the 3D TFT $\mathcal{T}$, they can be moved off $M_3$. The inherent lines of the 3D TFT are associated with open strings running between the 7-branes and D3-branes modulo this screening effect.

With this the line defects of $\mathcal{T}$ are simply the line defects of the 7-brane stack modulo $N$. Let us study the lines of $\mathcal{T}$ in absence of the D3-brane flux, i.e., we do not impose screening modulo $N$.

In order to study the spin (in the sense of \cite{Hsin:2018vcg, Gukov:2020btk}) of such lines, consider the purely geometric M-/F-theory dual setup of an isolated stack of such 7-branes consisting of a local K3 surface $Z\rightarrow \mathbb{C}$. The boundary $\partial Z\rightarrow S^1$ has first homology $H_1(\partial Z)\cong \mathbb{Z}\oplus \textnormal{coker}(\rho-1)$ and line defects charged under the center symmetry of the system are constructed by wrapping M2-branes on cones over torsional one-cycles in $H_1(\partial Z)$. Let us assume that this homology group  is isomorphic to $\mathbb{Z}_k$ with generator $\gamma$. This torsional one-cycle is contained in the elliptic fiber, we write $\gamma=r \sigma_{a}+ s \sigma_{b}$ (in the obvious notation) for one of its representatives. This is dual to a string of charge $Q=[s,r]$. We have $k\gamma=0$ and $k$ copies of the corresponding string are screened in the F-theory setup. 

The spin of the lines associated with $\gamma$ is now determined by the refined self-linking number of $\gamma$ in $\partial Z$ as given by
\begin{equation}
    \ell(\gamma,\gamma)=\frac{1}{2k}\;\!\gamma \cdot \Sigma_\gamma\quad \mathrm{mod~1}
    \end{equation}
where $\partial \Sigma_\gamma=k\gamma$, determines the spin $h[L_\gamma]=\ell(L_\gamma,L_\gamma)=m/2k$ of the line $L_\gamma$ (see table \ref{tab:Fibs}). For a stack of $k$ $(p,q)$ 7-branes we have $m=k-1$.

Let us now take the screening effects due to the D3-brane flux into account. In this case the lines of 3D TFT $\mathcal{T}$ trivialize modulo $N$ and $k$, and therefore give rise to a 1-form symmetry $\mathbb{Z}_K$ with charged lines $L_\gamma$ and $K=\textnormal{gcd}(k,N)$. Whenever $mK\in 2\mathbb{Z}$ and $\textnormal{gcd}(m,K)=1$ it follows from the general discussion in \cite{Hsin:2018vcg} that the lines $L_\gamma$ form a consistent minimal abelian TFT denoted $\mathcal{A}^{K,m}$ \footnote{$\mathcal{A}^{K,m}$ is defined as a minimal 3D TFT with $\mathbb{Z}_K^{(1)}$ symmetry, whose symmetry lines have spins given by $h[a^s] \equiv \frac{ms^2}{2K} \mod 1$, where $a$ is a generating symmetry line such that $a^K = 1$.}. For branes of constant axio-dilaton and stacks of $(p,q)$ 7-branes we have $\textnormal{gcd}(k,m)=1$ and therefore $\textnormal{gcd}(K,m)=1$ follows. When these two conditions are met we have
\begin{equation}\label{eq:TFT}
    \mathcal{T}[B]=\mathcal{A}^{K,m}[B] \otimes \mathcal{T}'
\end{equation}
where $\mathcal{T}'$ is a decoupled TFT with lines neutral under the $\mathbb{Z}_K$ 1-form symmetry and $B$ is a 2-form background field for the 1-form symmetry which follows from the coupling of the open strings running between the D3-branes and 7-branes to $(B_2,C_2)$ and is $B=Q_i \mathcal{B}^i|_{\textnormal{7-brane}}$ which is $SL(2,\mathbb{Z})$ invariant where $Q$ is the charge vector of the strings.

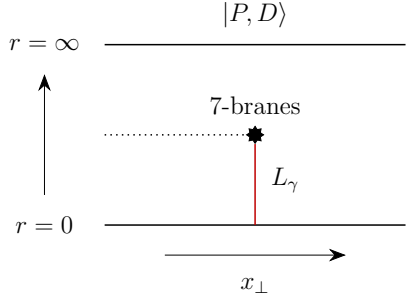
\begin{figure}[t]
    \centering
    \scalebox{0.8}{\begin{tikzpicture}
	\begin{pgfonlayer}{nodelayer}
		\node [style=none] (0) at (-2.5, 1.5) {};
		\node [style=none] (1) at (2.5, 1.5) {};
		\node [style=none] (2) at (-2.5, -1.5) {};
		\node [style=none] (3) at (2.5, -1.5) {};
		\node [style=Star] (4) at (0, 0) {};
		\node [style=none] (5) at (-2.5, 0) {};
		\node [style=none] (6) at (0, 2) {$\ket{P,D}$};
		\node [style=none] (7) at (-1.5, -2) {};
		\node [style=none] (8) at (1.5, -2) {};
		\node [style=none] (9) at (0, -2.5) {$x_\perp$};
		\node [style=none] (10) at (0, 0.5) {7-branes};
		\node [style=none] (17) at (0, -1.5) {};
		\node [style=none] (18) at (-3.5, -1.5) {$r=0$};
		\node [style=none] (19) at (-3.5, 1.5) {$r=\infty$};
		\node [style=none] (20) at (-3.5, 1) {};
		\node [style=none] (21) at (-3.5, -1) {};
		\node [style=none] (22) at (0.5, -0.75) {$L_\gamma$};
	\end{pgfonlayer}
	\begin{pgfonlayer}{edgelayer}
		\draw [style=ThickLine] (0.center) to (1.center);
		\draw [style=ThickLine] (3.center) to (2.center);
		\draw [style=DottedLine] (5.center) to (4);
		\draw [style=ArrowLineRight] (7.center) to (8.center);
		\draw [style=RedLine] (4) to (17.center);
		\draw [style=ArrowLineRight] (21.center) to (20.center);
	\end{pgfonlayer}
\end{tikzpicture}
}
    \caption{Line operators of $\mathcal{T}$ coupled to the 4D theory are open strings running between the 7-branes and the D3-brane stack. The lines $L_\gamma^n$ organize into the minimal abelian TFT $\mathcal{A}^{K,m}$ where $K,m$ follow from the elliptic data of the 7-brane.}
    \label{fig:MinimalLines}
\end{figure}

The charge vector $Q_i$ associated with $\gamma$ is defined modulo vectors in the image of $\rho-1$. Consider the charge vector $Q'_i=Q_i+(q (\rho-1))_i$ in the same coset. Then, $B$ changes as
\begin{equation}
    B=Q_i\mathcal{B}^i~\rightarrow ~ Q_i'\mathcal{B}^i=Q_i\mathcal{B}^i+(q(\rho-1))_i\mathcal{B}^i=Q_i\mathcal{B}^i+q_i((\rho-1)\mathcal{B})^i
\end{equation}
and for this coupling to be well-defined we require the profile of $\mathcal{B}$ to lie in the kernel of $\rho-1$ modulo $K$. 

For example, consider $N=2$ with insertion of a 7-brane of type $\mathfrak{F}=III^{\ast}$. We compute $k=2$ and therefore $K=2$. The eigenvector is $\mathcal{B}_2^\rho=B_2^\rho[1,1]^t$ and the background in \eqref{eq:TFT} is $B=B_2^\rho$. We have $\mathcal{T}[B_2^\rho]=\mathcal{A}^{2,1}[B_2^\rho]\otimes \mathcal{T}'$.

When $K=1$ no eigenvector exists, $B_2^\rho$ vanishes and $\mathcal{T}$ has no lines coupling to the bulk. In this case $\mathcal{T}$ does not absorb an anomaly. 

We note that this discussion of the TFT $\mathcal{T}$ is independent of branch cut choice and therefore extends to cases 2,3,4. 

Next, let us discuss case 2 with vertically running branch cut terminating on the 0-form operator $\mathcal{D}(M_3
',B_2^\rho)$ contained in the D3-worldvolume. This is another defect possible supporting a 3D TFT. We claim that for all parameter values $\mathcal{D}(M_3
',B_2^\rho)$ does not support a TFT interacting with the bulk. As discussed previously, the operator $\mathcal{D}(M_3
',B_2^\rho)$ realizes a gluing condition on the enriched Neumann boundary condition set by the D3-brane stack and, is not realized by a 7-brane. Therefore, no strings end on $M_3'$ and it does not support a defect group of its own, in contrast to the 7-brane insertion. With this we conjecture that the TFT living at the intersection is trivial or at least does not interact with the bulk.

\section{Example: $\mathcal{N} = 4$ SCFTs}
\label{sec:N4}
In the previous section we presented a general discussion of how 7-branes implement duality defects in systems obtained from D3-branes probing an isolated Calabi-Yau singularity. In particular, we saw that the structure of the branch cuts leads to distinct implementations for various sorts of duality defects (as well as triality defects). In this section we show that this matches to the available results in the literature for $\mathcal{N} = 4$ SYM theory, in particular the case where the gauge algebra is $\mathfrak{su}(2)$. Our construction readily generalizes to higher rank Lie algebras, and (deferring the classification of possible global realizations of the gauge group) this provides a uniform perspective for duality defects in other $\mathcal{N} = 4$ SCFTs realized by probe D3-branes. Combining this with the discussion in Appendix \ref{app:orbo}, we anticipate that the same considerations will also apply to the full set of $\mathcal{N} = 4$ SCFTs.

Before proceeding, let us briefly spell out our notational conventions, which essentially follow those of references 
\cite{Aharony:2013hda,Kaidi:2022uux}. All possible global forms are given by $SU(2)_i$ and $SO(3)_{\pm, i}$, $i = 0, 1$. Here $SU(2)$ is the electric polarization where only the Wilson lines with $(z_e, z_m) = (1, 0)$ are present; $SO(3)_+$ is the global form in which only the 't Hooft lines with $(z_e, z_m) = (0, 1)$ are present, and for $SO(3)_-$ only the dyonic lines with $(z_e, z_m) = (1, 1)$ are present. On top of that, we use $i = 0, 1$ to specify the absence or presence of a counterterm $\delta S = -\int \frac{\mathcal{P}(B)}{2}$, where $B$ is the background gauge field for the $\mathbb{Z}_2^{(1)}$ 1-form symmetry.

\subsection{Duality Defects in \texorpdfstring{$\mathfrak{su}(2)$}{} SYM\texorpdfstring{$_{\mathcal{N} = 4}$}{} Theory}

Duality defects for 4D $\mathcal{N}=4$ $\mathfrak{su}(2)$ supersymmetric Yang-Mills theory can be constructed field theoretically via the constructions in \cite{Kaidi:2021xfk} for Kramers-Wannier-like defects and half-space gauging \cite{Choi:2022zal}. We present the string theory construction for both emphasizing differences and similarities between the two approaches.

\subsubsection*{KOZ construction of Kramers-Wannier-like Duality Defects}
Let us first discuss the construction of \cite{Kaidi:2021xfk} for Kramers-Wannier-like duality defects and its string theory realization. We start with a lighting review of the field-theoretic construction and refer the reader to  \cite{Kaidi:2021xfk} for more details.  Consider the $SO(3)_-$ theory, with 1-form symmetry background field $B$ and mixed anomaly
\begin{equation}
    \pi \int_{M_5} A^{(1)}\cup \frac{\mathcal{P}(B)}{2}
\end{equation}
where $A^{(1)}$ is the background field for a $\mathbb{Z}_2$ 0-form symmetry, here S-duality on local operators at $\tau=i$ . Denote by $\mathcal{D}(M_3,B)$ the codimension one topological operator realizing this $\mathbb{Z}_2$ symmetry operator in the presence of a 1-form background $B$. The mixed anomaly implies that
\begin{equation}
    \mathcal{D}(M_3',B)\exp\left( i\pi \int_{\mathbb{\mathbb{H}}'}\frac{\mathcal{P}(B)}{2} \right)
\end{equation}
is invariant under background transformations of $B$. Here, $\mathbb{H}'\subset M_4$ is a half-space of the spacetime $M_4$ with $\partial \mathbb{H}'=M_3'$. Similarly, the minimal abelian TFT $\mathcal{A}^{2,1}$ is an edge mode, and the combination:
\begin{equation}
    \mathcal{A}^{2,1}(M_3,B)\exp\left( i\pi \int_{\mathbb{H}}\frac{\mathcal{P}(B)}{2} \right)
\end{equation}
is also invariant under background transformations of $B$, here $\partial \mathbb{H}=M_3$. We can therefore consider
\begin{equation}\label{eq:Fun}
    \mathcal{A}^{2,1}(M_3,B)\exp\left( i\pi \int_{\mathbb{H}''}\frac{\mathcal{P}(B)}{2} \right) \mathcal{D}(M_3',B)
\end{equation}
with $\partial \mathbb{H}''=M_3-M_3'$ and an invertible codimension one defect in $SO(3)_-$ theory is constructed contracting $\mathbb{H}''$ and setting $M_3=M_3'$. Gauging $B$ to the global form $SU(2)$, this defect becomes non-invertible.

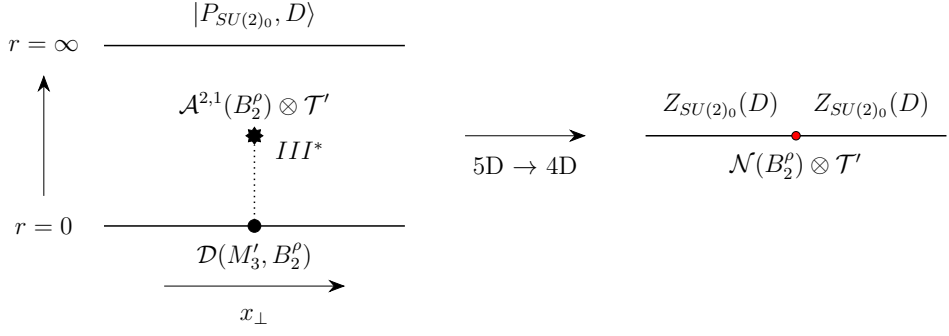
\begin{figure}
    \centering
    \scalebox{0.8}{
    \begin{tikzpicture}
	\begin{pgfonlayer}{nodelayer}
		\node [style=none] (0) at (-2.5, 1.5) {};
		\node [style=none] (1) at (2.5, 1.5) {};
		\node [style=none] (2) at (-2.5, -1.5) {};
		\node [style=none] (3) at (2.5, -1.5) {};
		\node [style=Star] (4) at (0, 0) {};
		\node [style=none] (5) at (0, -1.5) {};
		\node [style=none] (7) at (0.75, -0.25) {$III^*$};
		\node [style=none] (8) at (0, 2) {$\ket{P_{SU(2)_0},D}$};
		\node [style=none] (9) at (-1.5, -2.5) {};
		\node [style=none] (10) at (1.5, -2.5) {};
		\node [style=none] (11) at (0, -3) {$x_\perp$};
		\node [style=Circle] (12) at (0, -1.5) {};
		\node [style=none] (13) at (0, -2) {$\mathcal{D}(M_3',B_2^\rho)$};
		\node [style=none] (14) at (-3.5, -1.5) {$r=0$};
		\node [style=none] (15) at (-3.5, 1.5) {$r=\infty$};
		\node [style=none] (16) at (-3.5, 1) {};
		\node [style=none] (17) at (-3.5, -1) {};
        \node [style=none] (29) at (0, 0.5) {$\mathcal{A}^{2,1}(B_2^\rho)\otimes \mathcal{T}'$};
        \node [style=none] (18) at (3.5, 0) {};
		\node [style=none] (19) at (5.5, 0) {};
		\node [style=none] (20) at (6.5, 0) {};
		\node [style=none] (21) at (11.5, 0) {};
		\node [style=none] (22) at (9, 0) {};
		\node [style=none] (23) at (4.5, -0.5) {5D $\rightarrow$ 4D};
		\node [style=none] (24) at (9, -0.5) {$\mathcal{N}(B_2^\rho)\otimes \mathcal{T}'$};
		\node [style=CircleRed] (25) at (9, 0) {};
		\node [style=none] (26) at (7.75, 0.5) {$Z_{SU(2)_0}(D)$};
		\node [style=none] (28) at (10.25, 0.5) {$Z_{SU(2)_0}(D)$};
	\end{pgfonlayer}
	\begin{pgfonlayer}{edgelayer}
		\draw [style=ThickLine] (0.center) to (1.center);
		\draw [style=ThickLine] (3.center) to (2.center);
		\draw [style=DottedLine] (5.center) to (4);
		\draw [style=ArrowLineRight] (9.center) to (10.center);
		\draw [style=ArrowLineRight] (17.center) to (16.center);
        \draw [style=ThickLine] (20.center) to (21.center);
		\draw [style=ArrowLineRight] (18.center) to (19.center);
	\end{pgfonlayer}
\end{tikzpicture}
    }
    \caption{Insertion of a 7-brane of type $III^*$ into the 5D symmetry TFT of 4D $\mathcal{N}=4$ SYM with vertically running branch cut and global structure $SU(2)_0$ constructs the symmetry operator $\mathcal{N}(M_3,B_2^\rho)=\mathcal{A}^{2,1}(M_3,B_2^\rho)\otimes \mathcal{D}(M_3,B_2^\rho)$ introduced in \cite{Kaidi:2021xfk}. Here tensor product denotes stacking as a consequence of contracting the branch cut. The background $B_2^\rho$ is the background field for the global form $SO(3)_-$. In particular it is not the background field for $SU(2)$ and therefore $\mathcal{N}$ couples dynamically to the theories on both half-spaces.}
    \label{fig:KOZ}
\end{figure}

Now we turn to the string theory realization of the duality defects. We introduce a 7-brane of type $III^*$ wrapped on $M_3\times S^5$ with the branch cut intersecting with D3-branes at $M_3'$, separating the 4D spacetime into two parts. From table \ref{tab:Fibs}, $\tau=i$ is the fixed value of $III^*$ monodromy so we are able to have $\tau=i$ on the full worldvolume of the D3-branes and realize the operator $\mathcal{D}(M_3',B_2^\rho)$ on $M_3'$. The 3D TFT on $M_3$ is determined by the type of 7-brane, which can be read from table \ref{tab:Fibs} for type $III^*$ is $\mathcal{A}^{2,1}[B_2^\rho] \otimes \mathcal{T}'$. The left picture in figure \ref{fig:KOZ} illustrates this construction in terms of the 5D symmetry TFT slab.

Contracting the 5D slab to 4D then stacks $\mathcal{A}^{2,1}(M_{3}, B_{2}^{\rho})$ and $\mathcal{D}(M_{3}^{\prime}, B_{2}^{\rho})$, which gives rise to the 3D duality defect $\mathcal{N}(M_{3},B_2^\rho)$ within the 4D spacetime. Note that the whole construction so far is independent of the choice of the global structure of the theory. In other words, one can choose any boundary condition at $r=\infty$ for the 5D symmetry TFT and then contract the slab. In the case of $SU(2)$ and $SO(3)_+$, the $B_2^\rho$ is a dynamical field in the 4D bulk so $\mathcal{N}(B_2^\rho)$ is a non-invertible defect since it non-trivially couples to the 4D theory. In the case of $SO(3)_-$, $B_2^\rho$ is a background and non-dynamical, therefore $\mathcal{N}(B_2^\rho)$ is invertible. This perfectly matches the result from the field theory perspective and is identified as a non-intrinsic defect in \cite{Kaidi:2022cpf}.

\subsubsection*{Half-space gauging construction}\label{subsec:half-space gauging su(2) duality}
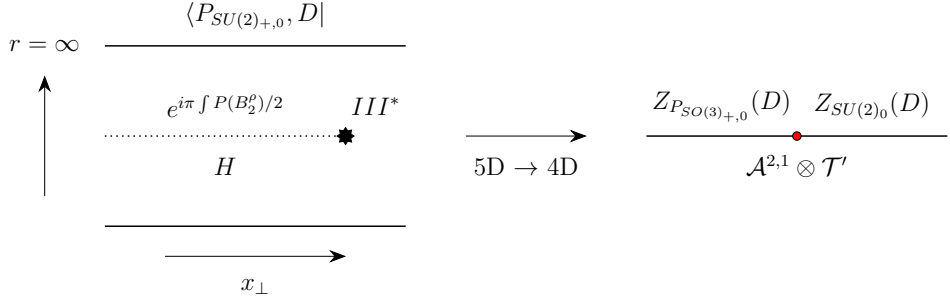
\begin{figure}
    \centering
    \scalebox{0.8}{\begin{tikzpicture}
	\begin{pgfonlayer}{nodelayer}
		\node [style=none] (5) at (-6, 1.5) {};
		\node [style=none] (6) at (-1, 1.5) {};
		\node [style=none] (7) at (-6, -1.5) {};
		\node [style=none] (8) at (-1, -1.5) {};
		\node [style=Star] (10) at (-2, 0) {};
		\node [style=none] (11) at (-6, 0) {};
		\node [style=none] (13) at (-1.5, 0.5) {$III^*$};
		\node [style=none] (18) at (-3.5, 2) {$\bra{P_{SU(2)_{+,0}},D}$};
		\node [style=none] (20) at (-7, 1.5) {$r=\infty$};
		\node [style=none] (21) at (-7, 1) {};
		\node [style=none] (22) at (-7, -1) {};
		\node [style=none] (26) at (-5, -2) {};
		\node [style=none] (27) at (-2, -2) {};
		\node [style=none] (28) at (-3.5, -2.5) {$x_\perp$};
		\node [style=none] (29) at (-4, -0.5) {$H$};
		\node [style=none] (31) at (0, 0) {};
		\node [style=none] (32) at (2, 0) {};
		\node [style=none] (33) at (3, 0) {};
		\node [style=none] (34) at (8, 0) {};
		\node [style=none] (35) at (5.5, 0) {};
		\node [style=none] (36) at (1, -0.5) {5D $\rightarrow$ 4D};
		\node [style=none] (37) at (5.5, -0.5) {$\mathcal{A}^{2,1}\otimes \mathcal{T}'$};
		\node [style=CircleRed] (38) at (5.5, 0) {};
		\node [style=none] (39) at (4.25, 0.5) {$Z_{P_{SO(3)_{+,0}}}(D)$};
		\node [style=none] (40) at (6.75, 0.5) {$Z_{SU(2)_{0}}(D)$};
		\node [style=none] (41) at (-4, 0.5) {$e^{i\pi \int P(B_2^\rho)/2}$};
	\end{pgfonlayer}
	\begin{pgfonlayer}{edgelayer}
		\draw [style=ThickLine] (5.center) to (6.center);
		\draw [style=ThickLine] (8.center) to (7.center);
		\draw [style=DottedLine] (11.center) to (10);
		\draw [style=ArrowLineRight] (22.center) to (21.center);
		\draw [style=ArrowLineRight] (26.center) to (27.center);
		\draw [style=ThickLine] (33.center) to (34.center);
		\draw [style=ArrowLineRight] (31.center) to (32.center);
	\end{pgfonlayer}
\end{tikzpicture}
    }
    \caption{Insertion of a 7-brane of type $III^*$ into the 5d symmetry TFT of 4D $\mathcal{N}=4$ SYM with horizontal running branch cut and topological boundary condition $SU(2)_0$ gives rise to half-space gauging as in \cite{Choi:2022zal}.}
    \label{fig:ShaoEtAl}
\end{figure}

Let us now give the string theoretic setup for the half-space gauging construction in \cite{Choi:2022zal}. For example, we insert a 7-brane of type $\mathfrak{F}=III^*$ into the bulk and with a horizontally running branch cut which funnels the anomaly sourced by this insertion to infinity (see figure \ref{fig:ShaoEtAl}).

\noindent We claim that this realizes half-space gauging. This follows since $B_2^\rho$ is oriented along the polarization of the global form $SO(3)_-$, we therefore have:
\begin{equation}\label{eq:LongComp}\begin{aligned}
&~~~\, \bra{P_{SU(2)_0},D}\exp\left( i\pi \int \mathcal{P}(B_2^\rho)/2 \right)\\  &=  \sum_d \braket{P_{SU(2)_0},D\,|\,P_{SO(3)_{-,0}},d}\bra{P_{SO(3)_{-,0}},d}\exp\left( i\pi \int \mathcal{P}(d)/2 \right)\\
 &=  \sum_d \braket{P_{SU(2)_1},D\,|\,P_{SO(3)_{-,0}},d}\bra{P_{SO(3)_{-,0}},d}\exp\left( i\pi \int \mathcal{P}(D)/2+\mathcal{P}(d)/2 \right)\\
  &=  \sum_d \bra{P_{SO(3)_{-,0}},d}\exp\left( i\pi \int \mathcal{P}(D)/2 +D\cup d+\mathcal{P}(d)/2 \right)\\
&=\sum_d \bra{P_{SO(3)_{-,1}},d}\exp\left( i\pi \int D\cup d\right)\exp\left( i\pi \int \mathcal{P}(D)/2 \right)
  \\
   &= \bra{P_{SO(3)_{+,1}},D}\exp\left( i\pi \int \mathcal{P}(D)/2 \right)\\
    &=\bra{P_{SO(3)_{+,0}},D} \\
\end{aligned}
\end{equation}
as anticipated by noting that the S-duality branch cut interchanges electric and magnetic lines. Here the sums run over $\mathbb{Z}_2$-valued 2-forms $d$ (see figure \ref{fig:ShaoEtAl}). 

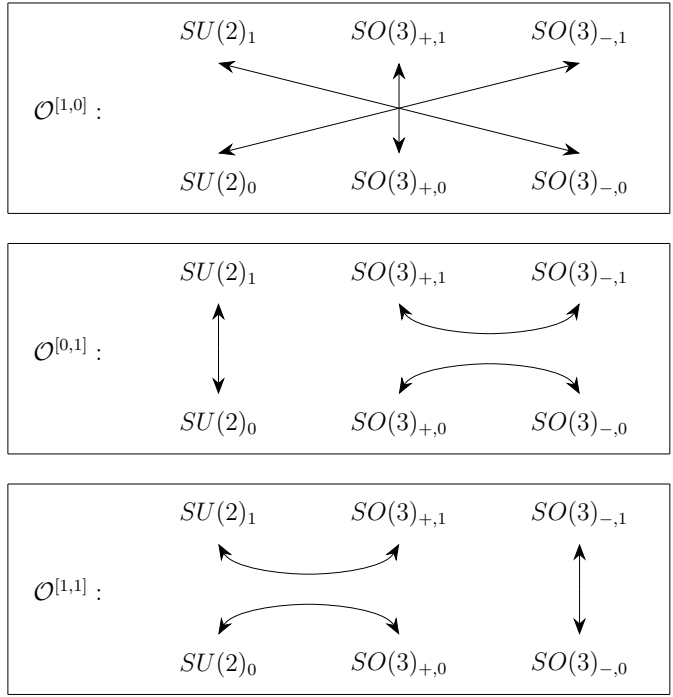
\begin{figure}
    \centering
    \scalebox{0.8}{\begin{tikzpicture}
	\begin{pgfonlayer}{nodelayer}
		\node [style=none] (0) at (-3, 4) {$SU(2)_1$};
		\node [style=none] (1) at (0, 4) {$SO(3)_{+,1}$};
		\node [style=none] (2) at (-3, 1.5) {$SU(2)_0$};
		\node [style=none] (3) at (0, 1.5) {$SO(3)_{+,0}$};
		\node [style=none] (4) at (3, 1.5) {$SO(3)_{-,0}$};
		\node [style=none] (5) at (3, 4) {$SO(3)_{-,1}$};
		\node [style=none] (12) at (-3, 3.5) {};
		\node [style=none] (13) at (0, 3.5) {};
		\node [style=none] (14) at (3, 3.5) {};
		\node [style=none] (15) at (-3, 2) {};
		\node [style=none] (16) at (0, 2) {};
		\node [style=none] (17) at (3, 2) {};
		\node [style=none] (18) at (-3, 0) {$SU(2)_1$};
		\node [style=none] (19) at (0, 0) {$SO(3)_{+,1}$};
		\node [style=none] (20) at (-3, -2.5) {$SU(2)_0$};
		\node [style=none] (21) at (0, -2.5) {$SO(3)_{+,0}$};
		\node [style=none] (22) at (3, -2.5) {$SO(3)_{-,0}$};
		\node [style=none] (23) at (3, 0) {$SO(3)_{-,1}$};
		\node [style=none] (24) at (-3, -0.5) {};
		\node [style=none] (25) at (0, -0.5) {};
		\node [style=none] (26) at (3, -0.5) {};
		\node [style=none] (27) at (-3, -2) {};
		\node [style=none] (28) at (0, -2) {};
		\node [style=none] (29) at (3, -2) {};
		\node [style=none] (30) at (-3, -4) {$SU(2)_1$};
		\node [style=none] (31) at (0, -4) {$SO(3)_{+,1}$};
		\node [style=none] (32) at (-3, -6.5) {$SU(2)_0$};
		\node [style=none] (33) at (0, -6.5) {$SO(3)_{+,0}$};
		\node [style=none] (34) at (3, -6.5) {$SO(3)_{-,0}$};
		\node [style=none] (35) at (3, -4) {$SO(3)_{-,1}$};
		\node [style=none] (38) at (3, -4.5) {};
		\node [style=none] (41) at (3, -6) {};
		\node [style=none] (42) at (-5.5, 2.75) {$\mathcal{O}^{[1,0]}:$};
		\node [style=none] (43) at (-5.5, -1.25) {$\mathcal{O}^{[0,1]}:$};
		\node [style=none] (44) at (-5.5, -5.25) {$\mathcal{O}^{[1,1]}:$};
		\node [style=none] (45) at (0, 2.75) {};
		\node [style=none] (48) at (-3, -1.25) {};
		\node [style=none] (49) at (1.5, -1.5) {};
		\node [style=none] (50) at (3, -5.25) {};
		\node [style=none] (51) at (1.5, -1) {};
		\node [style=none] (58) at (-3, -4.5) {};
		\node [style=none] (59) at (0, -4.5) {};
		\node [style=none] (60) at (-3, -6) {};
		\node [style=none] (61) at (0, -6) {};
		\node [style=none] (62) at (-1.5, -5.5) {};
		\node [style=none] (63) at (-1.5, -5) {};
		\node [style=none] (66) at (4.5, 4.5) {};
		\node [style=none] (67) at (4.5, 1) {};
		\node [style=none] (68) at (-6.5, 4.5) {};
		\node [style=none] (69) at (-6.5, 1) {};
		\node [style=none] (70) at (4.5, 0.5) {};
		\node [style=none] (71) at (4.5, -3) {};
		\node [style=none] (72) at (-6.5, 0.5) {};
		\node [style=none] (73) at (-6.5, -3) {};
		\node [style=none] (74) at (4.5, -3.5) {};
		\node [style=none] (75) at (4.5, -7) {};
		\node [style=none] (76) at (-6.5, -3.5) {};
		\node [style=none] (77) at (-6.5, -7) {};
	\end{pgfonlayer}
	\begin{pgfonlayer}{edgelayer}
		\draw [style=ArrowLineRight] (45.center) to (13.center);
		\draw [style=ArrowLineRight] (45.center) to (16.center);
		\draw [style=ArrowLineRight] (48.center) to (24.center);
		\draw [style=ArrowLineRight] (48.center) to (27.center);
		\draw [style=ArrowLineRight] (50.center) to (38.center);
		\draw [style=ArrowLineRight] (50.center) to (41.center);
		\draw [style=ArrowLineRight] (45.center) to (12.center);
		\draw [style=ArrowLineRight] (45.center) to (17.center);
		\draw [style=ArrowLineRight] (45.center) to (14.center);
		\draw [style=ArrowLineRight] (45.center) to (15.center);
		\draw [style=ArrowLineRight, in=60, out=-180, looseness=0.75] (49.center) to (28.center);
		\draw [style=ArrowLineRight, in=120, out=0, looseness=0.75] (49.center) to (29.center);
		\draw [style=ArrowLineRight, in=-120, out=0, looseness=0.75] (51.center) to (26.center);
		\draw [style=ArrowLineRight, in=-60, out=180, looseness=0.75] (51.center) to (25.center);
		\draw [style=ArrowLineRight, in=60, out=-180, looseness=0.75] (62.center) to (60.center);
		\draw [style=ArrowLineRight, in=120, out=0, looseness=0.75] (62.center) to (61.center);
		\draw [style=ArrowLineRight, in=-120, out=0, looseness=0.75] (63.center) to (59.center);
		\draw [style=ArrowLineRight, in=-60, out=180, looseness=0.75] (63.center) to (58.center);
		\draw (68.center) to (66.center);
		\draw (66.center) to (67.center);
		\draw (68.center) to (69.center);
		\draw (69.center) to (67.center);
		\draw (72.center) to (70.center);
		\draw (70.center) to (71.center);
		\draw (72.center) to (73.center);
		\draw (73.center) to (71.center);
		\draw (76.center) to (74.center);
		\draw (74.center) to (75.center);
		\draw (76.center) to (77.center);
		\draw (77.center) to (75.center);
	\end{pgfonlayer}
\end{tikzpicture}    }
    \caption{Operators $\mathcal{O}^{[q,p]}$ for 4D $\mathcal{N}=4$ $\mathfrak{su}(2)$ gauge theory.}
    \label{fig:Opq}
\end{figure}

Note that the operation of half-space gauging is not realized universally by one type of 7-brane. For example, in the above $\bra{P_{SO(3)_{-,r}}}$ is mapped to $\bra{P_{SO(3)_{-,r+1}}}$, with index $r$ mod 2, by branch cut stacking and the global structure is preserved. 

For this reason, it is instructive to study the other possible operators of type $\mathcal{O}^{[q,p]}$ as introduced in \eqref{eq:Opq}. The full generating set is: 
\begin{equation}\label{eq:definition of su(2) branch cut operators}
    \begin{aligned}
        \mathcal{O}^{[1,0]}&=\exp\left( i\pi \int \mathcal{P}(C_2)/2 \right)\\
        \mathcal{O}^{[0,1]}&=\exp\left( i\pi \int \mathcal{P}(B_2)/2 \right)\\
        \mathcal{O}^{[1,1]}&=\exp\left( i\pi \int \mathcal{P}(B_2+C_2)/2 \right)\\
    \end{aligned}
\end{equation}
and for the case of 7-branes of type $\mathfrak{F}=III^*$ we have $\mathcal{O}^{[1,1]}$ realized on the branch cut. Repeating computations similar to \eqref{eq:LongComp}, we find the results displayed in figure \ref{fig:Opq}. The global forms related by gauging are \cite{Kaidi:2022uux}:
\begin{equation}\label{eq:GlobalFormPair}\begin{aligned}
    SU(2)_0~&\leftrightarrow~ SO(3)_{+,0}\\
    SU(2)_1~&\leftrightarrow~ SO(3)_{-,0}\\
    SO(3)_{+,1}~&\leftrightarrow~ SO(3)_{-,1}
    \end{aligned}
\end{equation}

We can now give the completely general procedure. Pick a pair of global forms in \eqref{eq:GlobalFormPair} to be realized on two half-spaces. Then, determine the operator $\mathcal{O}^{[q,p]}$ connecting these
\begin{equation}\label{eq:GlobalFormPair2}\begin{aligned}
    \mathcal{O}^{[1,1]}\,:&\qquad SU(2)_0~\!\!\!\!\!\!&&\leftrightarrow~ SO(3)_{+,0}\\
    \mathcal{O}^{[1,0]}\,:&\qquad SU(2)_1~\!\!\!\!\!\!&&\leftrightarrow~ SO(3)_{-,0}\\
    \mathcal{O}^{[0,1]}\,:&\qquad SO(3)_{+,1}~\!\!\!\!\!\!&&\leftrightarrow~ SO(3)_{-,1}
    \end{aligned}
    \end{equation}
Next determine the 7-brane which supports $\mathcal{O}^{[q,p]}$ on their branch cut. These are for example $\mathfrak{F}=III^*,  I_1^{[0,1]},I_1^{[1,0]}$ respectively where the latter two are D7- and $[0,1]$-7-branes. Note that in all cases $B_2^\rho$ is neither the background field for the left nor for the right global form on each half-space. Consequently, the minimal abelian TFT supported on the 7-brane interacts with the degrees of freedom in both half-spaces. This leads to the defect realizing a non-invertible symmetry.

\subsection{Triality Defects in \texorpdfstring{$\mathfrak{su}(2)$}{} SYM\texorpdfstring{$_{\mathcal{N} = 4}$}{} Theory}
Let us move to triality defects in the $\mathfrak{su}(2)$ theory. We will see that in this case, only the half-space gauging construction works, which aligns with the fact that triality defects for $\mathfrak{su}(2)$ are intrinsic.

\subsubsection*{KOZ construction of Kramers-Wannier-like Triality Defects}

We first consider the construction following \cite{Kaidi:2021xfk} and show how it does not work. In this case the branch cut of the 7-brane is vertical and ends on the worldvolume of the D3-branes. 
Therefore, we are restricted in our construction to 7-branes whose monodromy has fixed points at $\tau=\frac{i\pi}{3}$ or $\tau=\frac{2\pi i}{3}$, which are values necessary for triality defects \cite{Choi:2022zal, Kaidi:2022uux}. 

Let us take the type $IV^*$ 7-brane as an example. Consider a similar setup as in Figure \ref{fig:KOZ}, but with the type $IV^*$ instead of $III^*$ 7-brane. Naively, one may expect 3D TFTs $\mathcal{A}^{3,2}(B_2^\rho)$ and $\mathcal{D}(M_3',B_2^\rho)$ living on the two ends of the branch cut, respectively. However, in the case of $\mathfrak{su}(2)$ theory, where $B_2$ and $C_2$ are both $\mathbb{Z}_2$ fields in the bulk, there is in fact no non-trivial $B_2^\rho$ preserved by the type $IV^*$ 7-brane monodromy and hence the 7-brane does not source an anomaly. So the setup with inserted 7-branes reduces to studying the branch cut. This is nicely aligned with our discussion on the mixed anomalies shown in \eqref{eq:mixed anomaly term in the bulk}. For triality defects of $\mathfrak{su}(2)$ theory, the mixed anomaly between the triality symmetry and the 1-form symmetry is trivial.

\subsubsection*{Half-space gauging construction}

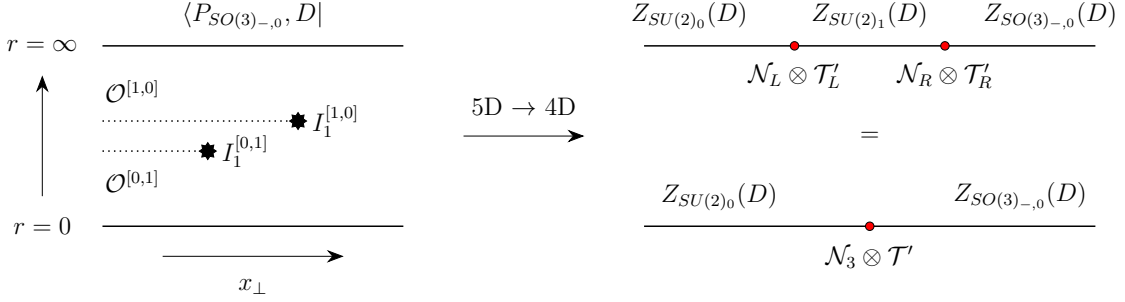
\begin{figure}
    \centering
    \scalebox{0.8}{\begin{tikzpicture}
	\begin{pgfonlayer}{nodelayer}
		\node [style=none] (0) at (-2.5, 1.5) {};
		\node [style=none] (1) at (2.5, 1.5) {};
		\node [style=none] (2) at (-2.5, -1.5) {};
		\node [style=none] (3) at (2.5, -1.5) {};
		\node [style=Star] (4) at (0.75, 0.25) {};
		\node [style=none] (6) at (0, 2) {$\bra{P_{SO(3)_{-,0}},D}$};
		\node [style=none] (8) at (-3.5, -1.5) {$r=0$};
		\node [style=none] (9) at (-3.5, 1.5) {$r=\infty$};
		\node [style=none] (10) at (-3.5, 1) {};
		\node [style=none] (11) at (-3.5, -1) {};
		\node [style=Star] (15) at (-0.75, -0.25) {};
		\node [style=none] (16) at (-2.5, 0.25) {};
		\node [style=none] (17) at (-2.5, -0.25) {};
		\node [style=none] (18) at (1.375, 0.25) {$I_1^{[1,0]}$};
		\node [style=none] (19) at (-0.125, -0.25) {$I_1^{[0,1]}$};
		\node [style=none] (20) at (-2, 0.75) {$\mathcal{O}^{[1,0]}$};
		\node [style=none] (21) at (-2, -0.75) {$\mathcal{O}^{[0,1]}$};
		\node [style=none] (22) at (3.5, 0) {};
		\node [style=none] (23) at (5.5, 0) {};
		\node [style=none] (24) at (6.5, 1.5) {};
		\node [style=none] (25) at (14, 1.5) {};
		\node [style=none] (26) at (10.25, 1.5) {};
		\node [style=none] (27) at (4.5, 0.5) {5D $\rightarrow$ 4D};
		\node [style=none] (28) at (9, 1) {$\mathcal{N}_L\otimes\mathcal{T}'_L$};
		\node [style=CircleRed] (29) at (9, 1.5) {};
		\node [style=none] (30) at (7.25, 2) {$Z_{SU(2)_0}(D)$};
		\node [style=none] (31) at (13.25, 2) {$Z_{SO(3)_{-,0}}(D)$};
		\node [style=CircleRed] (32) at (11.5, 1.5) {};
		\node [style=none] (33) at (10.25, 2) {$Z_{SU(2)_1}(D)$};
		\node [style=none] (34) at (11.5, 1) {$\mathcal{N}_R\otimes\mathcal{T}'_R$};
		\node [style=none] (35) at (10.25, 0) {$=$};
		\node [style=none] (36) at (6.5, -1.5) {};
		\node [style=none] (37) at (14, -1.5) {};
		\node [style=none] (38) at (10.25, -1.5) {};
		\node [style=none] (39) at (10.25, -2) {$\mathcal{N}_3\otimes\mathcal{T}'$};
		\node [style=CircleRed] (40) at (10.25, -1.5) {};
		\node [style=none] (41) at (7.75, -1) {$Z_{SU(2)_0}(D)$};
		\node [style=none] (42) at (12.75, -1) {$Z_{SO(3)_{-,0}}(D)$};
        \node [style=none] (43) at (-1.5, -2) {};
		\node [style=none] (44) at (1.5, -2) {};
		\node [style=none] (45) at (0, -2.5) {$x_\perp$};
	\end{pgfonlayer}
	\begin{pgfonlayer}{edgelayer}
		\draw [style=ThickLine] (0.center) to (1.center);
		\draw [style=ThickLine] (3.center) to (2.center);
		\draw [style=ArrowLineRight] (11.center) to (10.center);
		\draw [style=DottedLine] (16.center) to (4);
		\draw [style=DottedLine] (17.center) to (15);
		\draw [style=ThickLine] (24.center) to (25.center);
		\draw [style=ArrowLineRight] (22.center) to (23.center);
		\draw [style=ThickLine] (36.center) to (37.center);
        \draw [style=ArrowLineRight] (43.center) to (44.center);
	\end{pgfonlayer}
\end{tikzpicture}}
    \caption{Left: sketch of the 5D symmetry TFT for 4D $\mathcal{N}=4$ $\mathfrak{su}(2)$ theory. With two 7-brane insertions of type $\mathfrak{F}=I_1^{[1,0]},I_1^{[0,1]}$ and branch cut operators $\mathcal{O}^{[1,0]},\mathcal{O}^{[0,1]}$ respectively. Right: sketch of the 5D slab contracted to 4D, top and bottom are equivalent. The top, bottom figure shows the 4D theory when the 7-brane insertions are displaced, aligned along $x_\perp$ respectively. The latter results in the triality defect $\mathcal{N}_3$.   }
    \label{fig:Triality}
\end{figure}

All ingredients we need to build triality defects are already introduced in Section \ref{subsec:half-space gauging su(2) duality}. Let us take $SO(3)_{-,0}$ theory as an example. From (\ref{eq:GlobalFormPair2}) we know that $SO(3)_{-,0}$ turns into $SU(2)_1$ under half-space gauging, which can be realized by acting with the operator $\mathcal{O}^{[1,0]}$ living on the horizontal branch cut off a $(0,1)$-7-brane. We can then insert a D7-brane into the bulk, so that the operator $\mathcal{O}^{[0,1]}$ on the horizontal branch cut is introduced. This leads to adding a counterterm for the $SU(2)_1$ theory, so that it becomes the $SU(2)_0$ theory. According to \cite{Aharony:2013hda}, the $SU(2)_0$ theory is indeed dual to $SO(3)_{-,0}$ via the modular transformation $\mathbb{T}\cdot \mathbb{S}$. Therefore by contracting the 5D TFT slab, we indeed get a triality defect $\mathcal{N}_3$.

Furthermore, in this case we are able to determine the lines of the TFT $\mathcal{N}_3$. First, note that the insertion of a single $(p,q)$ 7-brane gives a 3D defect with no lines of its own. See the discussion in subsection \ref{sec:Proposal}. However, when inserting multiple 7-branes, as shown in figure \ref{fig:Triality}, the combined system can contain line defects which are constructed by $(p,q)$ string junctions terminating at the 7-brane insertion and the D3-brane locus. This generalizes the setup displayed in figure \ref{fig:MinimalLines}. These lines constitute the lines of the 3D TFT obtain from the fusion of the TFT supported at the individual 7-brane insertions. In the case of the triality defect we may have Y-shaped string junctions ending on $\mathcal{N}_{L},\mathcal{N}_{R}$ and at $r=0$ and these descend to the lines of the triality defect $\mathcal{N}_3$, the fusion of $\mathcal{N}_{L}$ with $\mathcal{N}_{R}$. 

The line defects of $\mathcal{N}_3$ are therefore determined by the total monodromy $\rho=\rho_L\rho_R$. In the case where $\mathcal{N}_{L},\mathcal{N}_{R}$ are respectively engineered by $I_1^{[0,1]},I_1^{[1,0]}$ type fibers the overall monodromy has trivial cokernel $\textnormal{coker}(\rho-1)$ and therefore there are no additional lines coupling to the fields $B_2,C_2$.

\begin{figure}
    \centering
    \scalebox{0.8}{\begin{tikzpicture}
	\begin{pgfonlayer}{nodelayer}
		\node [style=none] (0) at (-2.5, 1.5) {};
		\node [style=none] (1) at (2.5, 1.5) {};
		\node [style=none] (2) at (-2.5, -1.5) {};
		\node [style=none] (3) at (2.5, -1.5) {};
		\node [style=Star] (4) at (0.75, 0.25) {};
		\node [style=none] (6) at (0, 2) {$\bra{P,D}$};
		\node [style=none] (8) at (-3.5, -1.5) {$r=0$};
		\node [style=none] (9) at (-3.5, 1.5) {$r=\infty$};
		\node [style=none] (10) at (-3.5, 1) {};
		\node [style=none] (11) at (-3.5, -1) {};
		\node [style=Star] (15) at (-0.75, -0.25) {};
		\node [style=none] (16) at (-2.5, 0.25) {};
		\node [style=none] (17) at (-2.5, -0.25) {};
		\node [style=none] (18) at (1.375, 0.25) {$\mathfrak{F}_R$};
		\node [style=none] (19) at (-0.125, -0.25) {$\mathfrak{F}_L$};
		\node [style=none] (20) at (-2, 0.75) {$\mathcal{O}^{\mathfrak{F}_R}$};
		\node [style=none] (21) at (-2, -0.75) {$\mathcal{O}^{\mathfrak{F}_L}$};
		\node [style=none] (22) at (3.5, 0) {};
		\node [style=none] (23) at (5.5, 0) {};
		\node [style=none] (24) at (6.5, 1.5) {};
		\node [style=none] (25) at (14, 1.5) {};
		\node [style=none] (26) at (10.25, 1.5) {};
		\node [style=none] (27) at (4.5, 0.5) {5D $\rightarrow$ 4D};
		\node [style=none] (28) at (9, 1) {$\mathcal{N}_L\otimes\mathcal{T}'_L$};
		\node [style=CircleRed] (29) at (9, 1.5) {};
		\node [style=none] (30) at (7.25, 2) {$Z_{P_L}(D)$};
		\node [style=none] (31) at (13.25, 2) {$Z_{P_R}(D)$};
		\node [style=CircleRed] (32) at (11.5, 1.5) {};
		\node [style=none] (33) at (10.25, 2) {$Z_{P'}(D)$};
		\node [style=none] (34) at (11.5, 1) {$\mathcal{N}_R\otimes\mathcal{T}'_R$};
		\node [style=none] (35) at (10.25, 0) {$=$};
		\node [style=none] (36) at (6.5, -1.5) {};
		\node [style=none] (37) at (14, -1.5) {};
		\node [style=none] (38) at (10.25, -1.5) {};
		\node [style=none] (39) at (10.25, -2) {$\mathcal{N}_3\otimes\mathcal{T}'$};
		\node [style=CircleRed] (40) at (10.25, -1.5) {};
		\node [style=none] (41) at (7.75, -1) {$Z_{P_L}(D)$};
		\node [style=none] (42) at (12.75, -1) {$Z_{P_R}(D)$};
        \node [style=none] (43) at (-1.5, -2) {};
		\node [style=none] (44) at (1.5, -2) {};
		\node [style=none] (45) at (0, -2.5) {$x_\perp$};
	\end{pgfonlayer}
	\begin{pgfonlayer}{edgelayer}
		\draw [style=ThickLine] (0.center) to (1.center);
		\draw [style=ThickLine] (3.center) to (2.center);
		\draw [style=ArrowLineRight] (11.center) to (10.center);
		\draw [style=DottedLine] (16.center) to (4);
		\draw [style=DottedLine] (17.center) to (15);
		\draw [style=ThickLine] (24.center) to (25.center);
		\draw [style=ArrowLineRight] (22.center) to (23.center);
		\draw [style=ThickLine] (36.center) to (37.center);
        \draw [style=ArrowLineRight] (43.center) to (44.center);
	\end{pgfonlayer}
\end{tikzpicture}}
    \caption{Left: Triality defects construction for 4D $\mathcal{N}=4$ $\mathfrak{su}(2)$ theory from 
    the 5D symmetry TFT. With two 7-brane insertions of type $\mathfrak{F}_L,\mathfrak{F}_R$ and branch cut operators $\mathcal{O}^{\mathfrak{F}_L},\mathcal{O}^{\mathfrak{F}_R}$ respectively. }
    \label{fig:Trialityforallsu2}
\end{figure}
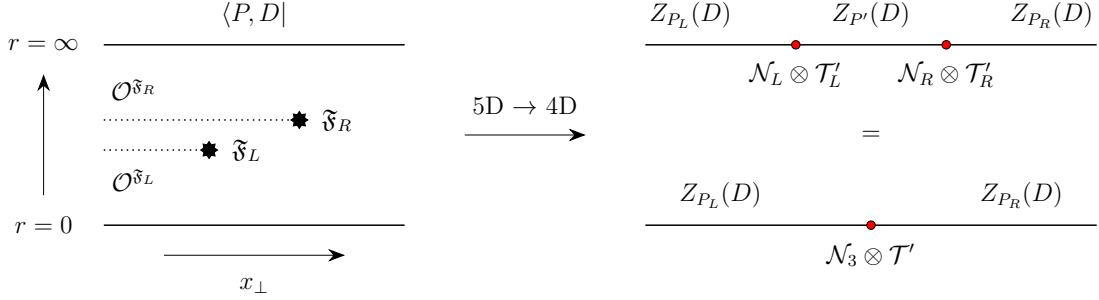

Triality defects for other global structures of $\mathfrak{su}(2)$ can be realized following the same steps. Figure \ref{fig:Trialityforallsu2} illustrates the generic construction, for which we specify ingredients for all cases in the following table.
{\renewcommand{\arraystretch}{1.35}
\begin{table}[H]
\begin{center}
\begin{tabular}{|c|c|c|}
\hline
   $P_{G_m}$  & $\mathfrak{F}_R, \mathfrak{F}_L$ & $Z_{P_L}(D), Z_{P'}(D)$, $Z_{P_R}(D)$  \\
   \hline 
   $SU(2)_0$  & $III^*, I_1^{[1,0]}$ & $Z_{SO(3)_{+,1}}(D), Z_{SO(3)_{+,0}}(D)$, $Z_{SU(2)_0}(D)$  \\
   \hline 
   $SU(2)_1$  & $I_1^{[1,0]},III^*$ & $Z_{SO(3)_{-,1}}(D), Z_{SO(3)_{-,0}}(D)$, $Z_{SU(2)_1}(D)$  \\
   \hline
   $SO(3)_{+,0}$  & $III^*,I_1^{[0,1]}$ & $Z_{SU(2)_1}(D), Z_{SU(2)_0}(D)$, $Z_{SO(3)_{+,0}}(D)$ \\
   \hline
   $SO(3)_{+,1}$  & $I_1^{[0,1]},III^*$ & $Z_{SO(3)_{-,0}}(D), Z_{SO(3)_{-,1}}(D)$, $Z_{SO(3)_{+,1}}(D)$\\
   \hline 
   $SO(3)_{-,0}$  & $I_1^{[1,0]},I_1^{[0,1]}$ & $Z_{SU(2)_{0}}(D), Z_{SU(2)_{1}}(D)$, $Z_{SO(3)_{-,0}}(D)$ \\
   \hline
   $SO(3)_{-,1}$  & $I_1^{[0,1]},I_1^{[1,0]}$ & $Z_{SO(3)_{+,0}}(D), Z_{SO(3)_{+,1}}(D)$, $Z_{SO(3)_{-,1}}(D)$ \\
   \hline
\end{tabular}
\end{center}
\caption{Triality defect constructions for all global structures of $\mathfrak{su}(2)$, with all ingredients illustrated in Figure \ref{fig:Trialityforallsu2}.}
    \label{tab:all triality defects for su(2)}
\end{table}}

\subsection{Defects in the \texorpdfstring{$\mathfrak{su}(N)$}{} Case}
With these examples in hand, we now generalize our discussion to 
construct duality and triality defects for $\mathfrak{su}(N)$ theories with $\mathcal{N} = 4$ supersymmetry. 
In this case, a complete treatment would first require the classification of possible global forms of the theory. 
Rather than proceed in this way, we mainly focus on how things work in the purely electric case, where the gauge group is $SU(N)$, as well as the purely magnetic case, where the gauge group is $SU(N) / \mathbb{Z}_N$.

First, consider the construction of Kramers-Wannier-like duality defects from 7-brane insertions with vertical branch cuts attaching to the D3-brane world volume. For duality/triality defects of $\mathfrak{su}(N)$ theories, if there is a non-trivial $B_2^\rho$ living on the vertical branch cut, one can always choose a topological boundary condition for the 5D TFT, i.e., a global structure of the 4D gauge theory, so that $B_2^\rho$ is the non-dynamical background gauge field. One then ends up with an invertible duality/triality defect $\mathcal{N}(M_3, B_2^\rho)$ in this duality frame, so it is a non-intrinsic duality/triality defect \cite{Kaidi:2022cpf}. On the other hand, if there is no non-trivial $B_2^\rho$ that can be defined on the branch cut, the mixed anomaly is trivial, hence the construction of \cite{Kaidi:2021xfk} does not work. This leads to an intrinsic duality/triality defect \cite{Kaidi:2022cpf} which can only be realized by half-space gauging.

From now on, we focus on providing a universal realization for the half-space gauging construction with $N$ generic.

\subsubsection*{Duality defects}
\begin{figure}
    \centering
    \scalebox{0.8}{\begin{tikzpicture}
	\begin{pgfonlayer}{nodelayer}
		\node [style=none] (0) at (-6.5, 1.5) {};
		\node [style=none] (1) at (-1.5, 1.5) {};
		\node [style=none] (2) at (-6.5, -1.5) {};
		\node [style=none] (3) at (-1.5, -1.5) {};
		\node [style=none] (4) at (-4, 2) {$\ket{P,D}$};
		\node [style=none] (5) at (-7.5, -1.5) {$r=0$};
		\node [style=none] (6) at (-7.5, 1.5) {$r=\infty$};
		\node [style=none] (7) at (-7.5, 1) {};
		\node [style=none] (8) at (-7.5, -1) {};
		\node [style=Star] (12) at (-4, 0) {};
		\node [style=none] (16) at (-6, 0) {};
		\node [style=none] (19) at (-6.5, 0) {};
		\node [style=none] (21) at (-3.25, 0.25) {$III^*$};
		\node [style=none] (22) at (1.5, 1.5) {};
		\node [style=none] (23) at (6.5, 1.5) {};
		\node [style=none] (24) at (1.5, -1.5) {};
		\node [style=none] (25) at (6.5, -1.5) {};
		\node [style=none] (26) at (4, 2) {$\ket{P,D}$};
		\node [style=Star] (34) at (5, 0) {};
		\node [style=none] (35) at (4, 0.5) {};
		\node [style=none] (36) at (4, -0.5) {};
		\node [style=none] (37) at (2.25, -0.5) {};
		\node [style=none] (38) at (2.25, 0) {};
		\node [style=none] (39) at (2.25, 0.5) {};
		\node [style=none] (40) at (1.5, -0.5) {$\mathcal{O}_A^6$};
		\node [style=none] (41) at (1.5, 0) {$\mathcal{O}_B$};
		\node [style=none] (42) at (1.5, 0.5) {$\mathcal{O}_C^2$};
		\node [style=none] (44) at (5.5, 0.5) {$III^*$};
		\node [style=none] (45) at (0, 0) {$=$};
		\node [style=none] (46) at (-4, -2) {};
		\node [style=none] (47) at (4, -2) {};
		\node [style=none] (48) at (-2, -3.5) {};
		\node [style=none] (49) at (2.5, -3.5) {};
		\node [style=none] (50) at (-4, -4.25) {};
		\node [style=none] (51) at (4, -4.25) {};
		\node [style=none] (52) at (0.25, -4.25) {};
		\node [style=CircleRed] (54) at (0, -4.25) {};
		\node [style=none] (55) at (-2, -4.75) {$Z_{P'}(D)=\mathbb{S}^{-1}Z_{P}(D)$};
		\node [style=none] (58) at (2, -4.75) {$Z_{P}(D)$};
		\node [style=none] (59) at (-5.25, -4.25) {$4D:$};
	\end{pgfonlayer}
	\begin{pgfonlayer}{edgelayer}
		\draw [style=ThickLine] (0.center) to (1.center);
		\draw [style=ThickLine] (3.center) to (2.center);
		\draw [style=ArrowLineRight] (8.center) to (7.center);
		\draw [style=DottedLine] (16.center) to (12);
		\draw [style=ThickLine] (22.center) to (23.center);
		\draw [style=ThickLine] (25.center) to (24.center);
		\draw [style=DottedLine] (39.center) to (35.center);
		\draw [style=DottedLine] (37.center) to (36.center);
		\draw [style=DottedLine] (35.center) to (34);
		\draw [style=DottedLine] (36.center) to (34);
		\draw [style=DottedLine] (38.center) to (34);
		\draw [style=ArrowLineRight] (46.center) to (48.center);
		\draw [style=ArrowLineRight] (47.center) to (49.center);
		\draw [style=ThickLine] (50.center) to (51.center);
	\end{pgfonlayer}
\end{tikzpicture}}
    \caption{Type $III^*$ 7-brane with monodromy inverse to S-duality. Left: the monodromy is localized onto a single branch cut. Right: the monodromy is localized onto three distinct branch cuts associated with stacks of $(p,q)$ 7-branes. Bottom: contraction of the 5D slab to 4D. }
    \label{fig:III*resolution}
\end{figure}
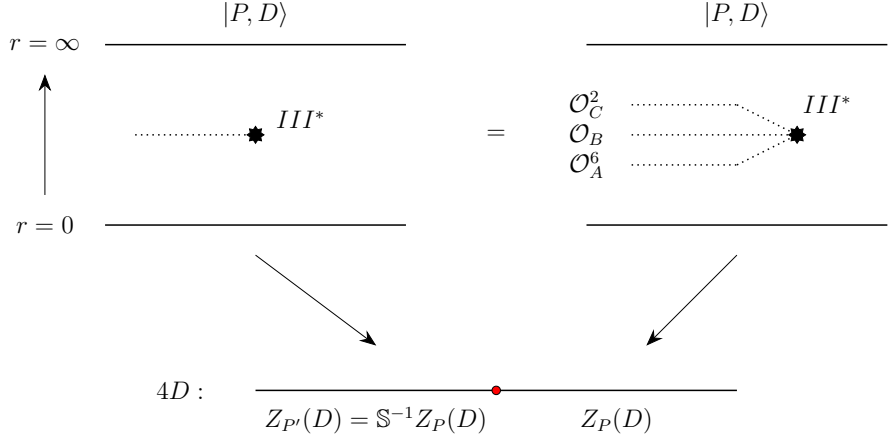

As we already discussed, 7-branes with the horizontal branch cut in the 5D TFT give rise to topological manipulations possibly changing the global structure, and / or adding counter-terms for the 4D theory. In order to build duality defects, we need 7-branes with topological manipulations which can be compensated exactly by that of the modular $\mathbb{S}$ or $\mathbb{S}^{-1}$ transformation. Recall the $\mathbb{S}$ transformation is defined by 
\begin{equation}
    \mathbb{S}: [q,p]\rightarrow [q,p]\begin{pmatrix} 0&1\\-1&0 \end{pmatrix}, \quad \tau \rightarrow -\frac{1}{\tau}
\end{equation}
where $p$ and $q$ are electric and magnetic charges respectively. In our string theory construction, $p$ and $q$ are charges for F1- and D1-strings respectively. Therefore, the candidate 7-branes for duality defects are type $III$ and $III^*$, whose monodromy matrices $\rho$ have same actions on the $[q,p]$ charges as $\mathbb{S}$ and $\mathbb{S}^{-1}$. Therefore, by inserting $III$ or $III^*$ 7-branes and contracting the 5D TFT slab, the 4D theory in the half-space acted by the branch cut is dual  to the original theory in the other half-space via a modular $\mathbb{S}$ or $\mathbb{S}^{-1}$ transformation. This realizes duality defects at $\tau=i$. See the left picture of Figure \ref{fig:III*resolution}.

Let us now discuss topological operators living on the branch cut. From now on, we will focus on the case of the type $III^\ast$ 7-brane, but the following discussion also works for the type $III$ 7-brane similarly. For generic values of $N$, i.e., $B_2$ and $C_2$ both $\mathbb{Z}_N$-valued, it is not always possible to define a $B_2^\rho$ which is invariant under the $III^\ast$ monodromy matrix. So it seems unclear how to build topological operators living on the branch cut. However, it is always possible to factorize the monodromy matrix into elements in $SL(2,\mathbb{Z}_N)$, such that the branch cut is correspondingly separated for each element where a non-trivial $B_2^\rho$ can be defined. In fact, the type $III^\ast$ 7-brane can be constructed by three simple types of $(p,q)$ 7-branes:
\begin{equation}\label{eq:definition ABC brane}
    A:(1,0), ~B:(3,1), ~C:(1,1),
\end{equation}
as 
\begin{equation}
    III^*: A^6BC^2.
\end{equation}
The monodromy matrix for a $(p,q)$ 7-brane is given by 
\begin{equation}
    \rho_{(p,q)}=\begin{pmatrix}1+pq&p^2\\-q^2&1-pq\end{pmatrix},
\end{equation}
under which $pB_2+qC_2$ is obviously always invariant. Therefore, based on our discussion in Section \ref{sec:SETUP}, on the branch cut with monodromy matrix $\rho_{(p,q)}$, we can define a 4D topological operator: 
\begin{equation}
    \mathcal{O}^{(p,q)}=\exp \left( {\frac{2\pi i}{N}\int \frac{\mathcal{P}(pB_2+qC_2)}{2}}\right)\,.
\end{equation}

Now we can separate the branch cut of the $III^*$ 7-brane into multiple ones corresponding to the monodromy matrices for $C^2,B$ and $A^6$, respectively. See the right picture of figure \ref{fig:III*resolution}. Stacking topological operators living on all branch cuts together, we end up with the topological operator for the type $III^\ast$ 7-brane:
\begin{equation}\label{eq:branchcut operator for e7}
    \mathcal{O}^{III^*}=\exp \left( \frac{2\pi i\times 6}{N}\int \frac{\mathcal{P}(B_2)}{2}  \right) \exp \left( \frac{2\pi i}{N}\int \frac{\mathcal{P}(3B_2+C_2)}{2}  \right)\exp \left( \frac{2\pi i\times 2}{N}\int \frac{\mathcal{P}(B_2+C_2)}{2}  \right)
\end{equation}
This is well-defined for $\mathbb{Z}_N$-valued $B_2$ and $C_2$ with $N$ generic. 

For the $\mathfrak{su}(2)$ theory, the first and the third factors in (\ref{eq:branchcut operator for e7}) are both trivial. The second term becomes 
$\exp \left( i\pi \int \mathcal{P}(B_2+C_2)/2  \right)$ because $B_2$ is $\mathbb{Z}_2$-valued. Now $\mathcal{O}^{III^*}$ exactly reduces to the operator $\mathcal{O}^{[1,1]}$ we discussed around (\ref{eq:definition of su(2) branch cut operators}), realizing non-invertible duality defects for $SU(2)$ and $SO(3)_+$. For the $\mathfrak{su}(3)$ theory, the operator reduces to 
\begin{equation}
    \mathcal{O}^{III^*}_{N=3}=\exp \left( \frac{2\pi i}{3}\int \frac{\mathcal{P}(C_2)}{2}  \right)\exp \left( \frac{4\pi i}{3}\int \frac{\mathcal{P}(B_2+C_2)}{2}  \right). 
\end{equation}
Acting with this operator on, for example, the boundary conditions of the 5D TFT which result in the $SU(3)_0$ theory, one can do a  similar computation as in (\ref{eq:LongComp}) and reach the $\overline{PSU(3)}_{0,0}$ theory\footnote{Here we denote global structures with the notation in \cite{Kaidi:2022cpf}, with the two sub-indices for different background gauge fields and counterterms respectively. The overline means an opposite sign for the background field compared to that without an overline. Explicitly, $\overline{PSU(3)}_{0,0}$ means background field purely magnetic $-C_2$ without counterterm.}, which is dual to $SU(3)_0$ theory via the modular $\mathbb{S}$ transformation \cite{Kaidi:2022cpf}. Hence, this realizes non-invertible duality defects. Note the fact that duality defects for $\mathfrak{su}(2)$ and $\mathfrak{su}(3)$ are respectively non-intrinsic and intrinsic, but our construction provides a simple unified realization for them. In fact, our construction works for all values of $N$ regardless of its divisors.

\subsubsection*{Triality defects}
Triality defects from 7-branes can be built following similar steps as in duality defects. In order to find an order 3 topological manipulation, we need to consider 7-branes with monodromy matrices acting on $[q,p]$-string charges which can be compensated by modular transformations  $\mathbb{S}^{m}\cdot \mathbb{T}^{n}$ or $\mathbb{T}^{m}\cdot \mathbb{S}^{n}$, with $m=\pm1, n=\pm1$. The candidate 7-branes with this desired property include the Kodaiar types $II, II^\ast, IV$ and $IV^\ast$.

As in the case of duality defects, we will focus on one certain type of 7-brane and all other candidates work in a similar way. Let us take the $IV^\ast$ 7-brane as an example. In order to  define the topological operator living on its branch cut consistently for all $N$, we separate its branch cut into multiple ones. Each of these branch cuts provides a monodromy as an element in $SL(2,\mathbb{Z})$. In IIB string/F-theory, the $IV^\ast$ 7-brane admits a standard decomposition  
\begin{equation}
    \mathrm{Type} \, IV^\ast : A^5BC^2,
\end{equation}
where $A,B$ and $C$ are elementary $(p,q)$ 7-branes defined in (\ref{eq:definition ABC brane}). Therefore, we can now separate the branch cut of the $IV^*$ 7-brane into those corresponding to the monodromy matrices for $C^2, B$ and $A^5$ 7-branes, respectively. Stacking topological operators living on all branch cuts together, we end up with the operator for the type $IV^\ast$ 7-brane as: 
\begin{equation}\label{eq:branchcut operator for e6}
    \mathcal{O}^{IV^*}=\exp \left( \frac{2\pi i \times 5}{N}\int \frac{\mathcal{P}(B_2)}{2}  \right) \exp \left( \frac{2\pi i }{N}\int \frac{\mathcal{P}(3B_2+C_2)}{2}  \right)\exp \left( \frac{2\pi i \times 2}{N}\int \frac{\mathcal{P}(B_2+C_2)}{2}  \right).
\end{equation}

For the $\mathfrak{su}(2)$ theory, the third factor in (\ref{eq:branchcut operator for e6}) degenerates, so the operator reduces to 
\begin{equation}
    \mathcal{O}^{IV^\ast}_{N=2}=\exp \left( \pi i\int \frac{\mathcal{P}(B_2)}{2}  \right) \exp \left( \pi i \int \frac{\mathcal{P}(B_2+C_2)}{2}  \right),
\end{equation}
which reads $\mathcal{O}^{[0,1]}\cdot \mathcal{O}^{[1,1]}$ under the definition (\ref{eq:definition of su(2) branch cut operators}) for $\mathfrak{su}(2)$. One can easily see from Figure \ref{fig:Opq} that this corresponds to triality defects. For example, starting from $SO(3)_{-,0}$, this topological operator corresponds to first adding a counterterm and then gauging. The resulting theory is $SO(3)_{+,1}$, which is dual to $SO(3)_{-,0}$ via modular $\mathbb{S}\cdot \mathbb{T}$ transformation, thus realizing triality defects. For the $\mathfrak{su}(3)$ theory, the operator for $IV^\ast$ 7-brane reduces to
\begin{equation}
    \mathcal{O}^{IV^\ast}_{N=3}=\exp \left( \frac{4\pi i}{3}\int \mathcal{P}(B_2)/2  \right) \exp \left( \frac{2\pi i}{3}\int \mathcal{P}(C_2)/2  \right)\exp \left( \frac{4\pi i}{3}\int \mathcal{P}(B_2+C_2)/2  \right).
\end{equation}
Acting with this operator on, for example, the $SU(3)_0$ theory, one can do a similar computation as in (\ref{eq:LongComp}) and reach the $\overline{PSU(3)}_{1,0}$ theory, which is dual to $SU(3)_0$ theory via a modular $\mathbb{S}\cdot \mathbb{T}^{-1}$ transformation \cite{Kaidi:2022cpf}, thus realizing non-invertible duality defects. As in the case of duality defects, triality defects for $\mathfrak{su}(2)$ and $\mathfrak{su}(3)$ are of different types, namely intrinsic for $\mathfrak{su}(2)$ and non-intrinsic for $\mathfrak{su}(3)$. However, our construction does not depend on this difference and provides a simple unified realization for them. In fact, our construction works for all values of $N$ regardless of its divisors.

\section{Example: $\mathcal{N} = 1$ SCFTs}\label{sec:N=1}

In this section we show that the considerations of the previous sections readily extend to a broad class of $\mathcal{N} = 1$ SCFTs. We focus on the case of D3-brane probes of an isolated Calabi-Yau threefold singularity. It is well-known that this can be characterized in terms of a quiver gauge theory. In this characterization, the IIB axio-dilaton descends to a specific marginal coupling of the gauge theory. In particular, the IIB $SL(2,\mathbb{Z})$ duality group descends to a specific non-abelian duality group action at this special point in the moduli space. In particular, at the special points in moduli space given by $\tau_{IIB} = i$ and $\tau_{IIB} = \exp(2 \pi i / 6)$ we expect there to be duality / triality defects which act on the resulting $\mathcal{N} = 1$ SCFTs. Of course, at these points of strong coupling the sense in which the weakly coupled Lagrangian description in terms of a quiver gauge theory is actually available is less clear, but the definition of the field theory is still clear. An important feature of this more general class of SCFTs is that we typically have more 0-form symmetries, and we can also consider passing the associated defects charged under these symmetries through the  duality / triality defects.

To begin, let us recall some basic properties of these D3-brane probes of Calabi-Yau threefold singularities.\footnote{There is a vast literature on the subject of D-brane probes of singularities, see e.g., \cite{Douglas:1996sw, Franco:2005rj, Yamazaki:2008bt} for reviews.} In general terms, we get quiver gauge theories with gauge algebra $\mathfrak{su}(N_1)\times \mathfrak{su}(N_2)\times \cdots \times \mathfrak{su}(N_M)$, with $M$ the number of quiver gauge factors. Each gauge group factor has a complexified gauge coupling $\tau_i$ which specifies a marginal coupling of the SCFT. In an SCFT realized by a probe D3-brane, there exists a linear combination of these $\tau_{i}$ which corresponds to the IIB axio-dilaton:
\begin{equation}
\tau_{IIB} = \underset{i}{\sum} n_i\tau_{i},
\end{equation}
where the $n_i$ are positive integers which depend on the details of the geometry. There is a special subspace in the conformal manifold, computed in \cite{Halmagyi:2004ju}, at which the $SL(2,\mathbb{Z})$ duality group action of type IIB strings descends to the SCFT. 


In any case, the IIB $SL(2,\mathbb{Z})$ duality descends to a field-theoretic duality, much as in the $\mathcal{N}=4$ case, and just as there, this also changes the global structure of the theory. The 1-form symmetry depends on the global structure of the theory. For example, with gauge group $SU(N)^M$, the $\mathcal{N}=1$ theory has a purely electric $\mathbb{Z}_N$ 1-form symmetry, which is the diagonal center of all $SU(N)$ groups. The charged Wilson lines can be built, as in the $\mathcal{N}=4$ case, by F1-strings stretched between D3-branes and the asymptotic boundary $\partial X$ of Calabi-Yau threefold $X$.

The construction of duality / triality defects in $\mathcal{N}=4$ theories readily generalizes to $\mathcal{N}=1$ theories. Indeed, our analysis carries over essentially unchanged, with 7-branes wrapped on the boundary $\partial X$ of the Calabi-Yau cone $X$ probed by the stack of D3-branes. In this sense, we automatically implement duality / triality defects in these $\mathcal{N} = 1$ SCFTs just from the top down origin as D3-brane probe theories. 

However, this is not the end of the story. Compared to the $\mathcal{N} = 4$ case where $X = \mathbb{C}^3$, in the $\mathcal{N} = 1$ case, the boundary topology $\partial X$ will in general be more intricate. In particular, dimensional reduction of the 10D supergravity on $\partial X$ will result in a Symmetry TFT with additional fields and interactions terms. These additional fields indicate the presence of more symmetries for the 4D field theory, and these can also be stacked with the duality defect. Here we explain some of the main elements of this construction, focusing on how 0-form symmetry topological operators behave under crossing a duality defect.

\subsection{Symmetry TFTs via IIB String Theory}

We will now present a top down approach to extracting symmetry TFTs for $\mathcal{N}=1$ SCFTs via IIB on a Calabi-Yau threefold $X$. We will reduce the topological term of IIB action on the asymptotic boundary $\partial X$ of the internal manifold $X$, which is a Sasaki-Einstein 5-manifold. This procedure involves treating the various IIB supergravity fluxes as elements in differential cohomology (see e.g., \cite{Freed:2006yc, Freed:2006ya}). Reduction of the linking 5-manifold leads to a symmetry TFT, as explained in  \cite{Apruzzi:2021nmk} (see also \cite{Aharony:1998qu, Heckman:2017uxe}).
Compared with the $\mathcal{N}=4$ case, this reduction leads to additional fields and interaction terms.

More explicitly, consider D3-branes probing a non-compact Calabi-Yau threefold $X$ with an isolated singularity at the tip of the cone. For ease of exposition, we assume that the asymptotic boundary $\partial X$ has the cohomology classes:
\begin{equation}\label{eq:cohomology class of E-S}
    H^*(\partial X)=\{ \mathbb{Z},0,\mathbb{Z}^{b_2}\oplus \text{Tor}H^2(\partial X),\mathbb{Z}^{b_2},\text{Tor}H^4(\partial X),\mathbb{Z} \},
\end{equation}
which covers many cases of interest. In the above,
$b_2$ is the second Betti number. Although not exhaustive, the cases covered by line (\ref{eq:cohomology class of E-S}) includes an infinite family of Calabi-Yau singularities, including $\mathbb{C}^3/\Gamma$ with isolated singularities, as well as complex cones over del Pezzo surfaces. The relevant term in the IIB supergravity action descends from the IIB topological term:\footnote{Here we neglect a subtlety with the 5-form field strength being self-dual. This amounts to a choice of quadratic refinement in an auxiliary 11D spacetime. See \cite{Belov:2006jd, Belov:2006xj, Monnier:2012xd, Heckman:2017uxe}.}
\begin{equation}\label{eq:IIB TFT differential cohomology}
     -\int_{M_4\times X}C_4\wedge dB_2 \wedge dC_2 \rightarrow -\int_{M_4\times  X} \breve{F}_5\star \breve{H}_3\star \breve{G}_3.
\end{equation}
Based on (\ref{eq:cohomology class of E-S}), we can expand $\breve{F}_5, \breve{H}_3$ and $\breve{G}_3$ as 
\begin{equation}
\begin{split}
    \breve{F}_5=&\breve{f}_5\star \breve{1}+\sum_{\alpha=1}^{b_2}\breve{f}_3^{(\alpha)}\star \breve{u}_{2(\alpha)}+\sum_{\alpha=1}^{b_2}\breve{f}_{2(\alpha)}\star \breve{u}_3^{(\alpha)}+N\mathrm{\breve{v}ol}
    +\sum_{i}\breve{E}_3^{(i)}\star \breve{t}_{2(i)}+\sum_{i}\breve{E}_{1(i)}\star \breve{t}_4^{(i)},\\
    \breve{H}_3=&\breve{h}_3\star \breve{1}+\sum_{\alpha=1}^{b_2}\breve{h}_1^{(\alpha)}\star \breve{u}_{2(\alpha)}+\sum_{\alpha=1}^{b_2}\breve{h}_{0(\alpha)}\star \breve{u}_3^{(\alpha)}
    +\sum_{i}\breve{B}_1^{(i)}\star \breve{t}_{2(i)},\\
    \breve{G}_3=&\breve{g}_3\star \breve{1}+\sum_{\alpha=1}^{b_2}\breve{g}_1^{(\alpha)}\star \breve{u}_{2(\alpha)}+\sum_{\alpha=1}^{b_2}\breve{g}_{0(\alpha)}\star \breve{u}_3^{(\alpha)}
    +\sum_{i}\breve{C}_1^{(i)}\star \breve{t}_{2(i)}.
\end{split}
\end{equation}
In the above equations, $\breve{1}, \mathrm{\breve{v}ol}$ and $\breve{u}_{2(\alpha)},\breve{u}_{3}^{(\alpha)}$ correspond to the free part $\mathbb{Z}$ and $\mathbb{Z}^{b_2}$ of cohomology classes, respectively, whereas $\breve{t}_{2(i)}$ and $\breve{t}_4^{(i)}$ correspond to the torsional part $\text{Tor}H^2(\partial X)=\text{Tor}H^4(\partial X)$ respectively and $i$ runs over its generators. The index placement of $\breve{u}_{2(\alpha)}$ compared to $  \breve{u}_3^{(\alpha)}$ (and $\breve{t}_{2(i)}$ compared to $\breve{t}_{4}^{(i)}$) indicate that their star product have non-trivial integral over $\partial X$. 

In particular, since $\partial X$ has infinite volume, any fields arising as coefficients of free cocycles that are dual to free cycles of positive degree should be interpreted as infinitely massive dynamical fields, and thus should be set to zero:
\begin{equation}
  \breve{f}_3^{(\alpha)} = \breve{f}_{2(\alpha)} = \breve{h}_1^{(\alpha)} = \breve{h}_{0(\alpha)} =
    \breve{g}_1^{(\alpha)} = \breve{g}_{0(\alpha)} = 0.
\end{equation}
The IIB topological term (\ref{eq:IIB TFT differential cohomology}) can then be expanded as
\begin{equation}\label{eq:DC expansion of IIB TFT}
\begin{split}
    &-\int_{M_4\times X} \breve{F}_5\star \breve{H}_3\star \breve{G}_3 \\
    &=-\int_{\partial X} \mathrm{\breve{v}ol}\star \breve{1} \star \breve{1} \int_{M_4\times \mathbb{R}_{\ge 0}}N\breve{h}_3\star \breve{g}_3\\
    &-\sum_{i,j,k}\int_{\partial X}\breve{t}_{2(i)}\star \breve{t}_{2(j)}\star \breve{t}_{2(k)}\int_{M_4\times \mathbb{R}_{\ge 0}}\breve{E}_3^{(i)}\star \breve{B}_1^{(j)}\star \breve{C}_1^{(k)}\\
    &-\sum_{i,j}\int_{\partial X}\breve{t}_{2(i)}\star \breve{t}_4^{(j)}\int_{M_4\times \mathbb{R}_{\ge 0}} \breve{E}_{1(j)}\star \left( \breve{B}_1^{(i)}\star \breve{g}_3+\breve{h}_3\star \breve{C}_1^{(i)} \right).
\end{split}
\end{equation}

Carrying out the reduction on $\partial X$ now involves integrating over this space in differential cohomology. 
For non-torsional classes, we find:
\begin{equation}\label{eq:integral over vol and dual forms}
\int_{\partial X}\mathrm{\breve{v}ol}\star \breve{1}\star \breve{1}=1,
\end{equation}
since $\mathrm{\breve{v}ol}$ is the volume form of $\partial X$. Integrals of torsional generators over $\partial X$ are determined by linking numbers which can be derived from intersection numbers between divisors of $X$:
\begin{equation}\label{eq:integral from linking pairing}
\begin{split}
    &\mathcal{C}_{ijk}\equiv \int_{\partial X}\breve{t}_{2(i)}\star \breve{t}_{2(j)}\star \breve{t}_{2(k)},\\
    &\mathcal{C}_i^j\equiv \int_{\partial X}\breve{t}_{2(i)}\star \breve{t}_4^{(j)}.
\end{split}
\end{equation}

The IIB topological term then reduces to the 5D Symmetry TFT:
\begin{equation}\label{eq:TFT for generic N=1}
    \begin{split}
       \mathcal{S}_{5D}&=-\underset{M_4\times \mathbb{R}_{\ge 0}}{\int} \bigg\{ N\breve{h}_3\star \breve{g}_3 - \sum_{i,j,k}\mathcal{C}_{ijk}\breve{E}_3^{(i)}\star \breve{B}_1^{(j)}\star \breve{C}_1^{(k)}-\sum_{i,j}\mathcal{C}_{i}^j\breve{E}_{1(j)}\star \left( \breve{B}_1^{(i)}\star \breve{g}_3+\breve{h}_3\star \breve{C}_1^{(i)} \right) \bigg\}.
    \end{split}
\end{equation}
Here fields denoted by capital letters are from torsional classes and are background fields for discrete symmetries:
\begin{equation}
    \breve{E}_3^{(i)} \leftrightarrow G^{(2)},\quad  \breve{B}_1^{(i)} \leftrightarrow G^{(0)},\quad  \breve{C}_1^{(i)} \leftrightarrow \tilde{G}^{(0)}
\end{equation}
where $G^{(2)}  \cong G^{(0)} \cong \text{Tor} H^2(\partial X) \cong \text{Tor} H^4(\partial X)$. On the other hand, fields denoted by lowercase letters correspond to field strengths of background fields for $U(1)$ symmetries. However, due to the presence of the coefficient $N$ in the first term, they are effectively $\mathbb{Z}_N$ symmetries
\begin{equation}
    \breve{g}_3 \leftrightarrow \mathbb{Z}_{N(m)}^{(1)},\quad  \breve{h}_3 \leftrightarrow \mathbb{Z}_{N(e)}^{(1)},
\end{equation}

The correspondence between fields, global symmetries and charged operators from wrapped branes are presented in table \ref{tab:charged operators of N=1 scfts}.

\begin{table}[t!]
    \centering
    \begin{tabular}{|c|c|c|}
\hline Fields&Global symmetries&Charged operators\\
\hline $\breve{h}_3$&$\mathbb{Z}_N^{(1)}$&F1-strings along $\mathbb{R}_{\geq 0}$, D3-branes wrapping $\mathbb{R}_{\geq 0}\times \sigma_2$\\
\hline $\breve{g}_3$&$\mathbb{Z}_N^{(1)}$&D1-strings along $\mathbb{R}_{\geq 0}$, D3-branes wrapping $\mathbb{R}_{\geq 0}\times \sigma_2$\\
\hline $\breve{E}_3^{(i)}$&$[\text{Tor}H^4(\partial X)]^{(2)}$&D3-branes wrapping $\mathbb{R}_{\geq 0}\times \gamma_1^{(i)}$\\
\hline $\breve{E}_{1(i)}$&$[\text{Tor}H^2(\partial X)]^{(0)}$&D3-branes wrapping $\mathbb{R}_{\geq 0}\times \gamma_{3(i)}$\\
\hline $\breve{B}_{1}^{(i)}$&$[\text{Tor}H^4(\partial X)]^{(0)}$&F1-string wrapping $\mathbb{R}_{\geq 0}\times \gamma_1^{(i)}$\\
\hline $\breve{C}_{1}^{(i)}$&$[\text{Tor}H^4(\partial X)]^{(0)}$&D1-string wrapping $\mathbb{R}_{\geq 0}\times \gamma_1^{(i)}$\\
\hline
\end{tabular}
    \caption{Fields in the 5D TFT and their corresponding global symmetries in 4D SCFTs. The charged operators composed of various types of branes are also presented. $\sigma$ and $\gamma$ denote non-torsional and torsional cycles, respectively. We use an upper index for the 1-cycles and a lower index for the 3-cycles.}
    \label{tab:charged operators of N=1 scfts}
\end{table}

The first term in (\ref{eq:TFT for generic N=1}) is the same as (\ref{eq:5d TFT terms1}), and so in this sense, all of our analysis of the $\mathcal{N} = 4$ SYM case carries over directly to this more general setting. In the $\mathcal{N}=1$ quiver SCFTs with gauge algebra $\mathfrak{su}(N)^K$, $\breve{h}_3$ and $\breve{g}_3$ are field strengths of gauge fields for the diagonal $\mathbb{Z}_N$ electric and magnetic 1-form symmetries, respectively.  The second term encodes the mixed anomaly between a 2-form symmetry and two 0-form symmetries. The last two terms encode mixed anomalies between 1-form symmetries and two 0-form symmetries.

These additional contributions beyond $\breve{h}_3$ and $\breve{g}_3$ are the main distinction from the $\mathcal{N} = 4$ case. We now discuss in further detail their stringy origins as well some of their properties.

\subsection{Discrete 0-form symmetries}
Let us now investigate the discrete 0-form symmetries more carefully. Denote the three classes of 0-form symmetries as $\mathbf{E}_{(i)}, \mathbf{B}^{(i)}$ and $\mathbf{C}^{(i)}$, with respective background fields $\breve{E}_{1(i)},\breve{B}^{1(i)}$ and $\breve{C}^{1(i)}$. As we list in table \ref{tab:charged operators of N=1 scfts}, $\mathbf{B}$ and $\mathbf{C}$ act on charged local operators in the 4D SCFT constructed respectively by F1- and D1-strings wrapping the cone over a torsional one-cycle $\gamma_1^{(i)}$. According to \cite{Heckman:2022muc}, we can build their symmetry operators with the magnetic dual branes, i.e., NS5- and D5-branes wrapping $M_3\times \gamma_{3{(i)}}$ where $M_3\subset M_4$ are 3D symmetry defects inside the 4D spacetime. The symmetry operators are then given by:
\begin{equation}\label{eq:symmetry operators for BC}
\begin{split}
     \mathcal{U}_{\mathbf{B}^{(i)}}=\exp \left(  i \int_{M_3\times \gamma_{3(i)}} B_6+\cdots \right),\\
    \mathcal{U}_{\mathbf{C}^{(i)}}=\exp \left(  i \int_{M_3\times \gamma_{3(i)}} C_6 +\cdots \right),
\end{split}
\end{equation}
where the $\cdots$ indicate additional terms in the respective Wess-Zumino actions. We defer a treatment of these other terms to future work. These two 0-form symmetry operators are related to each other under the IIB $SL(2,\mathbb{Z})$ duality. For example, the leading order terms $B_6$ and $C_6$ transform as:
\begin{equation}\label{eq:sl2z for b6c6}
\left[
    \begin{array}
[c]{c}%
B_{6}\\
C_{6}%
\end{array}
\right]  \mapsto\left[
\begin{array}
[c]{cc}%
a & b\\
c & d
\end{array}
\right]  \left[
\begin{array}
[c]{c}%
B_{6}\\
C_{6}%
\end{array}
\right] ,
\end{equation}
which gives rise to
\begin{equation}\label{eq:s-duality between 0-form symmetries}
    \mathcal{U}_{\mathbf{B}^{(i)}}\rightarrow \mathcal{U}_{\mathbf{B}^{(i)}}^a\mathcal{U}_{\mathbf{C}^{(i)}}^b, ~\mathcal{U}_{\mathbf{C}^{(i)}}\rightarrow \mathcal{U}_{\mathbf{B}^{(i)}}^c\mathcal{U}_{\mathbf{C}^{(i)}}^d.
\end{equation}
The symmetry operator of $\mathbf{E}_{(i)}$ is built by D3-branes wrapping torsional 1-cycles:
\begin{equation}\label{eq:symmetry operator for E}
    \mathcal{U}_{\mathbf{E}_{(i)}}=\exp \left( i \int_{M_3\times \gamma_1^{(i)}}C_4 +\cdots \right),
\end{equation}
which is obviously self-dual under S-duality due to the self-dual property of D3-branes.

We comment that similar transformation rules were worked out in \cite{Gukov:1998kn} in the special case $X = \mathbb{C}^3 / \mathbb{Z}_3$ with boundary topology $S^5 / \mathbb{Z}_3$. See Appendix \ref{app:orbo} for further discussion on this point, and the relation between the present work and this analysis. As noted in \cite{Gukov:1998kn}, the symmetry generators for $\mathbf{B}^{(i)}$ and $\mathbf{C}^{(i)}$ do not quite commute in the presence of a D3-brane wrapped on a torsional 1-cycle. Essentially the same arguments used in this special case carry over to this more general case as well, and we refer the interested reader to \cite{Gukov:1998kn} for further details.

Let us now turn to the interplay of these 0-form symmetries with our duality defects. We have already seen that the $\mathbf{B}^{(i)}$ and $\mathbf{C}^{(i)}$ symmetry generators transform non-trivially under IIB dualities. As such, we expect that when passing the corresponding operators charged under these symmetries through a duality wall that the transformation will be non-trivial.

\begin{figure}[t!]
    \centering
    \scalebox{0.8}{\begin{tikzpicture}
	\begin{pgfonlayer}{nodelayer}
		\node [style=none] (0) at (-6.5, 1.5) {};
		\node [style=none] (1) at (-1.5, 1.5) {};
		\node [style=none] (2) at (-6.5, -1.5) {};
		\node [style=none] (3) at (-1.5, -1.5) {};
		\node [style=Star] (4) at (-4, 0) {};
		\node [style=none] (5) at (-6.5, 0) {};
		\node [style=none] (6) at (-4, 0.5) {7-branes};
		\node [style=none] (7) at (-4, 2) {$\ket{P,D}$};
		\node [style=none] (9) at (-5.5, -2) {};
		\node [style=none] (10) at (-2.5, -2) {};
		\node [style=none] (11) at (-4, -2.5) {$x_\perp$};
		\node [style=none] (14) at (-5.25, 1.5) {};
		\node [style=none] (15) at (-5.25, -1.5) {};
		\node [style=none] (16) at (-5.25, 0) {};
		\node [style=none] (18) at (-6, 0.75) {$[q,p]$};
		\node [style=none] (19) at (-6, -0.75) {$[q,p]\rho$};
		\node [style=none] (20) at (-7.5, -1.5) {$r=0$};
		\node [style=none] (21) at (-7.5, 1.5) {$r=\infty$};
		\node [style=none] (22) at (-7.5, 1) {};
		\node [style=none] (23) at (-7.5, -1) {};
		\node [style=none] (24) at (1.5, 1.5) {};
		\node [style=none] (25) at (6.5, 1.5) {};
		\node [style=none] (26) at (1.5, -1.5) {};
		\node [style=none] (27) at (6.5, -1.5) {};
		\node [style=Star] (28) at (3, 0) {};
		\node [style=none] (29) at (1.5, 0) {};
		\node [style=none] (30) at (3, 0.5) {7-branes};
		\node [style=none] (31) at (4, 2) {$\ket{P,D}$};
		\node [style=none] (33) at (2.5, -2) {};
		\node [style=none] (34) at (5.5, -2) {};
		\node [style=none] (35) at (4, -2.5) {$x_\perp$};
		\node [style=none] (38) at (5.5, 1.5) {};
		\node [style=none] (39) at (5.5, -1.5) {};
		\node [style=none] (40) at (5.5, 0) {};
		\node [style=none] (44) at (6.25, 0.75) {$[q,p]$};
		\node [style=none] (45) at (4.25, -0.5) {$[q,p]\rho$};
		\node [style=none] (46) at (6.75, -0.5) {$[q,p](\rho-1)$};
		\node [style=none] (47) at (0, 0) {$=$};
	\end{pgfonlayer}
	\begin{pgfonlayer}{edgelayer}
		\draw [style=ThickLine] (0.center) to (1.center);
		\draw [style=ThickLine] (3.center) to (2.center);
		\draw [style=DottedLine] (5.center) to (4);
		\draw [style=ArrowLineRight] (9.center) to (10.center);
		\draw [style=RedLine] (14.center) to (16.center);
		\draw [style=BlueLine] (16.center) to (15.center);
		\draw [style=ArrowLineRight] (23.center) to (22.center);
		\draw [style=ThickLine] (24.center) to (25.center);
		\draw [style=ThickLine] (27.center) to (26.center);
		\draw [style=DottedLine] (29.center) to (28);
		\draw [style=ArrowLineRight] (33.center) to (34.center);
		\draw [style=RedLine] (38.center) to (40.center);
		\draw [style=BlueLine] (40.center) to (39.center);
		\draw [style=ThickLine] (40.center) to (28);
	\end{pgfonlayer}
\end{tikzpicture}
}
    \caption{A $[q,p]$ string wrapping the cone of a torsional 1-cycle at $r=\infty$ corresponds to a local operator of the 4D SCFT at $r=0$. Passing the  duality defect $\mathcal{U}(M_3,\mathfrak{F})$ through the local operator from right to left creates a symmetry line operator of charge $[q,p](\rho-1)$ via a Hanany-Witten transition.}
    \label{fig:HW1}
\end{figure}
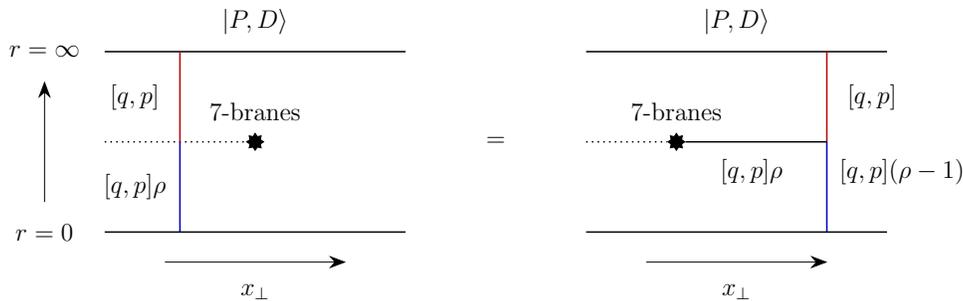

Recall that charged operators for $\mathbf{B}^{(i)}$ and $\mathbf{C}^{(i)}$ are respectively non-compact D3-branes, F1- and D1-strings which stretch along the radial direction of the Calabi-Yau cone. To answer what happens when we pull such a brane through the duality wall, it is enough to track the transformation of the respective branes. The action of duality defects on these charged local operators in the 4D SCFT can be read from the Hanany-Witten transition. For example, insert a 7-brane as the duality defect in 4D theory, and consider a string wrapping Cone$(\gamma_1)$ as a local operator with charged  respectively under $\mathbf{B}^{(i)}$ and $\mathbf{C}^{(i)}$. The Hanany-Witten transition between the 7-brane and the string creates a new string attaching them, which is a 1D symmetry operator in the 4D SCFT. See figure \ref{fig:HW1} for an illustration in the case of half-space gauging construction.

This is very similar to our discussion around Figure \ref{fig:HW} how charged line operators transformed through duality defects, which from the string theory point of view obviously have the same origin as the Hanany-Witten transition. As it is for local operators, this is also analogous to the case in 2D CFTs, e.g. the critical Ising model, where the non-invertible symmetry line maps the local spin operator to the disorder operator which lives at the edge of another symmetry line \cite{Verlinde:1988sn}.

One can in principle also consider passing the topological operators for these 0-form symmetries through a duality / triality defect as well. This case is more subtle since it involves an order of limits issue concerning how fast we send the 7-branes to infinity, and how the 5-branes (by)pass the corresponding branch cuts. It would be interesting to study this issue, but it is beyond the scope of the present work.

\section{Conclusions}
\label{sec:CONC}
In this paper we have presented a top down construction of duality defects for 4D QFTs engineered via D3-brane probes of isolated Calabi-Yau singularities. In this description, the duality group of type IIB strings descends to a duality of the localized QFT at a special point of the conformal manifold, and topological duality defects are implemented by suitable 7-branes wrapped ``at infinity'' which implement field theoretic duality topological symmetry operators. The branch cuts of the 7-branes directly descend to branch cuts in the 5D symmetry TFT which governs the anomalies of the 4D theory. This provides a uniform perspective on various ``bottom up'' approaches to realizing duality / triality defects from the gauging of 1-form symmetries in presence of a mixed anomaly \cite{Kaidi:2021xfk}
and half-space gauging constructions \cite{Choi:2021kmx, Choi:2022zal}. This uniform perspective applies both to $\mathcal{N} = 4$ SYM theory as well as a number of $\mathcal{N} = 1$ quiver gauge theories tuned to a point of strong coupling. In the remainder of this section we discuss a few avenues which would be natural to consider further.

One of the main items in our analysis is the important role of the branch cuts present in 7-branes, and how these dictate the structure of anomalies in the 5D symmetry TFT, as well as the resulting 3D TFTs localized on the duality defects. In Appendix \ref{app:minimalTFT7branes} we take some preliminary steps in reading this data off directly from dimensional reduction of a 7-brane. It would be interesting to perform further checks on this proposal, perhaps by considering dimensional reduction on other ``internal'' manifolds.

In our extension to $\mathcal{N} = 1$ quiver gauge theories, we focused on the special case where the singularity of the Calabi-Yau cone is isolated. This was more to obtain some technical simplifications rather than there being any fundamental obstacle to performing the same computations. It would be interesting to study the structure of the resulting 5D symmetry TFT, as well as the interplay between 7-branes (and their branch cuts) with non-isolated singularities.

It would also be quite interesting to consider other special points in the conformal manifold of 4D SCFTs realized via D3-brane probes of singularities. If these special points admit a non-trivial automorphism, one could hope to similarly lift this automorphism to a geometric object in a string construction, and thus realize a broader class of symmetry operators.

Our primary focus has been on QFTs realized via D3-branes at singularities, but we have also indicated in Appendix \ref{app:other} how these considerations can be generalized to other top down constructions. In particular, it would be interesting to study the structure of duality defects in theories which cannot be obtained from D3-brane probes of singularities.


\section*{Acknowledgements}

We thank M. Del Zotto, S. Franco, H.T. Lam, S.N. Meynet, R. Moscrop,  and S.-H. Shao for helpful discussions.
Some of this work was performed at the workshop, ``Generalized Global Symmetries, Quantum Field Theory, and Geometry''
held in 2022 at the Simons Center for Geometry and Physics, and we thank the center and organizers for kind hospitality.
XY also thanks the UPenn Theory Group for hospitality during part of this work.
The work of JJH and ET is supported by DOE (HEP) Award DE-SC0013528. The work
of MH and HYZ is supported by the Simons Foundation Collaboration grant
\#724069 on ``Special Holonomy in Geometry, Analysis and Physics''. 
XY acknowledges support from the James Arthur Graduate Associate fellowship.

\appendix


\section{Other Realizations of Defects}\label{app:other}

In this paper we have focussed on one particular realization of 4D QFTs via D3-brane probes of 
a transverse geometry. This realization is especially helpful for studying duality defects since the 
corresponding defect is realized in terms of conventional bound states of $(p,q)$ 7-branes wrapped ``at infinity'' in the 
transverse geometry. On the other hand, some structures are more manifest in other duality frames. Additionally, 
there are other ways to engineer 4D QFTs via string constructions as opposed to considering worldvolumes of probe D3-branes.

With this in mind, in this Appendix we discuss how some of the structures considered in this paper are represented in other top down constructions and how this can be used to obtain further generalizations. We begin by discussing how the defect group of $\mathcal{N} = 4 $ SYM is specified in different top down setups. After this, we discuss how duality defects are constructed in some of these alternative constructions.

\subsection{Defect Groups Revisited}

To begin, we recall that the ``defect group'' of a theory is obtained from the spectrum of heavy defects which are not screened by dynamical objects \cite{DelZotto:2015isa} (see also \cite{Tachikawa:2013hya, GarciaEtxebarria:2019caf, Albertini:2020mdx, Morrison:2020ool}). For example, in the context of a 6D SCFT engineered via F-theory on an elliptically fibered Calabi-Yau threefold $X \rightarrow B$ with base $B = \mathbb{C}^2 / \Gamma_{U(2)}$ a quotient by a finite subgroup of $U(2)$ (see \cite{Heckman:2013pva} and \cite{Heckman:2018jxk} for a review), we get extended surface defects (high tension effective strings) from D3-branes wrapped on non-compact 2-cycles of $\mathbb{C}^2 / \Gamma_{U(2)}$. The charge of these defects is screened by D3-branes wrapped on the collapsing cycles of the singularity. The corresponding quotient (as obtained from the relative homology exact sequence for $\mathbb{C}^2 / \Gamma_{U(2)}$ and its boundary geometry $S^3 / \Gamma_{U(2)}  = \partial B$) is:
\begin{equation}
0 \rightarrow H_{2}(B ) \rightarrow H_{2}(B , \partial{B}) \rightarrow H_{1}(\partial B) \rightarrow 0.
\end{equation}
In particular, the defect group for surface defects is simply given by $H_{1}(S^3 / \Gamma_{U(2)}) = \mathrm{Ab}[\Gamma]$, the abelianization of $\Gamma$.

Turning to the case of interest in this paper, observe that for the $\mathcal{N} = (2,0)$ theory obtained from compactification of IIB on the A-type singularity $\mathbb{C}^2 / \mathbb{Z}_{N}$, this leads, upon further reduction on a $T^2$, to the 4D $\mathcal{N} =  4$ SYM theory with Lie algebra $\mathfrak{su}(N)$. The surface defects of the 6D theory can be wrapped on 1-cycles of the $T^2$, and this leads to line defects of the 4D theory. In this case, the defect group for lines is just $\mathbb{Z}_N^{\mathrm{(elec)}} \times \mathbb{Z}^{\mathrm{(mag)}}_N$, and a choice of polarization serves to specify the global form of the gauge group. Similar considerations hold for theories with reduced supersymmetry, as obtained from general 6D SCFTs compactified on Riemann surfaces (see \cite{DelZotto:2015isa} for further discussion).

Returning to the case of D3-branes probing a transverse singularity, this would appear to pose a bit of a puzzle for the ``defect group picture''. To see why, observe that for D3-branes probing $\mathbb{C}^3$, the heavy line defects are obtained from F1- and D1-strings which stretch from the D3-brane out to infinity. On the other hand, the boundary topology of the $\partial \mathbb{C}^3 = S^5$ has no torsional homology.\footnote{That being said, provided one just considers the dimensional reduction to the 5D Symmetry TFT, one can still readily detect the electric and magnetic 1-form symmetries (as we did in this paper).} The physical puzzle, then, is to understand how the defect group is realized in this case.

The answer to this question relies on the fact that the precise notion of ``defect group'' is really specified by the Hilbert space of the string theory background, and quotienting heavy defects by dynamical states of the localized QFT. The general physical point is that in the boundary $S^5$, there is an asymptotic 5-form flux, and in the presence of this flux, the endpoints of the F1- and D1-strings will ``puff up'' to a finite size. The size is tracked by a single integer parameter, and the maximal value of this is precisely $N$.

To see how this flux picture works in more detail, it is helpful to work in a slightly different duality frame. Returning to the realization of $\mathcal{N} = 4$ SYM via type IIB on the background $\mathbb{R}^{3,1} \times T^2 \times \mathbb{C}^2 / \mathbb{Z}_N$, consider T-dualizing the circle fiber of the $\mathbb{C}^2 / \mathbb{Z}_N$ geometry. 
As explained in \cite{Ooguri:1995wj}, we then arrive in a type IIA background with $N$ NS5-branes filling $\mathbb{R}^{3,1} \times T^2$ and sitting at a point of the transverse $\mathbb{C}^2$. There is a dilaton gradient profile in the presence of the NS5-brane, but far away from it, the boundary geometry is simply an $S^3$ threaded by $N$ units of NSNS 3-form flux. In this realization of the 4D QFT, the line operators of interest are obtained from D2-branes which wrap a 1-cycle of the $T^2$ as well as the radial direction of the transverse geometry, ending at ``point'' of the boundary $S^3$ with flux. This appears to have the same puzzle already encountered in the case of the D3-brane realization of the QFT.

Again, the ordinary homology / K-theory for the $S^3$ is not torsional, but the \textit{twisted} K-theory is indeed torsional. Recall that the twisted K-theory for $S^3$ (see \cite{baraglia2015fourier}) involves the choice of a twist class $N \in H^3(S^3) \simeq \mathbb{Z}$, and for this choice of twist class, one gets:
\begin{equation}
K^{\ast}_{H}(S^3) \simeq \mathbb{Z}_{N}, 
\end{equation}
in the obvious notation. From a physical viewpoint, one can also see that the spectrum of ``point-like'' branes in this system is actually more involved. Indeed, the boundary $S^3$ is actually better described as an $SU(2)$ WZW model. Boundary states of the worldsheet CFT correspond to D-branes, and these are in turn characterized by fuzzy points of the geometry.\footnote{We are neglecting the additional boundary states provided by having a supersymmetric WZW model. This leads to additional extended objects / topological symmetry operators. For further discussion on these additional boundary states, see \cite{Maldacena:2001xj, Maldacena:2001ss}.} The interpretation of these boundary conditions can be visualized as ``fuzzy 2-spheres'' (see e.g., \cite{Alekseev:2000fd}), as specified by the non-commutative algebra:
\begin{equation}
[J^{a},J^{b} ] = i \varepsilon^{abc} J^c,
\end{equation}
for $a,b,c = 1,2,3$, namely a representation of $\mathfrak{su}(2)$. The size of the fuzzy 2-sphere is set by $J^{a} J^{a}$, the Casimir of the representation, and this leads to a finite list of admissible representations going from spin $j = 0,...,(N-1)/2$. Beyond this point, the stringy exclusion principle \cite{Maldacena:1998bw} is in operation, and cuts off the size of the fuzzy 2-sphere. The upshot of this is that the single point of ordinary boundary homology has now been supplemented by a whole collection of fuzzy 2-spheres, and these produce the required spectrum of heavy defects which cannot be screened by dynamical objects. Similar considerations clearly apply for topological symmetry operators generated by D4-branes wrapped on a 1-cycle of the $T^2$ and a fuzzy 2-sphere.

With this example in mind, we clearly see that similar considerations will apply in systems where the boundary geometry contains a non-trivial flux. In particular, in the D3-brane realization of $\mathcal{N} = 4$ SYM 
we can expect the F1- and D1-strings used to engineer heavy defects to also ``puff up'' to non-commutative cycles in the boundary $S^5$.

One point we wish to emphasize is that so long as we dimensionally reduce the boundary geometry to reach a lower-dimensional system (as we mainly did in this paper), then the end result of the flux can also be detected directly in the resulting 5D bulk SymTFT.

\subsection{Duality Defects Revisited}

In the previous subsection we presented a general proposal for how to identify the defect group in 
duality frames where asymptotic flux is present. Now, one of the main reasons we chose to focus on the D3-brane 
realization of our QFTs is that the top down identification of duality defects is relatively straightforward (even if the defect group computation is more subtle). Turning the discussion around, one might also ask how our top down duality defects are realized in other string backgrounds which realize the same QFT. For related discussion on this point, see the recent reference \cite{Bashmakov:2022uek}.

To illustrate, consider the IIB background $\mathbb{R}^{3,1} \times T^2 \times \mathbb{C}^2 / \mathbb{Z}_{N}$, 
in which the 4D QFT is realized purely in terms of geometry. In this case, a duality of the 4D field theory will be specified by a large diffeomorphism of the $T^2$, namely as an $SL(2,\mathbb{Z})$ transformation of the complex structure of the $T^2$.

Because the duality symmetry is now encoded purely in the geometry, the ``brane at infinity'' which implements a topological defect / interface will necessarily be a variation in the asymptotic profile of the 10D metric far from the location of the QFT. Since, however, only the topology of the configuration actually matters, it will be enough to specify how this works at the level of a holomorphic Weierstrass model.

Along these lines, we single out one of the directions $x_{\bot}$ along the $\mathbb{R}^{3,1}$ such that the duality defect / interface will be localized at $x_{\bot} = 0$ in the 4D spacetime. Combining this with the radial direction of $\mathbb{C}^2 / \mathbb{Z}_N$, we get a pair of coordinates which locally fill out a patch of the complex line $\mathbb{C}$. It is helpful to introduce the complex combination:
\begin{equation}
z = x_{\bot} + \frac{i}{r},
\end{equation}
where $r = 0$ and $r = \infty$ respectively indicate the location of the QFT and the conformal boundary, where we reach the $S^3 / \mathbb{Z}_{N}$ lens space. In terms of this local coordinate, we can now introduce a Weierstrass model with the prescribed Kodaira fiber type at $z = 0$, namely $x_{\bot} = 0$ and $r = \infty$. 
For example, a type $IV^{\ast}$ and type $III^{\ast}$ fiber would respectively be written as:
\begin{align}
\mathrm{type} \,\, III^{\ast}: \, & y^2 = x^3 + x z^3 \\
\mathrm{type} \,\, IV^{\ast}: \, & y^2 = x^3 + z^4.
\end{align}
This sort of asymptotic profile geometrizes the duality / triality defects we considered.

\section{3D TFTs from 7-branes}\label{app:minimalTFT7branes}

In the main body of this paper we showed how basic structure of 7-branes can account for duality / triality defects in 
QFTs engineered via D3-brane probes of a Calabi-Yau singularity $X$. In particular, we saw that anomaly inflow analyses constrain the resulting 3D TFT of the corresponding duality defect. Of course, given the fact that we are also claiming 
that these topological defects arise from 7-branes, it is natural to ask whether we can directly extract these terms from dimensional reduction of topological terms of the 7-brane. Our aim in this Appendix will be to show to what extent we can derive a 3D TFT living on the duality / triality defect whose 1-form anomalies match that of the appropriate minimal abelian 3D TFTs, $\mathcal{A}^{k,p}$. This is required due to in-flowing the mixed 't Hooft anomaly between the 0-form duality / triality symmetry and the 1-form symmetry of the 4D SCFT, as well as from the line operator linking arguments of Section \ref{ssec:linkingminimaltheory}. While we leave a proper match of these anomalies to future work, this appendix will highlight that, in general, that 3D TFTs on the 7-branes will differ from the minimal 3D TFTs due to the presence of a non-abelian gauge group. Additionally, we propose an 8D WZ term that allows us to determine the level of the 3D CS theory. 

We first review the case of a stack of $n$ D7-branes. The Wess-Zumino (WZ) terms are known from string perturbation theory to be \cite{Douglas:1995bn}\footnote{More generally $(p,q)$ 7-brane WZ topological actions can be inferred from $SL(2,\mathbb{Z})$ transformations of \eqref{eq:WZD7}.}:
\begin{equation}\label{eq:WZD7}
\begin{aligned}
    \mathcal{S}_{\text{WZ}} &= \int_{X_8} \left( \sum_{k}C_{2k} \mathrm{Tr} e^{ \mathcal{F}_2} \sqrt{\frac{\hat{\mathcal{A}}(R_T)}{\hat{\mathcal{A}}(R_N)}} \right)_{\text{8-form}}
\end{aligned}
\end{equation}
where 
\begin{equation}
\mathrm{Tr}\mathcal{F}_2=\mathrm{Tr}( F_2-i^{\ast}B_2)=nF^{U(1)}_2-n i^{\ast} B_2\,,
\end{equation}
with $i^{\ast}B_2$ denoting the pullback from the 10D bulk to the 8D worldvolume $X_8$ of the 7-brane, $F^{U(1)}_2$ is $U(1)$ the gauge curvature associated to factor in the numerator of the 7-brane gauge group $U(n)\simeq (U(1)\times SU(n))/\mathbb{Z}_n$. This precise combination is required because F1-strings can end on D7-branes. In particular, since F1-strings couple to the bulk 2-form, there is a gauge transformation 
$B_2 \rightarrow B_2 + d \lambda_1$ which is cancelled by introducing a compensating $U(1)$ curvature associated with the 
``center of mass'' of the 7-brane. $\mathcal{A}(R_T)$ and $\mathcal{A}(R_N)$ in \eqref{eq:WZD7} are the A-roof genera of the tangent and normal bundles which is given by the expansion
\begin{equation}
    \hat{\mathcal{A}}=1-\frac{1}{24}p_1+\frac{1}{5760}(7p^2_1-4p_2)+...
\end{equation}
where for completeness, we have included $p_i$, the $i^{\mathrm{th}}$ Pontryagin class of the tangent bundle / normal bundle. Since we are concerned with reducing the 7-brane on $S^5$, such contributions play little role in our analysis but could in principle play an important role in more intricate boundary geometries $\partial X$. Taking this into account, the only terms that concern us then are
\begin{equation}\label{eq:wzd7un}
    \int_{M_3\times S^5}\frac{1}{8\pi}C_4\wedge \mathrm{Tr}(F^2)+\frac{2\pi}{2}C_4\wedge \left(\mathrm{Tr}((F_2/2\pi)-B_2) \right)^2
\end{equation}
where we are now being careful with the overall factors of $2\pi$ and considering $\mathrm{Tr}(F)/2\pi$ as integrally quantized. Reducing \eqref{eq:wzd7un} on the $S^5$ surrounding $N$ D3s would then naively produce a level $N$ $U(n)$ 3D Chern-Simons theory living on $M_3$ with an additional coupling to the background $U(1)$ 1-form field $B_2$ that is proportional to $\int_{M_3}\mathrm{Tr}(A)\wedge B_2$. We say ``naive'' because one must first understand how the center-of-mass $U(1)$ of $n$ D7-branes is gapped out via the St\"uckelberg mechanism, i.e., how the gauge algebra reduces from $\mathfrak{u}(n)$ to $\mathfrak{su}(n)$. Indeed, observe that the coupling $C_6 \mathrm{Tr} F_2$ can gap out this $U(1)$ since integrating over $C_6$ produces the constraint\footnote{This is provided that we choose Neumann boundary conditions for $C_6$ along the D7-brane stack.}:
\begin{equation}\label{eq:Stukconstraint}
    \mathrm{Tr}(F_2)=nF^{U(1)}_2=0
\end{equation}
so $F^{U(1)}_2$ still survives as a $\mathbb{Z}_n$-valued 2-form field and is in fact equivalent to the generalized Stiefel-Whitney class \cite{Aharony:2013hda}
\begin{equation}
 F^{U(1)}_2  ~\rightarrow~ w_2\in H^2(X_8,\mathbb{Z}_n)\,.
\end{equation}
One sees this by supposing that there is a magnetic 4-brane monopole in the $U(n)$ gauge theory in the fundamental representation $\mathbf{n}_{+1}$. Then $\frac{1}{n}\int_{S^2}\mathrm{Tr}(F)$ measures a magnetic charge $+1$ with respect to the $U(1)$ in the numerator of $U(n) = (U(1) \times SU(n) ) / \mathbb{Z}_n$, where the $\mathbb{Z}_n$ embeds in the center of $SU(n)$ in the standard fashion. After gapping out the center of mass $U(1)$, we still measure a magnetic flux $1 \; \mathrm{mod}\; n$ around the monopole, which is a defining property of $w_2$. 

Similarly, when considering flat connections of the $U(n)$ gauge theory, we have that $\mathrm{Tr}(F_2)=d\mathrm{Tr}(A)$, and the constraint \eqref{eq:Stukconstraint} implies $n\mathrm{Tr}(A)=0$. Then, the integral $\int_{\gamma_1}\mathrm{Tr}A$ measures a $U(1)\subset U(n)$ monodromy around a 1-cycle $\gamma_1$ and becomes, after decoupling the center-of-mass $U(1)$, the value of $\int_{\gamma_1}w_1$ because this naturally measures the $\mathbb{Z}_n$ monodromy. A subtle distinction we should make is that while $w_2$ and $w_1$ are analogs of $\mathrm{Tr}(F)$ and $\mathrm{Tr}(A)$, the former are cohomology classes while the discrete remnants of the latter are discrete cocycles, i.e. members of $C^i(X_8,\mathbb{Z}_n)$. We therefore name these $\mathcal{J}_2$ and $a$, respectively, which satisfy $[\mathcal{J}_2]=w_2$ and $[a]=w_1$. Moreover, when $w_2=0$, we have that $\mathcal{J}_2=\delta a$ where $\delta$ is the coboundary operator. 

Generalizing to non-perturbative bound states of 7-branes,\footnote{Again, by this we mean 7-branes whose monodromy fixes $\tau$.} with some monodromy matrix $\rho$, we know from the main text that the analog of $B_2$ for the D7 case is generalized to $B^\rho_2$ which takes values in $\mathrm{ker}(\rho-1)$, it is natural then
ask whether there is an analogous discrete remnant of the ``center of mass $U(1)$'' for a general 7-brane 
of type $\mathfrak{F}$, namely the analog of the specific combination $\mathcal{F}_2 = \mathrm{Tr}(F_2 - B_2)$. From the discussion below \eqref{eq:Stukconstraint}, we can already make a reasonable guess that $\mathrm{Tr}(F_2)$ should be replaced by $\mathcal{J}_2\in C^2(M_3\times \partial X_6,\mathrm{ker}(\rho-1))$. In the case of perturbative IIB D7-branes, this involves the specific decomposition $U(n) = (SU(n) \times U(1)) / \mathbb{Z}_n$. In the case of constant axio-dilaton profiles, all of these cases can be obtained from the specific subgroups:\footnote{For example, we have the following subgroups of $E_8$: $(E_7 \times U(1)) / \mathbb{Z}_2$ and $(E_6 \times U(1)^2) / \mathbb{Z}_3$.} $E_8 \supset (G_{\mathfrak{F}} \times U(1)^m) / \mathbb{Z}_{k} $ with $n + m = 8$, where maximal torus of $G_{\mathfrak{F}}$ has dimension $n$, and $G_{\mathfrak{F}}$ has center $\mathbb{Z}_k$. We then see that all of the remarks below \eqref{eq:Stukconstraint} equally apply here if we start with an $E_8$ 7-brane and Higgs down to another 7-brane with constant axio-dilaton. Extending the treatment in \cite{Kapustin:2014gua} for A-type Lie groups, 
we introduce the gauge field:
\begin{equation}
    \mathbf{A}=A+\frac{1}{k}\widehat{a}
\end{equation}
and its field strength $\mathbf{F}=d\mathbf{A}+\mathbf{A}\wedge \mathbf{A}$ where the connections $A,\widehat{a}$ take values in the Lie algebras of $G_{\mathfrak{F}}$ and $U(1)^m$, respectively. As in our D7-brane discussion, the center of mass $U(1)\subset U(1)^m$ is gapped out up to a discrete $\mathbb{Z}_k$-valued gauge field $a$. The curvature on the 7-brane worldvolume therefore takes the form 
\begin{equation}
   \mathbf{F}=  F+\mathcal{J}_2
\end{equation} 
where $F$ is the traceless curvature of $A$ and $\mathcal{J}_2$ is the discrete remnant of the center of mass mode valued in $\Gamma=\mathbb{Z}_k$. Note that $\Gamma$ coincides with the center of the 7-brane gauge group (with electric polarization) which then nicely matches our guess that $\mathcal{J}_2$ should serve as our analog of $\mathrm{Tr}(F)$. Moreover, when we take electric polarization on the brane, $w_2$ is trivial for gauge bundles and therefore we have that $\mathcal{J}_2=\delta a$ with $a\in C^1(M_3\times \partial X_6, \mathrm{ker}(\rho-1))$

We now conjecture a non-perturbative generalization of the WZ terms in \eqref{eq:wzd7un} which applies to all types of 7-brane stacks (labelled by gauge algebra $\mathfrak{g}$) as listed in table \ref{tab:Fibs}:\footnote{A brief comment on the normalization of the instanton density: We have adopted a convention where $\mathrm{Tr} F^2 = \frac{1}{h_{G}^{\vee}} \mathrm{Tr}_{\mathrm{adj}} F^2$, with the latter a trace over the adjoint representation, and $h_{G}^{\vee}$ the dual Coxeter number of the Lie group $G$. Moreover, in our conventions, we have that for a single instanton on a compact four-manifold, $\frac{1}{4} \mathrm{Tr} F^2$ integrates to $1$. }
\begin{equation}\begin{aligned}\label{eq:WZtermsu3}
 \mathcal{S}_{WZ}^{\mathrm{(7)}} &\supset  \int_{M_3\times \partial X_6} \left(  C_4 \wedge \textnormal{tr}\exp \left(\mathbf{F}-B_2^\rho  \right)\right) \\
   &=  \int_{M_3\times \partial X_6} \left(  \frac{1}{8\pi} C_4 \wedge \mathrm{Tr}\,{F}^2 + C_4 \cup \frac{1}{2} (\mathcal{J}_2-B^\rho_2 )^2 \right)
\end{aligned}
\end{equation}
Here we have chosen normalizations in \eqref{eq:WZtermsu3} such that $\exp(i \mathcal{S}_{WZ})$ appears in the 7-brane path integral and $B_2^\rho, \mathcal{J}_2$ are $U(1)$ valued. The first term in \eqref{eq:WZtermsu3} is the D3-brane instanton density term. Namely, it can be obtained by considering a single D3-brane inside a 7-brane and viewing it as a charge-1 instanton which sets the normalization. The second term is a generalization of the term 
\begin{equation}
   \int_{D7}C_4\wedge n B_2\wedge \mathrm{Tr}F_2 = \int_{D7}C_4\wedge n B_2\wedge nF^{U(1)}_2
   \end{equation}
appearing in \eqref{eq:WZD7} by replacing $B_2$ with the more general $B_2^\rho$. The coefficient of $1$ for the second term of \eqref{eq:WZtermsu3} follows from the standard substitutions $B_2\rightarrow \frac{1}{n}B_2$ and $F^{U(1)}_2\rightarrow \frac{1}{n}F^{\mathbb{Z}_n}_2$ when one converts a $U(1)$-valued field to its  remnant field valued in $\mathbb{Z}_n\subset U(1)$.

After reducing on an $S^5$ with flux $\int F_5= N$, \eqref{eq:WZtermsu3} produces a 3D $(G_{\mathfrak{F}})_N$ CS theory\footnote{Subscript denotes the level.} along with a coupling to its electric 1-form symmetry background. In other words, our 3D action becomes:
\begin{equation}\label{eq:3dCS}
  \int_{M_3}   N \cdot CS(A) + 
\frac{N}{2 \pi}a \cup B_2^\rho+ N\cdot CS(a)\\
\end{equation}
where fields are treated as elements in $U(1)$ (suitably restricted): the background $B_2^\rho$ can be normalized to take values in $\mathbb{Z}_{\mathrm{gcd}(k,N)}$ (recall $k$ is the order of the monodromy matrix of the non-perturbative 7-brane), and similar considerations apply for $a$. Note also that a priori, the 3D TFT we get in this way need not match the minimal TFT of type $\mathcal{A}^{K,m}$, since anomaly inflow considerations do not fully fix the form of the 3D TFT. It would be interesting to carry out a complete match with the analysis presented in the main text, but we leave this for future work.

\section{D3-Brane Probe of $\mathbb{C}^3 / \mathbb{Z}_3$} \label{app:orbo}

\begin{figure}[H]
    \centering
    \includegraphics[width=7cm]{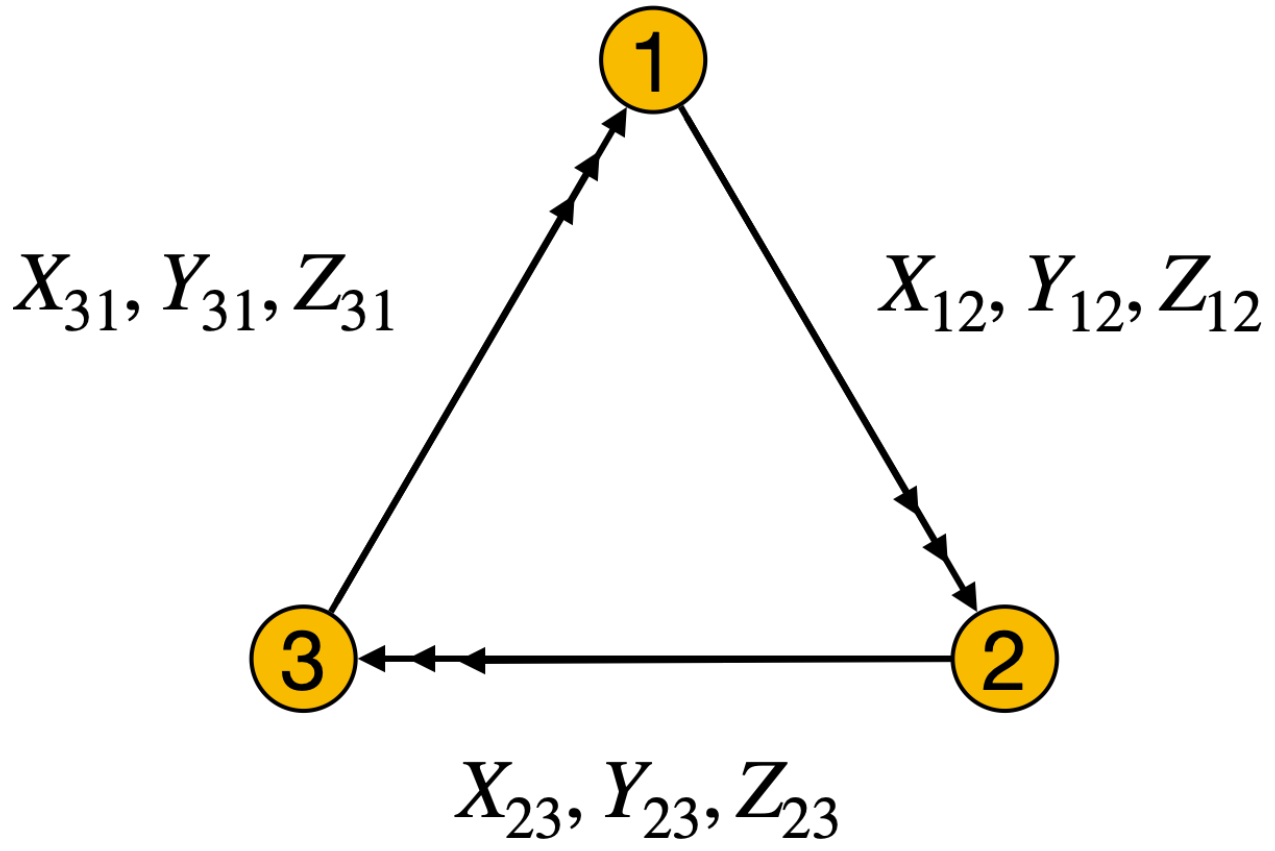}
    \caption{The quiver diagram of $\mathcal{N}=1$ theory on D3-branes probing $\mathbb{C}^3/\mathbb{Z}_3$.}
    \label{fig:quiverc3z3}
\end{figure}

In this Appendix we present further details on the case of a D3-brane probing the orbifold singularity $\mathbb{C}^3 / \mathbb{Z}_3$. The orbifold group action on $\mathbb{C}^3$ is defined by
\begin{equation}
    (z_1,z_2,z_3)\rightarrow (\zeta z_1, \zeta z_2, \zeta z_3),~  \zeta ^3=1.
\end{equation}
Following the procedure in \cite{Douglas:1996sw, Lawrence:1998ja, Kachru:1998ys}, 
the field content of the resulting 4D theory is given in Figure \ref{fig:quiverc3z3}. 
The superpotential of the theory is:
\begin{equation}
    W= \kappa \text{Tr}\left[ X_{12}Y_{23}Z_{31}-X_{12}Z_{23}Y_{31}+X_{23}Y_{31}Z_{12}-X_{23}Z_{31}Y_{12}+X_{31}Y_{12}Z_{23}-X_{31}Z_{12}Y_{23} \right].
\end{equation}

The 5D boundary $\partial \mathbb{C}^3 / \mathbb{Z}_3 = S^5/\mathbb{Z}_3$ 
of the orbifold singularity has the cohomology classes:
\begin{equation}
    H^{\ast}(S^5/\mathbb{Z}_3)=\{ \mathbb{Z},0,\mathbb{Z}_3,0,\mathbb{Z}_3,\mathbb{Z}   \},
\end{equation}
with linking numbers: 
\begin{equation}
    \int_{S^5/\mathbb{Z}_3}t_2\star t_2 \star t_2=\int_{S^5/\mathbb{Z}_3}t_2\star t_4=-\frac{E_0^3}{3\cdot 3\cdot 3}=-\frac{1}{3}.
\end{equation}
$E_0^3$ is the triple self-intersection number of the compact divisor of $O(-3)\rightarrow \mathbb{P}^2$, which can be read from the toric data.\footnote{We refer the reader to \cite{hori2003mirror} for technical details on computing intersection numbers of toric varieties.}

The 5D TFT with the generic form (\ref{eq:TFT for generic N=1}) now reduces to 
\begin{equation}\label{eq:TFT for c3z3}
    \begin{split}
        S_{\text{symTFT}}&=-\int_{M_4\times \mathbb{R}_{\ge 0}} \bigg\{ N\breve{h}_3\star \breve{g}_3 - \frac{1}{3}\breve{E}_3\star \breve{B}_1\star \breve{C}_1-\frac{1}{3}\breve{E}_{1}\star \left( \breve{B}_1\star \breve{g}_3+\breve{h}_3\star \breve{C}_1 \right) \bigg\}.
    \end{split}
\end{equation}
Let us now identify the correspondence between the symmetries in the 4D SCFT and the 5D symmetry TFT fields. The first term in (\ref{eq:TFT for c3z3}) is obvious. It is just the differential cohomology version of the familiar $N\int B_2\wedge dC_2$ term which also appears in the $\mathcal{N}=4$ SYM case. In the $\mathcal{N}=1$ quiver gauge theory, $\breve{h}_3$ (resp. $\breve{g}_3$) corresponds to $B_2^{\text{diag}}$ (resp. $C_2^{\text{diag}}$) for the diagonal $\mathbb{Z}_N$ electric (resp. magnetic) 1-form symmetry of the $\mathfrak{su}(N)^3$ theory.

Based on our previous discussion, we know $E_1, B_1$ and $C_1$ are background gauge fields for $\mathbb{Z}_3$ 0-form symmetries. In fact, as explained in \cite{Gukov:1998kn}, there are indeed three candidate $\mathbb{Z}_3$ symmetries in the 4D SCFT which act on the fields of the quiver gauge theory as follows:
\begin{equation}
\begin{split}
    &\mathbf{B}:(X_{ij},Y_{ij},Z_{ij})\rightarrow (X_{i-1,j-1},Y_{i-1,j-1},Z_{i-1,j-1})\\
    &\mathbf{C}:(X_{ij},Y_{ij},Z_{ij})\rightarrow (\zeta X_{ij}, \zeta^{-1} Y_{ij},Z_{ij}),\\
    &\mathbf{E}:(X_{ij},Y_{ij},Z_{ij})\rightarrow (\zeta^{-2} X_{ij},\zeta Y_{ij},\zeta Z_{ij}).
\end{split}
\end{equation}
where $i$ and $j$ are mod 3 numbers denoting gauge factors. These symmetry generators transform non-trivially under IIB dualities, and their transformations are the same as those already stated in section \ref{sec:N=1}.

\bibliographystyle{utphys}
\bibliography{DualityDefects}

\providecommand{\href}[2]{#2}\begingroup\raggedright\begin{thebibliography}{100}

\bibitem{Gaiotto:2014kfa}
D.~Gaiotto, A.~Kapustin, N.~Seiberg, and B.~Willett, ``{Generalized Global
  Symmetries},'' \href{http://dx.doi.org/10.1007/JHEP02(2015)172}{{\em JHEP}
  {\bfseries 02} (2015) 172}, \href{http://arxiv.org/abs/1412.5148}{{\ttfamily
  arXiv:1412.5148 [hep-th]}}.

\bibitem{Gaiotto:2010be}
D.~Gaiotto, G.~W. Moore, and A.~Neitzke, ``{Framed BPS States},''
  \href{http://dx.doi.org/10.4310/ATMP.2013.v17.n2.a1}{{\em Adv. Theor. Math.
  Phys.} {\bfseries 17} no.~2, (2013) 241--397},
  \href{http://arxiv.org/abs/1006.0146}{{\ttfamily arXiv:1006.0146 [hep-th]}}.

\bibitem{Kapustin:2013qsa}
A.~Kapustin and R.~Thorngren, ``{Topological Field Theory on a Lattice,
  Discrete Theta-Angles and Confinement},''
  \href{http://dx.doi.org/10.4310/ATMP.2014.v18.n5.a4}{{\em Adv. Theor. Math.
  Phys.} {\bfseries 18} no.~5, (2014) 1233--1247},
  \href{http://arxiv.org/abs/1308.2926}{{\ttfamily arXiv:1308.2926 [hep-th]}}.

\bibitem{Kapustin:2013uxa}
A.~Kapustin and R.~Thorngren, ``{Higher symmetry and gapped phases of gauge
  theories},'' \href{http://arxiv.org/abs/1309.4721}{{\ttfamily arXiv:1309.4721
  [hep-th]}}.

\bibitem{Aharony:2013hda}
O.~Aharony, N.~Seiberg, and Y.~Tachikawa, ``{Reading between the lines of
  four-dimensional gauge theories},''
  \href{http://dx.doi.org/10.1007/JHEP08(2013)115}{{\em JHEP} {\bfseries 08}
  (2013) 115}, \href{http://arxiv.org/abs/1305.0318}{{\ttfamily arXiv:1305.0318
  [hep-th]}}.

\bibitem{DelZotto:2015isa}
M.~Del~Zotto, J.~J. Heckman, D.~S. Park, and T.~Rudelius, ``{On the Defect
  Group of a 6D SCFT},''
  \href{http://dx.doi.org/10.1007/s11005-016-0839-5}{{\em Lett. Math. Phys.}
  {\bfseries 106} no.~6, (2016) 765--786},
  \href{http://arxiv.org/abs/1503.04806}{{\ttfamily arXiv:1503.04806
  [hep-th]}}.

\bibitem{Sharpe:2015mja}
E.~Sharpe, ``{Notes on generalized global symmetries in QFT},''
  \href{http://dx.doi.org/10.1002/prop.201500048}{{\em Fortsch. Phys.}
  {\bfseries 63} (2015) 659--682},
  \href{http://arxiv.org/abs/1508.04770}{{\ttfamily arXiv:1508.04770
  [hep-th]}}.

\bibitem{Heckman:2017uxe}
J.~J. Heckman and L.~Tizzano, ``{6D Fractional Quantum Hall Effect},''
  \href{http://dx.doi.org/10.1007/JHEP05(2018)120}{{\em JHEP} {\bfseries 05}
  (2018) 120}, \href{http://arxiv.org/abs/1708.02250}{{\ttfamily
  arXiv:1708.02250 [hep-th]}}.

\bibitem{Tachikawa:2017gyf}
Y.~Tachikawa, ``{On gauging finite subgroups},''
  \href{http://dx.doi.org/10.21468/SciPostPhys.8.1.015}{{\em SciPost Phys.}
  {\bfseries 8} no.~1, (2020) 015},
  \href{http://arxiv.org/abs/1712.09542}{{\ttfamily arXiv:1712.09542
  [hep-th]}}.

\bibitem{Cordova:2018cvg}
C.~C\'ordova, T.~T. Dumitrescu, and K.~Intriligator, ``{Exploring 2-Group
  Global Symmetries},'' \href{http://dx.doi.org/10.1007/JHEP02(2019)184}{{\em
  JHEP} {\bfseries 02} (2019) 184},
  \href{http://arxiv.org/abs/1802.04790}{{\ttfamily arXiv:1802.04790
  [hep-th]}}.

\bibitem{Benini:2018reh}
F.~Benini, C.~C\'ordova, and P.-S. Hsin, ``{On 2-Group Global Symmetries and
  their Anomalies},'' \href{http://dx.doi.org/10.1007/JHEP03(2019)118}{{\em
  JHEP} {\bfseries 03} (2019) 118},
  \href{http://arxiv.org/abs/1803.09336}{{\ttfamily arXiv:1803.09336
  [hep-th]}}.

\bibitem{Hsin:2018vcg}
P.-S. Hsin, H.~T. Lam, and N.~Seiberg, ``{Comments on One-Form Global
  Symmetries and Their Gauging in 3d and 4d},''
  \href{http://dx.doi.org/10.21468/SciPostPhys.6.3.039}{{\em SciPost Phys.}
  {\bfseries 6} no.~3, (2019) 039},
  \href{http://arxiv.org/abs/1812.04716}{{\ttfamily arXiv:1812.04716
  [hep-th]}}.

\bibitem{Wan:2018bns}
Z.~Wan and J.~Wang, ``{Higher anomalies, higher symmetries, and cobordisms I:
  classification of higher-symmetry-protected topological states and their
  boundary fermionic/bosonic anomalies via a generalized cobordism theory},''
  \href{http://dx.doi.org/10.4310/AMSA.2019.v4.n2.a2}{{\em Ann. Math. Sci.
  Appl.} {\bfseries 4} no.~2, (2019) 107--311},
  \href{http://arxiv.org/abs/1812.11967}{{\ttfamily arXiv:1812.11967
  [hep-th]}}.

\bibitem{Thorngren:2019iar}
R.~Thorngren and Y.~Wang, ``{Fusion Category Symmetry I: Anomaly In-Flow and
  Gapped Phases},'' \href{http://arxiv.org/abs/1912.02817}{{\ttfamily
  arXiv:1912.02817 [hep-th]}}.

\bibitem{GarciaEtxebarria:2019caf}
I.~Garcia~Etxebarria, B.~Heidenreich, and D.~Regalado, ``{IIB flux
  non-commutativity and the global structure of field theories},''
  \href{http://dx.doi.org/10.1007/JHEP10(2019)169}{{\em JHEP} {\bfseries 10}
  (2019) 169}, \href{http://arxiv.org/abs/1908.08027}{{\ttfamily
  arXiv:1908.08027 [hep-th]}}.

\bibitem{Eckhard:2019jgg}
J.~Eckhard, H.~Kim, S.~Schafer-Nameki, and B.~Willett, ``{Higher-Form
  Symmetries, Bethe Vacua, and the 3d-3d Correspondence},''
  \href{http://dx.doi.org/10.1007/JHEP01(2020)101}{{\em JHEP} {\bfseries 01}
  (2020) 101}, \href{http://arxiv.org/abs/1910.14086}{{\ttfamily
  arXiv:1910.14086 [hep-th]}}.

\bibitem{Wan:2019soo}
Z.~Wan, J.~Wang, and Y.~Zheng, ``{Higher anomalies, higher symmetries, and
  cobordisms II: Lorentz symmetry extension and enriched bosonic / fermionic
  quantum gauge theory},''
  \href{http://dx.doi.org/10.4310/AMSA.2020.v5.n2.a2}{{\em Ann. Math. Sci.
  Appl.} {\bfseries 05} no.~2, (2020) 171--257},
  \href{http://arxiv.org/abs/1912.13504}{{\ttfamily arXiv:1912.13504
  [hep-th]}}.

\bibitem{Bergman:2020ifi}
O.~Bergman, Y.~Tachikawa, and G.~Zafrir, ``{Generalized symmetries and
  holography in ABJM-type theories},''
  \href{http://dx.doi.org/10.1007/JHEP07(2020)077}{{\em JHEP} {\bfseries 07}
  (2020) 077}, \href{http://arxiv.org/abs/2004.05350}{{\ttfamily
  arXiv:2004.05350 [hep-th]}}.

\bibitem{Morrison:2020ool}
D.~R. Morrison, S.~Schafer-Nameki, and B.~Willett, ``{Higher-Form Symmetries in
  5d},'' \href{http://dx.doi.org/10.1007/JHEP09(2020)024}{{\em JHEP} {\bfseries
  09} (2020) 024}, \href{http://arxiv.org/abs/2005.12296}{{\ttfamily
  arXiv:2005.12296 [hep-th]}}.

\bibitem{Albertini:2020mdx}
F.~Albertini, M.~Del~Zotto, I.~Garcia~Etxebarria, and S.~S. Hosseini, ``{Higher
  Form Symmetries and M-theory},''
  \href{http://dx.doi.org/10.1007/JHEP12(2020)203}{{\em JHEP} {\bfseries 12}
  (2020) 203}, \href{http://arxiv.org/abs/2005.12831}{{\ttfamily
  arXiv:2005.12831 [hep-th]}}.

\bibitem{Hsin:2020nts}
P.-S. Hsin and H.~T. Lam, ``{Discrete theta angles, symmetries and
  anomalies},'' \href{http://dx.doi.org/10.21468/SciPostPhys.10.2.032}{{\em
  SciPost Phys.} {\bfseries 10} no.~2, (2021) 032},
  \href{http://arxiv.org/abs/2007.05915}{{\ttfamily arXiv:2007.05915
  [hep-th]}}.

\bibitem{Bah:2020uev}
I.~Bah, F.~Bonetti, and R.~Minasian, ``{Discrete and higher-form symmetries in
  SCFTs from wrapped M5-branes},''
  \href{http://dx.doi.org/10.1007/JHEP03(2021)196}{{\em JHEP} {\bfseries 03}
  (2021) 196}, \href{http://arxiv.org/abs/2007.15003}{{\ttfamily
  arXiv:2007.15003 [hep-th]}}.

\bibitem{DelZotto:2020esg}
M.~Del~Zotto, I.~Garcia~Etxebarria, and S.~S. Hosseini, ``{Higher form
  symmetries of Argyres-Douglas theories},''
  \href{http://dx.doi.org/10.1007/JHEP10(2020)056}{{\em JHEP} {\bfseries 10}
  (2020) 056}, \href{http://arxiv.org/abs/2007.15603}{{\ttfamily
  arXiv:2007.15603 [hep-th]}}.

\bibitem{Hason:2020yqf}
I.~Hason, Z.~Komargodski, and R.~Thorngren, ``{Anomaly Matching in the Symmetry
  Broken Phase: Domain Walls, CPT, and the Smith Isomorphism},''
  \href{http://dx.doi.org/10.21468/SciPostPhys.8.4.062}{{\em SciPost Phys.}
  {\bfseries 8} no.~4, (2020) 062},
  \href{http://arxiv.org/abs/1910.14039}{{\ttfamily arXiv:1910.14039
  [hep-th]}}.

\bibitem{Bhardwaj:2020phs}
L.~Bhardwaj and S.~Sch\"afer-Nameki, ``{Higher-form symmetries of 6d and 5d
  theories},'' \href{http://dx.doi.org/10.1007/JHEP02(2021)159}{{\em JHEP}
  {\bfseries 02} (2021) 159}, \href{http://arxiv.org/abs/2008.09600}{{\ttfamily
  arXiv:2008.09600 [hep-th]}}.

\bibitem{Apruzzi:2020zot}
F.~Apruzzi, M.~Dierigl, and L.~Lin, ``{The fate of discrete 1-form symmetries
  in 6d},'' \href{http://dx.doi.org/10.21468/SciPostPhys.12.2.047}{{\em SciPost
  Phys.} {\bfseries 12} no.~2, (2022) 047},
  \href{http://arxiv.org/abs/2008.09117}{{\ttfamily arXiv:2008.09117
  [hep-th]}}.

\bibitem{Cordova:2020tij}
C.~Cordova, T.~T. Dumitrescu, and K.~Intriligator, ``{2-Group Global Symmetries
  and Anomalies in Six-Dimensional Quantum Field Theories},''
  \href{http://dx.doi.org/10.1007/JHEP04(2021)252}{{\em JHEP} {\bfseries 04}
  (2021) 252}, \href{http://arxiv.org/abs/2009.00138}{{\ttfamily
  arXiv:2009.00138 [hep-th]}}.

\bibitem{Thorngren:2020aph}
R.~Thorngren, ``{Topological quantum field theory, symmetry breaking, and
  finite gauge theory in 3+1D},''
  \href{http://dx.doi.org/10.1103/PhysRevB.101.245160}{{\em Phys. Rev. B}
  {\bfseries 101} no.~24, (2020) 245160},
  \href{http://arxiv.org/abs/2001.11938}{{\ttfamily arXiv:2001.11938
  [cond-mat.str-el]}}.

\bibitem{DelZotto:2020sop}
M.~Del~Zotto and K.~Ohmori, ``{2-Group Symmetries of 6D Little String Theories
  and T-Duality},'' \href{http://dx.doi.org/10.1007/s00023-021-01018-3}{{\em
  Annales Henri Poincare} {\bfseries 22} no.~7, (2021) 2451--2474},
  \href{http://arxiv.org/abs/2009.03489}{{\ttfamily arXiv:2009.03489
  [hep-th]}}.

\bibitem{BenettiGenolini:2020doj}
P.~Benetti~Genolini and L.~Tizzano, ``{Instantons, symmetries and anomalies in
  five dimensions},'' \href{http://dx.doi.org/10.1007/JHEP04(2021)188}{{\em
  JHEP} {\bfseries 04} (2021) 188},
  \href{http://arxiv.org/abs/2009.07873}{{\ttfamily arXiv:2009.07873
  [hep-th]}}.

\bibitem{Yu:2020twi}
M.~Yu, ``{Symmetries and anomalies of (1+1)d theories: 2-groups and symmetry
  fractionalization},'' \href{http://dx.doi.org/10.1007/JHEP08(2021)061}{{\em
  JHEP} {\bfseries 08} (2021) 061},
  \href{http://arxiv.org/abs/2010.01136}{{\ttfamily arXiv:2010.01136
  [hep-th]}}.

\bibitem{Bhardwaj:2020ymp}
L.~Bhardwaj, Y.~Lee, and Y.~Tachikawa, ``{$SL(2,\mathbb{Z})$ action on QFTs
  with $\mathbb{Z}_2$ symmetry and the Brown-Kervaire invariants},''
  \href{http://dx.doi.org/10.1007/JHEP11(2020)141}{{\em JHEP} {\bfseries 11}
  (2020) 141}, \href{http://arxiv.org/abs/2009.10099}{{\ttfamily
  arXiv:2009.10099 [hep-th]}}.

\bibitem{DeWolfe:2020uzb}
O.~DeWolfe and K.~Higginbotham, ``{Generalized symmetries and 2-groups via
  electromagnetic duality in $AdS/CFT$},''
  \href{http://dx.doi.org/10.1103/PhysRevD.103.026011}{{\em Phys. Rev. D}
  {\bfseries 103} no.~2, (2021) 026011},
  \href{http://arxiv.org/abs/2010.06594}{{\ttfamily arXiv:2010.06594
  [hep-th]}}.

\bibitem{Gukov:2020btk}
S.~Gukov, P.-S. Hsin, and D.~Pei, ``{Generalized global symmetries of $T[M]$
  theories. Part I},'' \href{http://dx.doi.org/10.1007/JHEP04(2021)232}{{\em
  JHEP} {\bfseries 04} (2021) 232},
  \href{http://arxiv.org/abs/2010.15890}{{\ttfamily arXiv:2010.15890
  [hep-th]}}.

\bibitem{Iqbal:2020lrt}
N.~Iqbal and N.~Poovuttikul, ``{2-group global symmetries, hydrodynamics and
  holography},'' \href{http://arxiv.org/abs/2010.00320}{{\ttfamily
  arXiv:2010.00320 [hep-th]}}.

\bibitem{Hidaka:2020izy}
Y.~Hidaka, M.~Nitta, and R.~Yokokura, ``{Global 3-group symmetry and 't Hooft
  anomalies in axion electrodynamics},''
  \href{http://dx.doi.org/10.1007/JHEP01(2021)173}{{\em JHEP} {\bfseries 01}
  (2021) 173}, \href{http://arxiv.org/abs/2009.14368}{{\ttfamily
  arXiv:2009.14368 [hep-th]}}.

\bibitem{Brennan:2020ehu}
T.~D. Brennan and C.~Cordova, ``{Axions, higher-groups, and emergent
  symmetry},'' \href{http://dx.doi.org/10.1007/JHEP02(2022)145}{{\em JHEP}
  {\bfseries 02} (2022) 145}, \href{http://arxiv.org/abs/2011.09600}{{\ttfamily
  arXiv:2011.09600 [hep-th]}}.

\bibitem{Komargodski:2020mxz}
Z.~Komargodski, K.~Ohmori, K.~Roumpedakis, and S.~Seifnashri, ``{Symmetries and
  strings of adjoint QCD$_{2}$},''
  \href{http://dx.doi.org/10.1007/JHEP03(2021)103}{{\em JHEP} {\bfseries 03}
  (2021) 103}, \href{http://arxiv.org/abs/2008.07567}{{\ttfamily
  arXiv:2008.07567 [hep-th]}}.

\bibitem{Closset:2020afy}
C.~Closset, S.~Giacomelli, S.~Schafer-Nameki, and Y.-N. Wang, ``{5d and 4d
  SCFTs: Canonical Singularities, Trinions and S-Dualities},''
  \href{http://dx.doi.org/10.1007/JHEP05(2021)274}{{\em JHEP} {\bfseries 05}
  (2021) 274}, \href{http://arxiv.org/abs/2012.12827}{{\ttfamily
  arXiv:2012.12827 [hep-th]}}.

\bibitem{Thorngren:2020yht}
R.~Thorngren and Y.~Wang, ``{Anomalous symmetries end at the boundary},''
  \href{http://dx.doi.org/10.1007/JHEP09(2021)017}{{\em JHEP} {\bfseries 09}
  (2021) 017}, \href{http://arxiv.org/abs/2012.15861}{{\ttfamily
  arXiv:2012.15861 [hep-th]}}.

\bibitem{Closset:2020scj}
C.~Closset, S.~Schafer-Nameki, and Y.-N. Wang, ``{Coulomb and Higgs Branches
  from Canonical Singularities: Part 0},''
  \href{http://dx.doi.org/10.1007/JHEP02(2021)003}{{\em JHEP} {\bfseries 02}
  (2021) 003}, \href{http://arxiv.org/abs/2007.15600}{{\ttfamily
  arXiv:2007.15600 [hep-th]}}.

\bibitem{Bhardwaj:2021pfz}
L.~Bhardwaj, M.~Hubner, and S.~Schafer-Nameki, ``{1-form Symmetries of 4d
  $\mathcal{N}=2$ Class S Theories},''
  \href{http://dx.doi.org/10.21468/SciPostPhys.11.5.096}{{\em SciPost Phys.}
  {\bfseries 11} (2021) 096}, \href{http://arxiv.org/abs/2102.01693}{{\ttfamily
  arXiv:2102.01693 [hep-th]}}.

\bibitem{Nguyen:2021naa}
M.~Nguyen, Y.~Tanizaki, and M.~\"Unsal, ``{Noninvertible 1-form symmetry and
  Casimir scaling in 2D Yang-Mills theory},''
  \href{http://dx.doi.org/10.1103/PhysRevD.104.065003}{{\em Phys. Rev. D}
  {\bfseries 104} no.~6, (2021) 065003},
  \href{http://arxiv.org/abs/2104.01824}{{\ttfamily arXiv:2104.01824
  [hep-th]}}.

\bibitem{Heidenreich:2021xpr}
B.~Heidenreich, J.~McNamara, M.~Montero, M.~Reece, T.~Rudelius, and
  I.~Valenzuela, ``{Non-invertible global symmetries and completeness of the
  spectrum},'' \href{http://dx.doi.org/10.1007/JHEP09(2021)203}{{\em JHEP}
  {\bfseries 09} (2021) 203}, \href{http://arxiv.org/abs/2104.07036}{{\ttfamily
  arXiv:2104.07036 [hep-th]}}.

\bibitem{Apruzzi:2021phx}
F.~Apruzzi, M.~van Beest, D.~S.~W. Gould, and S.~Sch\"afer-Nameki,
  ``{Holography, 1-form symmetries, and confinement},''
  \href{http://dx.doi.org/10.1103/PhysRevD.104.066005}{{\em Phys. Rev. D}
  {\bfseries 104} no.~6, (2021) 066005},
  \href{http://arxiv.org/abs/2104.12764}{{\ttfamily arXiv:2104.12764
  [hep-th]}}.

\bibitem{Apruzzi:2021vcu}
F.~Apruzzi, L.~Bhardwaj, J.~Oh, and S.~Schafer-Nameki, ``{The Global Form of
  Flavor Symmetries and 2-Group Symmetries in 5d SCFTs},''
  \href{http://arxiv.org/abs/2105.08724}{{\ttfamily arXiv:2105.08724
  [hep-th]}}.

\bibitem{Hosseini:2021ged}
S.~S. Hosseini and R.~Moscrop, ``{Maruyoshi-Song flows and defect groups of $
  {\mathrm{D}}_{\mathrm{p}}^{\mathrm{b}} $(G) theories},''
  \href{http://dx.doi.org/10.1007/JHEP10(2021)119}{{\em JHEP} {\bfseries 10}
  (2021) 119}, \href{http://arxiv.org/abs/2106.03878}{{\ttfamily
  arXiv:2106.03878 [hep-th]}}.

\bibitem{Cvetic:2021sxm}
M.~Cvetic, M.~Dierigl, L.~Lin, and H.~Y. Zhang, ``{Higher-form symmetries and
  their anomalies in M-/F-theory duality},''
  \href{http://dx.doi.org/10.1103/PhysRevD.104.126019}{{\em Phys. Rev. D}
  {\bfseries 104} no.~12, (2021) 126019},
  \href{http://arxiv.org/abs/2106.07654}{{\ttfamily arXiv:2106.07654
  [hep-th]}}.

\bibitem{Buican:2021xhs}
M.~Buican and H.~Jiang, ``{1-form symmetry, isolated $ \mathcal{N} $ = 2 SCFTs,
  and Calabi-Yau threefolds},''
  \href{http://dx.doi.org/10.1007/JHEP12(2021)024}{{\em JHEP} {\bfseries 12}
  (2021) 024}, \href{http://arxiv.org/abs/2106.09807}{{\ttfamily
  arXiv:2106.09807 [hep-th]}}.

\bibitem{Bhardwaj:2021zrt}
L.~Bhardwaj, M.~Hubner, and S.~Schafer-Nameki, ``{Liberating confinement from
  Lagrangians: 1-form symmetries and lines in 4d $\mathcal{N}=1$ from 6d
  $\mathcal{N}=(2,0)$},''
  \href{http://dx.doi.org/10.21468/SciPostPhys.12.1.040}{{\em SciPost Phys.}
  {\bfseries 12} no.~1, (2022) 040},
  \href{http://arxiv.org/abs/2106.10265}{{\ttfamily arXiv:2106.10265
  [hep-th]}}.

\bibitem{Iqbal:2021rkn}
N.~Iqbal and J.~McGreevy, ``{Mean string field theory: Landau-Ginzburg theory
  for 1-form symmetries},'' \href{http://arxiv.org/abs/2106.12610}{{\ttfamily
  arXiv:2106.12610 [hep-th]}}.

\bibitem{Braun:2021sex}
A.~P. Braun, M.~Larfors, and P.-K. Oehlmann, ``{Gauged 2-form symmetries in 6D
  SCFTs coupled to gravity},''
  \href{http://dx.doi.org/10.1007/JHEP12(2021)132}{{\em JHEP} {\bfseries 12}
  (2021) 132}, \href{http://arxiv.org/abs/2106.13198}{{\ttfamily
  arXiv:2106.13198 [hep-th]}}.

\bibitem{Cvetic:2021maf}
M.~Cveti\v{c}, J.~J. Heckman, E.~Torres, and G.~Zoccarato, ``{Reflections on
  the Matter of 3D $\mathcal{N} = 1$ Vacua and Local $Spin(7)$
  Compactifications},''
  \href{http://dx.doi.org/10.1103/PhysRevD.105.026008}{{\em Phys. Rev. D}
  {\bfseries 105} no.~2, (2022) 026008},
  \href{http://arxiv.org/abs/2107.00025}{{\ttfamily arXiv:2107.00025
  [hep-th]}}.

\bibitem{Closset:2021lhd}
C.~Closset and H.~Magureanu, ``{The $U$-plane of rank-one 4d $\mathcal{N}=2$ KK
  theories},'' \href{http://dx.doi.org/10.21468/SciPostPhys.12.2.065}{{\em
  SciPost Phys.} {\bfseries 12} no.~2, (2022) 065},
  \href{http://arxiv.org/abs/2107.03509}{{\ttfamily arXiv:2107.03509
  [hep-th]}}.

\bibitem{Thorngren:2021yso}
R.~Thorngren and Y.~Wang, ``{Fusion Category Symmetry II: Categoriosities at
  $c$ = 1 and Beyond},'' \href{http://arxiv.org/abs/2106.12577}{{\ttfamily
  arXiv:2106.12577 [hep-th]}}.

\bibitem{Sharpe:2021srf}
E.~Sharpe, ``{Topological operators, noninvertible symmetries and
  decomposition},'' \href{http://arxiv.org/abs/2108.13423}{{\ttfamily
  arXiv:2108.13423 [hep-th]}}.

\bibitem{Bhardwaj:2021wif}
L.~Bhardwaj, ``{2-Group symmetries in class S},''
  \href{http://dx.doi.org/10.21468/SciPostPhys.12.5.152}{{\em SciPost Phys.}
  {\bfseries 12} no.~5, (2022) 152},
  \href{http://arxiv.org/abs/2107.06816}{{\ttfamily arXiv:2107.06816
  [hep-th]}}.

\bibitem{Hidaka:2021mml}
Y.~Hidaka, M.~Nitta, and R.~Yokokura, ``{Topological axion electrodynamics and
  4-group symmetry},''
  \href{http://dx.doi.org/10.1016/j.physletb.2021.136762}{{\em Phys. Lett. B}
  {\bfseries 823} (2021) 136762},
  \href{http://arxiv.org/abs/2107.08753}{{\ttfamily arXiv:2107.08753
  [hep-th]}}.

\bibitem{Lee:2021obi}
Y.~Lee and Y.~Zheng, ``{Remarks on compatibility between conformal symmetry and
  continuous higher-form symmetries},''
  \href{http://dx.doi.org/10.1103/PhysRevD.104.085005}{{\em Phys. Rev. D}
  {\bfseries 104} no.~8, (2021) 085005},
  \href{http://arxiv.org/abs/2108.00732}{{\ttfamily arXiv:2108.00732
  [hep-th]}}.

\bibitem{Lee:2021crt}
Y.~Lee, K.~Ohmori, and Y.~Tachikawa, ``{Matching higher symmetries across
  Intriligator-Seiberg duality},''
  \href{http://dx.doi.org/10.1007/JHEP10(2021)114}{{\em JHEP} {\bfseries 10}
  (2021) 114}, \href{http://arxiv.org/abs/2108.05369}{{\ttfamily
  arXiv:2108.05369 [hep-th]}}.

\bibitem{Hidaka:2021kkf}
Y.~Hidaka, M.~Nitta, and R.~Yokokura, ``{Global 4-group symmetry and
  \textquoteright{}t Hooft anomalies in topological axion electrodynamics},''
  \href{http://dx.doi.org/10.1093/ptep/ptab150}{{\em PTEP} {\bfseries 2022}
  no.~4, (2022) 04A109}, \href{http://arxiv.org/abs/2108.12564}{{\ttfamily
  arXiv:2108.12564 [hep-th]}}.

\bibitem{Koide:2021zxj}
M.~Koide, Y.~Nagoya, and S.~Yamaguchi, ``{Non-invertible topological defects in
  4-dimensional $\mathbb {Z}_2$ pure lattice gauge theory},''
  \href{http://dx.doi.org/10.1093/ptep/ptab145}{{\em PTEP} {\bfseries 2022}
  no.~1, (2022) 013B03}, \href{http://arxiv.org/abs/2109.05992}{{\ttfamily
  arXiv:2109.05992 [hep-th]}}.

\bibitem{Apruzzi:2021mlh}
F.~Apruzzi, L.~Bhardwaj, D.~S.~W. Gould, and S.~Schafer-Nameki, ``{2-Group
  symmetries and their classification in 6d},''
  \href{http://dx.doi.org/10.21468/SciPostPhys.12.3.098}{{\em SciPost Phys.}
  {\bfseries 12} no.~3, (2022) 098},
  \href{http://arxiv.org/abs/2110.14647}{{\ttfamily arXiv:2110.14647
  [hep-th]}}.

\bibitem{Kaidi:2021xfk}
J.~Kaidi, K.~Ohmori, and Y.~Zheng, ``{Kramers-Wannier-like duality defects in
  $(3+1)$d gauge theories},''
  \href{http://dx.doi.org/10.1103/PhysRevLett.128.111601}{{\em Phys. Rev.
  Lett.} {\bfseries 128} no.~11, (2022) 111601},
  \href{http://arxiv.org/abs/2111.01141}{{\ttfamily arXiv:2111.01141
  [hep-th]}}.

\bibitem{Choi:2021kmx}
Y.~Choi, C.~Cordova, P.-S. Hsin, H.~T. Lam, and S.-H. Shao, ``{Non-Invertible
  Duality Defects in 3+1 Dimensions},''
  \href{http://dx.doi.org/10.1103/PhysRevD.105.125016}{{\em Phys. Rev. D}
  {\bfseries 105} no.~12, (2022) 125016},
  \href{http://arxiv.org/abs/2111.01139}{{\ttfamily arXiv:2111.01139
  [hep-th]}}.

\bibitem{Bah:2021brs}
I.~Bah, F.~Bonetti, E.~Leung, and P.~Weck, ``{M5-branes Probing Flux
  Backgrounds},'' \href{http://arxiv.org/abs/2111.01790}{{\ttfamily
  arXiv:2111.01790 [hep-th]}}.

\bibitem{Gukov:2021swm}
S.~Gukov, D.~Pei, C.~Reid, and A.~Shehper, ``{Symmetries of 2d TQFTs and
  Equivariant Verlinde Formulae for General Groups},''
  \href{http://arxiv.org/abs/2111.08032}{{\ttfamily arXiv:2111.08032
  [hep-th]}}.

\bibitem{Closset:2021lwy}
C.~Closset, S.~Sch\"afer-Nameki, and Y.-N. Wang, ``{Coulomb and Higgs branches
  from canonical singularities. Part I. Hypersurfaces with smooth Calabi-Yau
  resolutions},'' \href{http://dx.doi.org/10.1007/JHEP04(2022)061}{{\em JHEP}
  {\bfseries 04} (2022) 061}, \href{http://arxiv.org/abs/2111.13564}{{\ttfamily
  arXiv:2111.13564 [hep-th]}}.

\bibitem{Yu:2021zmu}
M.~Yu, ``{Gauging Categorical Symmetries in 3d Topological Orders and Bulk
  Reconstruction},'' \href{http://arxiv.org/abs/2111.13697}{{\ttfamily
  arXiv:2111.13697 [hep-th]}}.

\bibitem{Apruzzi:2021nmk}
F.~Apruzzi, F.~Bonetti, I.~Garcia~Etxebarria, S.~S. Hosseini, and
  S.~Schafer-Nameki, ``{Symmetry TFTs from String Theory},''
  \href{http://arxiv.org/abs/2112.02092}{{\ttfamily arXiv:2112.02092
  [hep-th]}}.

\bibitem{Beratto:2021xmn}
E.~Beratto, N.~Mekareeya, and M.~Sacchi, ``{Zero-form and one-form symmetries
  of the ABJ and related theories},''
  \href{http://dx.doi.org/10.1007/JHEP04(2022)126}{{\em JHEP} {\bfseries 04}
  (2022) 126}, \href{http://arxiv.org/abs/2112.09531}{{\ttfamily
  arXiv:2112.09531 [hep-th]}}.

\bibitem{Bhardwaj:2021mzl}
L.~Bhardwaj, S.~Giacomelli, M.~H\"ubner, and S.~Sch\"afer-Nameki, ``{Relative
  Defects in Relative Theories: Trapped Higher-Form Symmetries and Irregular
  Punctures in Class S},'' \href{http://arxiv.org/abs/2201.00018}{{\ttfamily
  arXiv:2201.00018 [hep-th]}}.

\bibitem{Debray:2021vob}
A.~Debray, M.~Dierigl, J.~J. Heckman, and M.~Montero, ``{The anomaly that was
  not meant IIB},'' \href{http://arxiv.org/abs/2107.14227}{{\ttfamily
  arXiv:2107.14227 [hep-th]}}.

\bibitem{Wang:2021vki}
J.~Wang and Y.-Z. You, ``{Gauge Enhanced Quantum Criticality Between Grand
  Unifications: Categorical Higher Symmetry Retraction},''
  \href{http://arxiv.org/abs/2111.10369}{{\ttfamily arXiv:2111.10369
  [hep-th]}}.

\bibitem{Cvetic:2022uuu}
M.~Cveti\v{c}, M.~Dierigl, L.~Lin, and H.~Y. Zhang, ``{All eight- and
  nine-dimensional string vacua from junctions},''
  \href{http://dx.doi.org/10.1103/PhysRevD.106.026007}{{\em Phys. Rev. D}
  {\bfseries 106} no.~2, (2022) 026007},
  \href{http://arxiv.org/abs/2203.03644}{{\ttfamily arXiv:2203.03644
  [hep-th]}}.

\bibitem{DelZotto:2022fnw}
M.~Del~Zotto, J.~J. Heckman, S.~N. Meynet, R.~Moscrop, and H.~Y. Zhang,
  ``{Higher Symmetries of 5d Orbifold SCFTs},''
  \href{http://arxiv.org/abs/2201.08372}{{\ttfamily arXiv:2201.08372
  [hep-th]}}.

\bibitem{Cvetic:2022imb}
M.~Cveti\v{c}, J.~J. Heckman, M.~H\"ubner, and E.~Torres, ``{0-Form, 1-Form and
  2-Group Symmetries via Cutting and Gluing of Orbifolds},''
  \href{http://arxiv.org/abs/2203.10102}{{\ttfamily arXiv:2203.10102
  [hep-th]}}.

\bibitem{DelZotto:2022joo}
M.~Del~Zotto, I.~Garcia~Etxebarria, and S.~Schafer-Nameki, ``{2-Group
  Symmetries and M-Theory},'' \href{http://arxiv.org/abs/2203.10097}{{\ttfamily
  arXiv:2203.10097 [hep-th]}}.

\bibitem{DelZotto:2022ras}
M.~Del~Zotto and I.~Garcia~Etxebarria, ``{Global Structures from the
  Infrared},'' \href{http://arxiv.org/abs/2204.06495}{{\ttfamily
  arXiv:2204.06495 [hep-th]}}.

\bibitem{Bhardwaj:2022yxj}
L.~Bhardwaj, L.~Bottini, S.~Schafer-Nameki, and A.~Tiwari, ``{Non-Invertible
  Higher-Categorical Symmetries},''
  \href{http://arxiv.org/abs/2204.06564}{{\ttfamily arXiv:2204.06564
  [hep-th]}}.

\bibitem{Hayashi:2022fkw}
Y.~Hayashi and Y.~Tanizaki, ``{Non-invertible self-duality defects of
  Cardy-Rabinovici model and mixed gravitational anomaly},''
  \href{http://arxiv.org/abs/2204.07440}{{\ttfamily arXiv:2204.07440
  [hep-th]}}.

\bibitem{Kaidi:2022uux}
J.~Kaidi, G.~Zafrir, and Y.~Zheng, ``{Non-Invertible Symmetries of
  $\mathcal{N}=4$ SYM and Twisted Compactification},''
  \href{http://arxiv.org/abs/2205.01104}{{\ttfamily arXiv:2205.01104
  [hep-th]}}.

\bibitem{Roumpedakis:2022aik}
K.~Roumpedakis, S.~Seifnashri, and S.-H. Shao, ``{Higher Gauging and
  Non-invertible Condensation Defects},''
  \href{http://arxiv.org/abs/2204.02407}{{\ttfamily arXiv:2204.02407
  [hep-th]}}.

\bibitem{Choi:2022jqy}
Y.~Choi, H.~T. Lam, and S.-H. Shao, ``{Non-invertible Global Symmetries in the
  Standard Model},'' \href{http://arxiv.org/abs/2205.05086}{{\ttfamily
  arXiv:2205.05086 [hep-th]}}.

\bibitem{Choi:2022zal}
Y.~Choi, C.~Cordova, P.-S. Hsin, H.~T. Lam, and S.-H. Shao, ``{Non-invertible
  Condensation, Duality, and Triality Defects in 3+1 Dimensions},''
  \href{http://arxiv.org/abs/2204.09025}{{\ttfamily arXiv:2204.09025
  [hep-th]}}.

\bibitem{Arias-Tamargo:2022nlf}
G.~Arias-Tamargo and D.~Rodriguez-Gomez, ``{Non-Invertible Symmetries from
  Discrete Gauging and Completeness of the Spectrum},''
  \href{http://arxiv.org/abs/2204.07523}{{\ttfamily arXiv:2204.07523
  [hep-th]}}.

\bibitem{Cordova:2022ieu}
C.~Cordova and K.~Ohmori, ``{Non-Invertible Chiral Symmetry and Exponential
  Hierarchies},'' \href{http://arxiv.org/abs/2205.06243}{{\ttfamily
  arXiv:2205.06243 [hep-th]}}.

\bibitem{Bhardwaj:2022dyt}
L.~Bhardwaj, M.~Bullimore, A.~E.~V. Ferrari, and S.~Schafer-Nameki,
  ``{Anomalies of Generalized Symmetries from Solitonic Defects},''
  \href{http://arxiv.org/abs/2205.15330}{{\ttfamily arXiv:2205.15330
  [hep-th]}}.

\bibitem{Benedetti:2022zbb}
V.~Benedetti, H.~Casini, and J.~M. Magan, ``{Generalized symmetries and
  Noether's theorem in QFT},''
  \href{http://arxiv.org/abs/2205.03412}{{\ttfamily arXiv:2205.03412
  [hep-th]}}.

\bibitem{Bhardwaj:2022scy}
L.~Bhardwaj and D.~S.~W. Gould, ``{Disconnected 0-Form and 2-Group
  Symmetries},'' \href{http://arxiv.org/abs/2206.01287}{{\ttfamily
  arXiv:2206.01287 [hep-th]}}.

\bibitem{Antinucci:2022eat}
A.~Antinucci, G.~Galati, and G.~Rizi, ``{On Continuous 2-Category Symmetries
  and Yang-Mills Theory},'' \href{http://arxiv.org/abs/2206.05646}{{\ttfamily
  arXiv:2206.05646 [hep-th]}}.

\bibitem{Carta:2022spy}
F.~Carta, S.~Giacomelli, N.~Mekareeya, and A.~Mininno, ``{Dynamical
  consequences of 1-form symmetries and the exceptional Argyres-Douglas
  theories},'' \href{http://dx.doi.org/10.1007/JHEP06(2022)059}{{\em JHEP}
  {\bfseries 06} (2022) 059}, \href{http://arxiv.org/abs/2203.16550}{{\ttfamily
  arXiv:2203.16550 [hep-th]}}.

\bibitem{Apruzzi:2022dlm}
F.~Apruzzi, ``{Higher Form Symmetries TFT in 6d},''
  \href{http://arxiv.org/abs/2203.10063}{{\ttfamily arXiv:2203.10063
  [hep-th]}}.

\bibitem{Heckman:2022suy}
J.~J. Heckman, C.~Lawrie, L.~Lin, H.~Y. Zhang, and G.~Zoccarato, ``{6d SCFTs,
  Center-Flavor Symmetries, and Stiefel--Whitney Compactifications},''
  \href{http://arxiv.org/abs/2205.03411}{{\ttfamily arXiv:2205.03411
  [hep-th]}}.

\bibitem{Baume:2022cot}
F.~Baume, J.~J. Heckman, and C.~Lawrie, ``{Super-Spin Chains for 6D SCFTs},''
  \href{http://arxiv.org/abs/2208.02272}{{\ttfamily arXiv:2208.02272
  [hep-th]}}.

\bibitem{Choi:2022rfe}
Y.~Choi, H.~T. Lam, and S.-H. Shao, ``{Non-invertible Time-reversal
  Symmetry},'' \href{http://arxiv.org/abs/2208.04331}{{\ttfamily
  arXiv:2208.04331 [hep-th]}}.

\bibitem{Bhardwaj:2022lsg}
L.~Bhardwaj, S.~Schafer-Nameki, and J.~Wu, ``{Universal Non-Invertible
  Symmetries},'' \href{http://arxiv.org/abs/2208.05973}{{\ttfamily
  arXiv:2208.05973 [hep-th]}}.

\bibitem{Lin:2022xod}
L.~Lin, D.~Robbins, and E.~Sharpe, ``{Decomposition, condensation defects, and
  fusion},'' \href{http://arxiv.org/abs/2208.05982}{{\ttfamily arXiv:2208.05982
  [hep-th]}}.

\bibitem{Bartsch:2022mpm}
T.~Bartsch, M.~Bullimore, A.~E.~V. Ferrari, and J.~Pearson, ``{Non-invertible
  Symmetries and Higher Representation Theory I},''
  \href{http://arxiv.org/abs/2208.05993}{{\ttfamily arXiv:2208.05993
  [hep-th]}}.

\bibitem{Apruzzi:2022rei}
F.~Apruzzi, I.~Bah, F.~Bonetti, and S.~Schafer-Nameki, ``{Non-Invertible
  Symmetries from Holography and Branes},''
  \href{http://arxiv.org/abs/2208.07373}{{\ttfamily arXiv:2208.07373
  [hep-th]}}.

\bibitem{GarciaEtxebarria:2022vzq}
I.~Garcia~Etxebarria, ``{Branes and Non-Invertible Symmetries},''
  \href{http://arxiv.org/abs/2208.07508}{{\ttfamily arXiv:2208.07508
  [hep-th]}}.

\bibitem{Cherman:2022eml}
A.~Cherman, T.~Jacobson, and M.~Neuzil, ``{1-form symmetry versus large N
  QCD},'' \href{http://arxiv.org/abs/2209.00027}{{\ttfamily arXiv:2209.00027
  [hep-th]}}.

\bibitem{Heckman:2022muc}
J.~J. Heckman, M.~H\"ubner, E.~Torres, and H.~Y. Zhang, ``{The Branes Behind
  Generalized Symmetry Operators},''
  \href{http://arxiv.org/abs/2209.03343}{{\ttfamily arXiv:2209.03343
  [hep-th]}}.

\bibitem{Lu:2022ver}
D.-C. Lu and Z.~Sun, ``{On Triality Defects in 2d CFT},''
  \href{http://arxiv.org/abs/2208.06077}{{\ttfamily arXiv:2208.06077
  [hep-th]}}.

\bibitem{Niro:2022ctq}
P.~Niro, K.~Roumpedakis, and O.~Sela, ``{Exploring Non-Invertible Symmetries in
  Free Theories},'' \href{http://arxiv.org/abs/2209.11166}{{\ttfamily
  arXiv:2209.11166 [hep-th]}}.

\bibitem{Kaidi:2022cpf}
J.~Kaidi, K.~Ohmori, and Y.~Zheng, ``{Symmetry TFTs for Non-Invertible
  Defects},'' \href{http://arxiv.org/abs/2209.11062}{{\ttfamily
  arXiv:2209.11062 [hep-th]}}.

\bibitem{Mekareeya:2022spm}
N.~Mekareeya and M.~Sacchi, ``{Mixed Anomalies, Two-groups, Non-Invertible
  Symmetries, and 3d Superconformal Indices},''
  \href{http://arxiv.org/abs/2210.02466}{{\ttfamily arXiv:2210.02466
  [hep-th]}}.

\bibitem{vanBeest:2022fss}
M.~van Beest, D.~S.~W. Gould, S.~Schafer-Nameki, and Y.-N. Wang, ``{Symmetry
  TFTs for 3d QFTs from M-theory},''
  \href{http://arxiv.org/abs/2210.03703}{{\ttfamily arXiv:2210.03703
  [hep-th]}}.

\bibitem{Antinucci:2022vyk}
A.~Antinucci, F.~Benini, C.~Copetti, G.~Galati, and G.~Rizi, ``{The holography
  of non-invertible self-duality symmetries},''
  \href{http://arxiv.org/abs/2210.09146}{{\ttfamily arXiv:2210.09146
  [hep-th]}}.

\bibitem{Giaccari:2022xgs}
S.~Giaccari and R.~Volpato, ``{A fresh view on string orbifolds},''
  \href{http://arxiv.org/abs/2210.10034}{{\ttfamily arXiv:2210.10034
  [hep-th]}}.

\bibitem{Bashmakov:2022uek}
V.~Bashmakov, M.~Del~Zotto, A.~Hasan, and J.~Kaidi, ``{Non-invertible
  Symmetries of Class $\mathcal{S}$ Theories},''
  \href{http://arxiv.org/abs/2211.05138}{{\ttfamily arXiv:2211.05138
  [hep-th]}}.

\bibitem{Cordova:2022fhg}
C.~Cordova, S.~Hong, S.~Koren, and K.~Ohmori, ``{Neutrino Masses from
  Generalized Symmetry Breaking},''
  \href{http://arxiv.org/abs/2211.07639}{{\ttfamily arXiv:2211.07639
  [hep-ph]}}.

\bibitem{GarciaEtxebarria:2022jky}
I.~Garcia~Etxebarria and N.~Iqbal, ``{A Goldstone theorem for continuous
  non-invertible symmetries},''
  \href{http://arxiv.org/abs/2211.09570}{{\ttfamily arXiv:2211.09570
  [hep-th]}}.

\bibitem{Choi:2022fgx}
Y.~Choi, H.~T. Lam, and S.-H. Shao, ``{Non-invertible Gauss Law and Axions},''
  \href{http://arxiv.org/abs/2212.04499}{{\ttfamily arXiv:2212.04499
  [hep-th]}}.

\bibitem{Robbins:2022wlr}
D.~Robbins, E.~Sharpe, and T.~Vandermeulen, ``{Decomposition, Trivially-Acting
  Symmetries, and Topological Operators},''
  \href{http://arxiv.org/abs/2211.14332}{{\ttfamily arXiv:2211.14332
  [hep-th]}}.

\bibitem{Bhardwaj:2022kot}
L.~Bhardwaj, S.~Schafer-Nameki, and A.~Tiwari, ``{Unifying Constructions of
  Non-Invertible Symmetries},''
  \href{http://arxiv.org/abs/2212.06159}{{\ttfamily arXiv:2212.06159
  [hep-th]}}.

\bibitem{Bhardwaj:2022maz}
L.~Bhardwaj, L.~E. Bottini, S.~Schafer-Nameki, and A.~Tiwari, ``{Non-Invertible
  Symmetry Webs},'' \href{http://arxiv.org/abs/2212.06842}{{\ttfamily
  arXiv:2212.06842 [hep-th]}}.

\bibitem{Bartsch:2022ytj}
T.~Bartsch, M.~Bullimore, A.~E.~V. Ferrari, and J.~Pearson, ``{Non-invertible
  Symmetries and Higher Representation Theory II},''
  \href{http://arxiv.org/abs/2212.07393}{{\ttfamily arXiv:2212.07393
  [hep-th]}}.

\bibitem{Gaiotto:2020iye}
D.~Gaiotto and J.~Kulp, ``{Orbifold groupoids},''
  \href{http://dx.doi.org/10.1007/JHEP02(2021)132}{{\em JHEP} {\bfseries 02}
  (2021) 132}, \href{http://arxiv.org/abs/2008.05960}{{\ttfamily
  arXiv:2008.05960 [hep-th]}}.

\bibitem{Agrawal:2015dbf}
P.~Agrawal, J.~Fan, B.~Heidenreich, M.~Reece, and M.~Strassler, ``{Experimental
  Considerations Motivated by the Diphoton Excess at the LHC},''
\href{http://arxiv.org/abs/1512.05775}{{\ttfamily arXiv:1512.05775 [hep-ph]}}.

\bibitem{Robbins:2021ibx}
D.~G. Robbins, E.~Sharpe, and T.~Vandermeulen, ``{Quantum symmetries in
  orbifolds and decomposition},''
  \href{http://dx.doi.org/10.1007/JHEP02(2022)108}{{\em JHEP} {\bfseries 02}
  (2022) 108}, \href{http://arxiv.org/abs/2107.12386}{{\ttfamily
  arXiv:2107.12386 [hep-th]}}.

\bibitem{Robbins:2021xce}
D.~G. Robbins, E.~Sharpe, and T.~Vandermeulen, ``{Anomaly resolution via
  decomposition},'' \href{http://dx.doi.org/10.1142/S0217751X21502201}{{\em
  Int. J. Mod. Phys. A} {\bfseries 36} no.~29, (2021) 2150220},
  \href{http://arxiv.org/abs/2107.13552}{{\ttfamily arXiv:2107.13552
  [hep-th]}}.

\bibitem{Huang:2021zvu}
T.-C. Huang, Y.-H. Lin, and S.~Seifnashri, ``{Construction of two-dimensional
  topological field theories with non-invertible symmetries},''
  \href{http://dx.doi.org/10.1007/JHEP12(2021)028}{{\em JHEP} {\bfseries 12}
  (2021) 028}, \href{http://arxiv.org/abs/2110.02958}{{\ttfamily
  arXiv:2110.02958 [hep-th]}}.

\bibitem{Inamura:2021szw}
K.~Inamura, ``{On lattice models of gapped phases with fusion category
  symmetries},'' \href{http://dx.doi.org/10.1007/JHEP03(2022)036}{{\em JHEP}
  {\bfseries 03} (2022) 036}, \href{http://arxiv.org/abs/2110.12882}{{\ttfamily
  arXiv:2110.12882 [cond-mat.str-el]}}.

\bibitem{Cherman:2021nox}
A.~Cherman, T.~Jacobson, and M.~Neuzil, ``{Universal Deformations},''
  \href{http://dx.doi.org/10.21468/SciPostPhys.12.4.116}{{\em SciPost Phys.}
  {\bfseries 12} no.~4, (2022) 116},
  \href{http://arxiv.org/abs/2111.00078}{{\ttfamily arXiv:2111.00078
  [hep-th]}}.

\bibitem{Sharpe:2022ene}
E.~Sharpe, ``{An introduction to decomposition},''
  \href{http://arxiv.org/abs/2204.09117}{{\ttfamily arXiv:2204.09117
  [hep-th]}}.

\bibitem{Bashmakov:2022jtl}
V.~Bashmakov, M.~Del~Zotto, and A.~Hasan, ``{On the 6d Origin of Non-invertible
  Symmetries in 4d},'' \href{http://arxiv.org/abs/2206.07073}{{\ttfamily
  arXiv:2206.07073 [hep-th]}}.

\bibitem{Inamura:2022lun}
K.~Inamura, ``{Fermionization of fusion category symmetries in 1+1
  dimensions},'' \href{http://arxiv.org/abs/2206.13159}{{\ttfamily
  arXiv:2206.13159 [cond-mat.str-el]}}.

\bibitem{Damia:2022bcd}
J.~A. Damia, R.~Argurio, and E.~Garcia-Valdecasas, ``{Non-Invertible Defects in
  5d, Boundaries and Holography},''
  \href{http://arxiv.org/abs/2207.02831}{{\ttfamily arXiv:2207.02831
  [hep-th]}}.

\bibitem{Lin:2022dhv}
Y.-H. Lin, M.~Okada, S.~Seifnashri, and Y.~Tachikawa, ``{Asymptotic density of
  states in 2d CFTs with non-invertible symmetries},''
  \href{http://arxiv.org/abs/2208.05495}{{\ttfamily arXiv:2208.05495
  [hep-th]}}.

\bibitem{Burbano:2021loy}
I.~M. Burbano, J.~Kulp, and J.~Neuser, ``{Duality Defects in $E_8$},''
  \href{http://arxiv.org/abs/2112.14323}{{\ttfamily arXiv:2112.14323
  [hep-th]}}.

\bibitem{Damia:2022rxw}
J.~A. Damia, R.~Argurio, and L.~Tizzano, ``{Continuous Generalized Symmetries
  in Three Dimensions},'' \href{http://arxiv.org/abs/2206.14093}{{\ttfamily
  arXiv:2206.14093 [hep-th]}}.

\bibitem{Cordova:2022ruw}
C.~Cordova, T.~T. Dumitrescu, K.~Intriligator, and S.-H. Shao, ``{Snowmass
  White Paper: Generalized Symmetries in Quantum Field Theory and Beyond},'' in
  {\em {2022 Snowmass Summer Study}}.
\newblock 5, 2022.
\newblock \href{http://arxiv.org/abs/2205.09545}{{\ttfamily arXiv:2205.09545
  [hep-th]}}.

\bibitem{Aharony:1998qu}
O.~Aharony and E.~Witten, ``{Anti-de Sitter space and the center of the gauge
  group},'' \href{http://dx.doi.org/10.1088/1126-6708/1998/11/018}{{\em JHEP}
  {\bfseries 11} (1998) 018},
  \href{http://arxiv.org/abs/hep-th/9807205}{{\ttfamily arXiv:hep-th/9807205}}.

\bibitem{Klebanov:1998hh}
I.~R. Klebanov and E.~Witten, ``{Superconformal field theory on three-branes at
  a Calabi-Yau singularity},''
  \href{http://dx.doi.org/10.1016/S0550-3213(98)00654-3}{{\em Nucl. Phys. B}
  {\bfseries 536} (1998) 199--218},
  \href{http://arxiv.org/abs/hep-th/9807080}{{\ttfamily arXiv:hep-th/9807080}}.

\bibitem{Uranga:1998vf}
A.~M. Uranga, ``{Brane configurations for branes at conifolds},''
  \href{http://dx.doi.org/10.1088/1126-6708/1999/01/022}{{\em JHEP} {\bfseries
  01} (1999) 022}, \href{http://arxiv.org/abs/hep-th/9811004}{{\ttfamily
  arXiv:hep-th/9811004}}.

\bibitem{Aharony:1997ju}
O.~Aharony and A.~Hanany, ``{Branes, superpotentials and superconformal fixed
  points},'' \href{http://dx.doi.org/10.1016/S0550-3213(97)00472-0}{{\em Nucl.
  Phys. B} {\bfseries 504} (1997) 239--271},
  \href{http://arxiv.org/abs/hep-th/9704170}{{\ttfamily arXiv:hep-th/9704170}}.

\bibitem{Pantev:2009de}
T.~Pantev and M.~Wijnholt, ``{Hitchin's Equations and M-Theory
  Phenomenology},''
  \href{http://dx.doi.org/10.1016/j.geomphys.2011.02.014}{{\em J. Geom. Phys.}
  {\bfseries 61} (2011) 1223--1247},
  \href{http://arxiv.org/abs/0905.1968}{{\ttfamily arXiv:0905.1968 [hep-th]}}.

\bibitem{Tachikawa:2018njr}
Y.~Tachikawa and K.~Yonekura, ``{Why are fractional charges of orientifolds
  compatible with Dirac quantization?},''
  \href{http://dx.doi.org/10.21468/SciPostPhys.7.5.058}{{\em SciPost Phys.}
  {\bfseries 7} no.~5, (2019) 058},
  \href{http://arxiv.org/abs/1805.02772}{{\ttfamily arXiv:1805.02772
  [hep-th]}}.

\bibitem{Dierigl:2022reg}
M.~Dierigl, J.~J. Heckman, M.~Montero, and E.~Torres, ``{IIB Explored:
  Reflection 7-Branes},'' \href{http://arxiv.org/abs/2212.05077}{{\ttfamily
  arXiv:2212.05077 [hep-th]}}.

\bibitem{Lawrence:1998ja}
A.~E. Lawrence, N.~Nekrasov, and C.~Vafa, ``{On conformal field theories in
  four-dimensions},''
  \href{http://dx.doi.org/10.1016/S0550-3213(98)00495-7}{{\em Nucl. Phys. B}
  {\bfseries 533} (1998) 199--209},
  \href{http://arxiv.org/abs/hep-th/9803015}{{\ttfamily arXiv:hep-th/9803015}}.

\bibitem{Kachru:1998ys}
S.~Kachru and E.~Silverstein, ``{4-D conformal theories and strings on
  orbifolds},'' \href{http://dx.doi.org/10.1103/PhysRevLett.80.4855}{{\em Phys.
  Rev. Lett.} {\bfseries 80} (1998) 4855--4858},
  \href{http://arxiv.org/abs/hep-th/9802183}{{\ttfamily arXiv:hep-th/9802183}}.

\bibitem{Garcia-Etxebarria:2016bpb}
I.~Garcia~Etxebarria and B.~Heidenreich, ``{S-duality in N=1 orientifold
  SCFTs},'' \href{http://dx.doi.org/10.1002/prop.201700013}{{\em Fortsch.
  Phys.} {\bfseries 65} no.~3-4, (2017) 1700013},
  \href{http://arxiv.org/abs/1612.00853}{{\ttfamily arXiv:1612.00853
  [hep-th]}}.

\bibitem{Freed:2012bs}
D.~S. Freed and C.~Teleman, ``{Relative quantum field theory},''
  \href{http://dx.doi.org/10.1007/s00220-013-1880-1}{{\em Commun. Math. Phys.}
  {\bfseries 326} (2014) 459--476},
  \href{http://arxiv.org/abs/1212.1692}{{\ttfamily arXiv:1212.1692 [hep-th]}}.

\bibitem{Freed:2022qnc}
D.~S. Freed, G.~W. Moore, and C.~Teleman, ``{Topological symmetry in quantum
  field theory},'' \href{http://arxiv.org/abs/2209.07471}{{\ttfamily
  arXiv:2209.07471 [hep-th]}}.

\bibitem{Belov:2006jd}
D.~Belov and G.~W. Moore, ``{Holographic Action for the Self-Dual Field},''
  \href{http://arxiv.org/abs/hep-th/0605038}{{\ttfamily arXiv:hep-th/0605038}}.

\bibitem{Belov:2006xj}
D.~M. Belov and G.~W. Moore, ``{Type II Actions from 11-Dimensional
  Chern-Simons Theories},''
  \href{http://arxiv.org/abs/hep-th/0611020}{{\ttfamily arXiv:hep-th/0611020}}.

\bibitem{Witten:1998wy}
E.~Witten, ``{AdS / CFT correspondence and topological field theory},''
  \href{http://dx.doi.org/10.1088/1126-6708/1998/12/012}{{\em JHEP} {\bfseries
  12} (1998) 012}, \href{http://arxiv.org/abs/hep-th/9812012}{{\ttfamily
  arXiv:hep-th/9812012}}.

\bibitem{Bergman:2022otk}
O.~Bergman and S.~Hirano, ``{The holography of duality in ${\cal N}=4$
  Super-Yang-Mills theory},'' \href{http://arxiv.org/abs/2208.09396}{{\ttfamily
  arXiv:2208.09396 [hep-th]}}.

\bibitem{Weigand:2018rez}
T.~Weigand, ``{F-theory},'' {\em PoS} {\bfseries TASI2017} (2018) 016,
  \href{http://arxiv.org/abs/1806.01854}{{\ttfamily arXiv:1806.01854
  [hep-th]}}.

\bibitem{Hubner:2022kxr}
M.~Hubner, D.~R. Morrison, S.~Schafer-Nameki, and Y.-N. Wang, ``{Generalized
  Symmetries in F-theory and the Topology of Elliptic Fibrations},''
  \href{http://arxiv.org/abs/2203.10022}{{\ttfamily arXiv:2203.10022
  [hep-th]}}.

\bibitem{Hanany:1996ie}
A.~Hanany and E.~Witten, ``{Type IIB superstrings, BPS monopoles, and
  three-dimensional gauge dynamics},''
  \href{http://dx.doi.org/10.1016/S0550-3213(97)00157-0}{{\em Nucl. Phys. B}
  {\bfseries 492} (1997) 152--190},
  \href{http://arxiv.org/abs/hep-th/9611230}{{\ttfamily arXiv:hep-th/9611230}}.

\bibitem{Kapustin:2014gua}
A.~Kapustin and N.~Seiberg, ``{Coupling a QFT to a TQFT and Duality},''
  \href{http://dx.doi.org/10.1007/JHEP04(2014)001}{{\em JHEP} {\bfseries 04}
  (2014) 001}, \href{http://arxiv.org/abs/1401.0740}{{\ttfamily arXiv:1401.0740
  [hep-th]}}.

\bibitem{Douglas:1996sw}
M.~R. Douglas and G.~W. Moore, ``{D-branes, quivers, and ALE instantons},''
  \href{http://arxiv.org/abs/hep-th/9603167}{{\ttfamily arXiv:hep-th/9603167}}.

\bibitem{Franco:2005rj}
S.~Franco, A.~Hanany, K.~D. Kennaway, D.~Vegh, and B.~Wecht, ``{Brane dimers
  and quiver gauge theories},''
  \href{http://dx.doi.org/10.1088/1126-6708/2006/01/096}{{\em JHEP} {\bfseries
  01} (2006) 096}, \href{http://arxiv.org/abs/hep-th/0504110}{{\ttfamily
  arXiv:hep-th/0504110}}.

\bibitem{Yamazaki:2008bt}
M.~Yamazaki, ``{Brane Tilings and Their Applications},''
  \href{http://dx.doi.org/10.1002/prop.200810536}{{\em Fortsch. Phys.}
  {\bfseries 56} (2008) 555--686},
  \href{http://arxiv.org/abs/0803.4474}{{\ttfamily arXiv:0803.4474 [hep-th]}}.

\bibitem{Halmagyi:2004ju}
N.~Halmagyi, C.~Romelsberger, and N.~P. Warner, ``{Inherited duality and quiver
  gauge theory},'' \href{http://dx.doi.org/10.4310/ATMP.2006.v10.n2.a1}{{\em
  Adv. Theor. Math. Phys.} {\bfseries 10} no.~2, (2006) 159--179},
  \href{http://arxiv.org/abs/hep-th/0406143}{{\ttfamily arXiv:hep-th/0406143}}.

\bibitem{Freed:2006yc}
D.~S. Freed, G.~W. Moore, and G.~Segal, ``{Heisenberg Groups and Noncommutative
  Fluxes},'' \href{http://dx.doi.org/10.1016/j.aop.2006.07.014}{{\em Annals
  Phys.} {\bfseries 322} (2007) 236--285},
  \href{http://arxiv.org/abs/hep-th/0605200}{{\ttfamily arXiv:hep-th/0605200}}.

\bibitem{Freed:2006ya}
D.~S. Freed, G.~W. Moore, and G.~Segal, ``{The Uncertainty of Fluxes},''
  \href{http://dx.doi.org/10.1007/s00220-006-0181-3}{{\em Commun. Math. Phys.}
  {\bfseries 271} (2007) 247--274},
  \href{http://arxiv.org/abs/hep-th/0605198}{{\ttfamily arXiv:hep-th/0605198}}.

\bibitem{Monnier:2012xd}
S.~Monnier, ``{Canonical quadratic refinements of cohomological pairings from
  functorial lifts of the Wu class},''
  \href{http://arxiv.org/abs/1208.1540}{{\ttfamily arXiv:1208.1540 [math.AT]}}.

\bibitem{Gukov:1998kn}
S.~Gukov, M.~Rangamani, and E.~Witten, ``{Dibaryons, strings and branes in AdS
  orbifold models},''
  \href{http://dx.doi.org/10.1088/1126-6708/1998/12/025}{{\em JHEP} {\bfseries
  12} (1998) 025}, \href{http://arxiv.org/abs/hep-th/9811048}{{\ttfamily
  arXiv:hep-th/9811048}}.

\bibitem{Verlinde:1988sn}
E.~P. Verlinde, ``{Fusion Rules and Modular Transformations in 2D Conformal
  Field Theory},'' \href{http://dx.doi.org/10.1016/0550-3213(88)90603-7}{{\em
  Nucl. Phys. B} {\bfseries 300} (1988) 360--376}.

\bibitem{Tachikawa:2013hya}
Y.~Tachikawa, ``{On the 6d origin of discrete additional data of 4d gauge
  theories},'' \href{http://dx.doi.org/10.1007/JHEP05(2014)020}{{\em JHEP}
  {\bfseries 05} (2014) 020}, \href{http://arxiv.org/abs/1309.0697}{{\ttfamily
  arXiv:1309.0697 [hep-th]}}.

\bibitem{Heckman:2013pva}
J.~J. Heckman, D.~R. Morrison, and C.~Vafa, ``{On the Classification of 6D
  SCFTs and Generalized ADE Orbifolds},''
  \href{http://dx.doi.org/10.1007/JHEP05(2014)028}{{\em JHEP} {\bfseries 05}
  (2014) 028}, \href{http://arxiv.org/abs/1312.5746}{{\ttfamily arXiv:1312.5746
  [hep-th]}}. [Erratum: JHEP 06, 017 (2015)].

\bibitem{Heckman:2018jxk}
J.~J. Heckman and T.~Rudelius, ``{Top Down Approach to 6D SCFTs},''
  \href{http://dx.doi.org/10.1088/1751-8121/aafc81}{{\em J. Phys. A} {\bfseries
  52} no.~9, (2019) 093001}, \href{http://arxiv.org/abs/1805.06467}{{\ttfamily
  arXiv:1805.06467 [hep-th]}}.

\bibitem{Ooguri:1995wj}
H.~Ooguri and C.~Vafa, ``{Two-dimensional black hole and singularities of CY
  manifolds},'' \href{http://dx.doi.org/10.1016/0550-3213(96)00008-9}{{\em
  Nucl. Phys. B} {\bfseries 463} (1996) 55--72},
  \href{http://arxiv.org/abs/hep-th/9511164}{{\ttfamily arXiv:hep-th/9511164}}.

\bibitem{baraglia2015fourier}
D.~Baraglia and P.~Hekmati, ``A fourier--mukai approach to the k-theory of
  compact lie groups,'' {\em Advances in Mathematics} {\bfseries 269} (2015)
  335--345.

\bibitem{Maldacena:2001xj}
J.~M. Maldacena, G.~W. Moore, and N.~Seiberg, ``{D-brane instantons and
  K-theory charges},''
  \href{http://dx.doi.org/10.1088/1126-6708/2001/11/062}{{\em JHEP} {\bfseries
  11} (2001) 062}, \href{http://arxiv.org/abs/hep-th/0108100}{{\ttfamily
  arXiv:hep-th/0108100}}.

\bibitem{Maldacena:2001ss}
J.~M. Maldacena, G.~W. Moore, and N.~Seiberg, ``{D-brane charges in five-brane
  backgrounds},'' \href{http://dx.doi.org/10.1088/1126-6708/2001/10/005}{{\em
  JHEP} {\bfseries 10} (2001) 005},
  \href{http://arxiv.org/abs/hep-th/0108152}{{\ttfamily arXiv:hep-th/0108152}}.

\bibitem{Alekseev:2000fd}
A.~Y. Alekseev, A.~Recknagel, and V.~Schomerus, ``{Brane dynamics in background
  fluxes and noncommutative geometry},''
  \href{http://dx.doi.org/10.1088/1126-6708/2000/05/010}{{\em JHEP} {\bfseries
  05} (2000) 010}, \href{http://arxiv.org/abs/hep-th/0003187}{{\ttfamily
  arXiv:hep-th/0003187}}.

\bibitem{Maldacena:1998bw}
J.~M. Maldacena and A.~Strominger, ``{AdS$_{3}$ Black Holes and a Stringy
  Exclusion Principle},''
  \href{http://dx.doi.org/10.1088/1126-6708/1998/12/005}{{\em JHEP} {\bfseries
  12} (1998) 005}, \href{http://arxiv.org/abs/hep-th/9804085}{{\ttfamily
  arXiv:hep-th/9804085}}.

\bibitem{Douglas:1995bn}
M.~R. Douglas, ``{Branes within branes},'' {\em NATO Sci. Ser. C} {\bfseries
  520} (1999) 267--275, \href{http://arxiv.org/abs/hep-th/9512077}{{\ttfamily
  arXiv:hep-th/9512077}}.

\bibitem{hori2003mirror}
K.~Hori, S.~Katz, A.~Klemm, R.~Pandharipande, R.~Thomas, C.~Vafa, R.~Vakil, and
  E.~Zaslow, {\em {Mirror symmetry}}, vol.~1 of {\em Clay mathematics
  monographs}.
\newblock AMS, Providence, USA, 2003.

\end{thebibliography}\endgroup

\end{document}